\documentclass[acmtog]{acmart}

\AtBeginDocument{%
  \providecommand\BibTeX{{%
    \normalfont B\kern-0.5em{\scshape i\kern-0.25em b}\kern-0.8em\TeX}}}

\acmJournal{TOG} 

\usepackage{xurl}
\usepackage{xcolor}
\usepackage{bbold}
\usepackage{tabularx}
\usepackage{multirow}
\usepackage{soul}
\usepackage{graphicx}
\usepackage{float}
\usepackage{copyrightbox}
\usepackage{subcaption}
\usepackage{enumitem}
\newcommand{\edit}[1]{#1}

\newcommand{\R}{\mathbb{R}}

\newcommand{\z}{\mathbf{z}}

\newcommand{\x}{\mathbf{x}}
\newcommand{\n}{\mathbf{n}}

\newcolumntype{Y}{>{\centering\arraybackslash}X}

\newcommand{\ourgithub}{{\url{https://github.com/hjoonpark/3d-sim-super-res.git}}}

\newcommand{\timeHiSim}{{6.22s}}
\newcommand{\timeLoSim}{{0.033s}}
\newcommand{\timeInfer}{{0.0209s}}
\newcommand{\fpsInfer}{{47.82 FPS}}
\newcommand{\fpsLoSim}{{30.06 FPS}}
\newcommand{\fpsHiSim}{{0.16 FPS}}
\newcommand{\fpsEndToEnd}{{18.46 FPS}}
\newcommand{\timeEndToEnd}{{0.054 FPS}}
\newcommand{\xfasterSim}{{188$\times$}}
\newcommand{\xfasterEndToEnd}{{115$\times$}}

\newcommand{\fpsEndToEndCoarser}{{28.04 FPS}}
\newcommand{\fpsSimCoarser}{{67.79 FPS}}
\newcommand{\fpsInferCoarser}{{47.82 FPS}}

\definecolor{peach}{rgb}{ 0.943, 0.188, 0.526}
\definecolor{plum}{rgb}{ 0.858, 0.188, 0.478}
\definecolor{muted_navy_blue}{RGB}{63, 75, 166}
\definecolor{muted_sky_blue}{RGB}{134,166,213}
\definecolor{federal_blue}{RGB}{0,96,240}
\definecolor{regulation_red}{RGB}{226, 20, 79}
\definecolor{federal_gold}{RGB}{240, 212, 14}
\definecolor{chartreuse}{RGB}{178, 204, 0}
\definecolor{blue}{RGB}{0, 0, 255}

\newcommand{\nvidia}{{}}

\begin{document}

\title{Near-realtime Facial Animation by Deep 3D Simulation Super-Resolution}


\author{Hyojoon Park}
\email{hpark376@wisc.edu}
\orcid{0000-0002-3796-9777}
\affiliation{%
  \institution{University of Wisconsin-Madison}
  \city{Madison}
  \state{Wisconsin}
  \country{USA}
}

\author{Sangeetha Grama Srinivasan}
\email{sgsrinivasa2@wisc.edu}
\orcid{0000-0001-6508-7256}
\affiliation{%
  \institution{University of Wisconsin-Madison}
  \city{Madison}
  \state{Wisconsin}
  \country{USA}
}

\author{Matthew Cong}
\email{mdcong@cs.stanford.edu}
\orcid{0000-0003-2956-2050}
\affiliation{%
  \institution{NVIDIA}
  \city{San Francisco}
  \state{California}
  \country{USA}
}

\author{Doyub Kim}
\email{doyubkim@gmail.com}
\orcid{0000-0002-8932-5519}
\affiliation{%
  \institution{NVIDIA}
  \city{San Francisco}
  \state{California}
  \country{USA}
}

\author{Byungsoo Kim}
\email{contact.byungsoo@gmail.com}
\orcid{0000-0003-4482-8363}
\affiliation{%
  \institution{NVIDIA}
  \country{Switzerland}
}

\author{Jonathan Swartz}
\email{jonathanswartz@gmail.com}
\orcid{0000-0003-1959-6396}
\affiliation{%
  \institution{NVIDIA}
  \city{San Francisco}
  \state{California}
  \country{USA}
}

\author{Ken Museth}
\email{ken.museth@gmail.com}
\orcid{0000-0002-9926-780X}
\affiliation{%
  \institution{NVIDIA}
  \city{San Francisco}
  \state{California}
  \country{USA}
}

\author{Eftychios Sifakis}
\email{sifakis@cs.wisc.edu}
\orcid{0000-0001-5608-3085}
\affiliation{%
  \institution{University of Wisconsin Madison}
  \city{San Francisco}
  \state{California}
  \country{USA}
}
\affiliation{%
  \institution{NVIDIA}
  \city{San Francisco}
  \state{California}
  \country{USA}
}

\begin{abstract}
  We present a neural network-based simulation super-resolution framework that can efficiently and realistically enhance a facial performance produced by a low-cost, real-time physics-based simulation to a level of detail that closely approximates that of a reference-quality off-line simulator with much higher resolution ($27\times$ element count in our examples) and accurate physical modeling. Our approach is rooted in our ability to construct a training set of paired frames, from the low- and high-resolution simulators respectively, that are in semantic correspondence with each other. We use face animation as an exemplar of such a simulation domain, where creating this semantic congruence is achieved by simply dialing in the same muscle actuation controls and skeletal pose in the two simulators. Our proposed neural network super-resolution framework generalizes from this training set to unseen expressions, compensates for modeling discrepancies between the two simulations due to limited resolution or cost-cutting approximations in the real-time variant, and does not require any semantic descriptors or parameters to be provided as input, other than the result of the real-time simulation. We evaluate the efficacy of our pipeline on a variety of expressive performances and provide comparisons and ablation experiments for plausible variations and alternatives to our proposed scheme. 
  Our code is available at \edit{\ourgithub{}.}
\end{abstract}




\begin{teaserfigure}
    \copyrightbox[r]
    {\includegraphics[width=\textwidth]{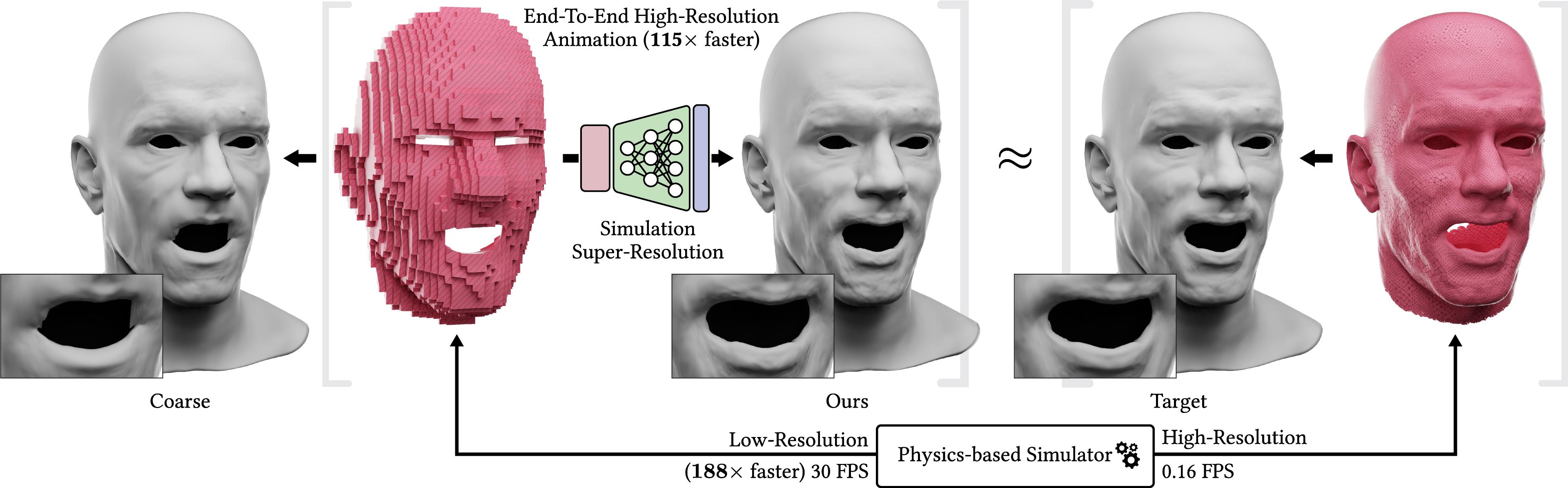}}
        {\nvidia{}}
  \caption{
  From left to right: Facial animation resulting from low-resolution simulation (Coarse), embedding low-resolution 3D mesh (red) simulating at \fpsLoSim{}, result of our simulation super-resolution framework (Ours), result from a corresponding off-line high-resolution simulation (Target), conforming high-resolution 3D mesh simulating at \fpsHiSim{}. Note the similarities between our result (Ours) and that from the high-resolution simulation (Target), which both differ from the result obtained by the low-resolution simulation (Coarse), especially around the mouth and chin area. Our simulation super-resolution achieves an effective \fpsEndToEnd{}, i.e. \xfasterEndToEnd{} faster than the high-resolution simulation. The low- and high-resolution meshes have 73 thousand and 1.9 million tetrahedra respectively, corresponding to a coarsening of $27\times$, and both simulations are accelerated with CUDA. \copyright NVIDIA
  }
\label{fig:teaser}
\end{teaserfigure}
\maketitle

\section{Introduction}
\label{sec:intro}
Physics-based simulation is widely used to drive animations of both human bodies and faces. However, in order to obtain the highest levels of visual quality and realism, traditional simulation pipelines based on anatomic first principles resort to costly design choices. Detailed specifications of geometry and materials are essential, including the muscle and tendon shapes and attachment; bone geometry and motion; and constitutive properties of soft tissue and skin. Collision and frictional contact are ubiquitous in faces, and the resolution of such effects is dependent on mesh detail and the sophistication of detection and response algorithms. Finally, recreating intricate local shapes to match performance details from real actors may impose further directability demands on the simulation pipeline. Such feature demands in conjunction with the sheer geometric mesh resolution necessary for detailed facial expressions often place reference-quality face simulation well beyond the cost that would allow for real-time performance. 

This paper explores an alternative approach to achieving faithful and accurate facial animation at a much reduced execution cost, ideally as close as possible to real-time. Our method (Figure \ref{fig:teaser}) seeks to convincingly approximate a full, high-resolution 3D simulation with the combination of a simulator that uses lower resolution and model simplifications, paired with a deep neural network that boosts the resolution, detail, and accuracy of this coarse simulated deformation. Our simulation super-resolution module is trained on a dataset of coordinated performances crafted using the high- and low-resolution face simulators and generalizes to novel performances by boosting the output of the low-resolution simulator to the quality anticipated from its high-resolution counterpart. 

We aspire to create the best preconditions for the success of such a super-resolution module by focusing our attention on types of physics-based simulations where it may be possible to craft animations from the low- and high-resolution simulators that have strong \emph{semantic correspondence} on a frame-by-frame basis. In other words, we look for types of simulation where it might be possible to infer -- at some level of abstraction -- what the fine-resolution simulation would want to do, by observing what the low-resolution simulator was able to do. Face simulation is a good exemplar of this concept; regardless of resolution, the same core drivers of deformation can be seen as being present in both cases: the action of muscles, and the kinematic state of skeletal bones and other collision objects. This allows us to create a training set by simply dialing in the same control parameters for these driving factors of simulations both in the low- and high-resolution models. Hence, we can hope that this semantic correspondence can be learned in a super-resolution neural network that generalizes this semantic correspondence between resolutions to unseen performances.

We highlight that even ``semantically corresponding'' simulated poses from the respective simulators described above can be quite different. In particular, the low-resolution result can deviate significantly from the mere downsampling of the high-resolution simulation, with discrepancies extending beyond high-frequency details. There are at least three core causes of such discrepancy: First, and most obvious, the reduced mesh resolution of the coarser simulation will be unable to resolve fine geometric features such as localized folds, wrinkles, and bulges that the fine-resolution mesh would capture. Second, the fact that governing physics and topology have to be represented using a coarser discretization may create bulk deviations from the expected behavior of the continuous medium. For example, the action of thin muscles might have to be dissipated over larger elements, reducing the crispness of their action. Fine topological features like the corners of the lips may be under-resolved, especially if at lower resolution we opt for an embedding simulation mesh that does not conform to the model boundary. Non-conforming embedded simulation offers well-conditioned elements and improved convergence that is attractive for real-time performance, but it also leads to a crude first-order approximation of the material volume for elements on the model boundary, leading to artificial stiffness and resistance to bending. The third and final contributor to bulk discrepancy between resolutions could be conscious design choices for the sake of interactive performance; for example, we may choose to perform elaborate contact/collision processing in our reference-quality simulation but forego collision processing altogether in the low-resolution simulator (as in our examples). Thus, our super-resolution module must account for much more than localized high-frequency deformation details and should compensate for all factors (mesh resolution, discretization non-convergence, and physical simplifications) of bulk differences between the two simulation resolutions. 

Our objective is to build a framework capable of producing high-accuracy animations without incurring the cost of simulations on high-resolution meshes. We achieve this by training a deep neural network to act as a super-resolution upsampler of simulations performed on a coarser 3D mesh. In practice, this allows for real-time simulations of facial animations that preserve many of the qualities associated with much slower high-resolution simulations.

We simulate a coarse low-resolution face mesh with significantly fewer mesh elements allowing for real-time simulations and reconstruct the high-resolution details learned from data. Our upsampling module accounts for both high-frequency details and bulk differences between resolutions, responses to dynamics and external forces, and can also approximate a degree of collision response even if collision handling is omitted from the low-resolution simulator. Our end-to-end animation attains near-realtime at \fpsEndToEnd{} from \fpsLoSim{} simulation and \fpsInfer{} upsampling. We also emphasize that true real-time end-to-end animation (i.e., 24 or more FPS) is attainable by scaling down to coarser representations at a modest sacrifice of upsampling accuracy (discussed more in Section \ref{appendix:coarser}).

Previous efforts to accelerate physics-based simulations of deforming elastic bodies have focused on building faster numerical methods \citep{hauth2001high, kharevych2006geometric, stern2009implicit, su2013energy}, employing alternative constraint-based formulations such as Position Based Dynamics \citep{muller2007position, macklin2016xpbd, bender2013position} and its variants \citep{bouaziz2014projective, liu2013fast, stam2009nucleus}, and other techniques such as adaptively computing higher resolutions only when needed \citep{bergou2007tracks}.
However, given the real-time performance afforded by regular, embedded models for low-resolution simulations and the fast inferencing time of deep models, our framework can reconstruct high-resolution facial expressions faster and with reduced developmental effort. 

\edit{We extend the concept of super-resolution to the domain of physics-based \emph{simulation}, contrasting with most prior applications of this process to purely geometric 3D models without regard to the fact the data originated from simulation.}
We summarize our core contributions as follows:

\textbf{$\bullet$} We demonstrate a neural network-based pipeline that can convincingly approximate a high-resolution facial simulation, using as input a real-time low-resolution approximate simulation and a fast inference step that performs the resolution boost.
We show that this pipeline can robustly compensate for discrepancies between the two simulation resolutions extending beyond localized high-frequency deformation details. 

\textbf{$\bullet$} We identify the opportunity to create a training set for our super-resolution module with a high degree of semantic correspondence between low- and high-resolution simulation frames, by giving the two simulators the same anatomical controls of muscle activations and bone kinematics.
    
\textbf{$\bullet$} We demonstrate near-realtime performance of the end-to-end pipeline, and a robust ability to generalize to expressions not in the training set.  We can even demonstrate this ability on deformations that extend beyond the parametric space used in the simulations that generated the training set (e.g. dynamics, external forces, collisions, or constraints not present in the training data).

\section{Related Work}
\label{sec:lit-survey}
\subsection{3D super-resolution}
Our framework shares the motivation (and also adopts the terminology) of \emph{super-resolution} approaches that operate in the domain of images. 
Super-resolution (SR) was initially introduced for 2D images to restore high-resolution images from their low-resolution observations \citep{nasrollahi2014super}. 
SR for 3D shapes shares similar characteristics with several relevant research areas. 

\paragraph{\textbf{Surface reconstruction}}
A closely related and widely studied area is a surface reconstruction from sampled points \citep{alexa2003computing}. Prior research can be classified into two groups: global and local methods. Global methods are more robust than local methods against noise and sparsity of the observations but at the cost of reconstruction accuracy, and vice versa. Global methods include, namely, the radial basis function (RBF) \citep{carr2001reconstruction, turk2002modelling, ohtake20053d} and Poisson problem \citep{kazhdan2006poisson, kazhdan2013screened}. On the other hand, local methods include MLS \citep{alexa2001point, alexa2003computing, fleishman2005robust}, fitting of piecewise functions \citep{ohtake2005sparse, nagai2009smoothing}, and construction of signed distance functions \citep{curless1996volumetric}. A comprehensive review of this topic can be found in \cite{berger2017survey}.

\paragraph{\textbf{Point cloud upsampling}}
Another widely studied area that resembles several aspects of our work is point cloud upsampling, which has been actively explored by both traditional and learning-based methods for many applications such as robotics, autonomous cars, and rendering \citep{zhang2022point}.
A pioneering approach is PU-Net \cite{yu2018pu} which operates on patches to learn per-point multi-level features and expands them through a multi-branch convolution network.
Follow-up works include EC-Net \citep{yu2018ec}, 3PU \cite{yifan2019patch}, PU-GAN \cite{li2019pu}, PUGeo-Net \cite{qian2020pugeo}, and PU-GCN \cite{qian2021pu}. While all the previous works supported only a fixed integer ratio of upsampling, Meta-PU \cite{ye2021meta} pioneered in adapting to arbitrary non-integer upsampling ratios.

 Although we similarly adopt point cloud representations, we do not assume the input and output points are from the same geometry which motivates us to carefully design the upsampling method to adapt to the geometric discrepancy between the low- and high-resolution points and arbitrary non-integer upsampling ratios (Section \ref{ss:upsampling} and more discussion in Section \ref{ss:abl_architecture}).

\paragraph{\textbf{3D face super-resolution}}
Existing works focusing on 3D face SR can be categorized as either method- or learning-based methods.
Method-based works include registration and filtering of the 3D acquisitions \citep{berretti2012superfaces, berretti2014face, bondi2016reconstructing}, whereas learning-based methods map from a low-resolution model to its high-resolution counterpart, namely, via intermediate cylindrical coordinate representations \citep{peng2005learning}, progressive resolution chain \citep{pan2006super}, database retrieval \citep{liang20143d}, curve fitting \citep{zhang20203d}, and mapping from a set of rig parameters to the 2D deformation maps \citep{bailey2020fast}.
Recently, the problem was formulated as a point cloud upsampling to predict $z$-coordinates of the high-resolution face point cloud given its $(x,y)$ coordinates; however, the upsampling ratio is fixed by a factor of 2, and each $(x,y)$ coordinate can only correspond to a unique $z$ coordinate \citep{li20213d}. 

In contrast to acquiring the low-resolution \textit{surface} data from a 3D scanner, depth camera, or multi-view fusion, our work is rooted in a fast but fully \textit{volumetric} physics-based simulator which allows us to provide as an input to our model a set of points that reach deep into the flesh volume and convey richer information about deformation and strain.

\subsection{3D Super-resolution in other domains}
3D super-resolution has also been actively explored in different simulation domains, namely, garments and fluids. Notably, garment surface upsampling by learning of per-vertex deformations \citep{zurdo2012animating} and 2D normal map representations \citep{zhang2021deep} have been explored.
For fluids, procedural \citep{kim2008wavelet} and GAN-based \citep{xie2018tempogan} methods have been explored to enhance the resolution of the simulated coarse turbulent flows.

\begin{figure*}[!ht]
    \centering
    \copyrightbox[r]
    {\includegraphics[width=\textwidth]{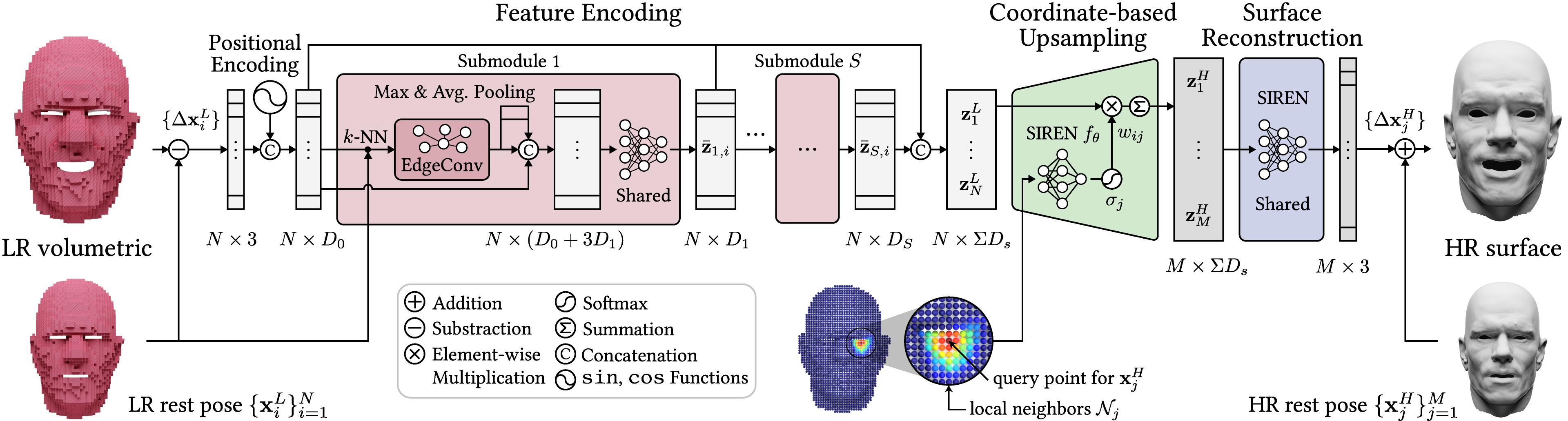}}{\nvidia{}}
    \caption{The overview of our pipeline for 3D simulation super-resolution aiming at learning a mapping from a low-resolution (LR) volumetric mesh to a high-resolution (HR) surface mesh. Our pipeline is comprised of (1) Feature Encoding, (2) Coordinate-based Upsampling, and (3) Surface Reconstruction modules. The input and output are sets of 3D displacement vectors from the LR and HR rest pose shapes, respectively. \copyright NVIDIA}
    \label{fig:pipeline}
\end{figure*}

\subsection{Coordinate-based MLPs}
We employ coordinate-based multilayer perceptrons (MLPs) \citep{tancik2020fourier} to model our upsampling (Section \ref{ss:upsampling}) and reconstruction modules (Section \ref{ss:recon}). Coordinate-based MLPs learn a continuous mapping from input coordinates to signals and have shown promising results for various visual tasks, such as 3D shape representation \citep{mescheder2019occupancy, park2019deepsdf, jiang2020local, saito2019pifu}, novel view synthesis \citep{mildenhall2021nerf, ma2021pixel, chan2021pi}, and super-resolution frameworks for images  \citep{chen2021learning}. 
\edit{Coordinate-based MLPs have also been employed to enforce physical constraints in the super-resolution framework for physics simulations and generate continuous grid-free high-resolution solutions from low-resolution data \citep{esmaeilzadeh2020meshfreeflownet}.}


Recently, SIREN \citep{sitzmann2020implicit} leverages periodic activation functions for implicit neural representations and has also demonstrated superior expressivity \edit{(with principled initialization scheme)} in modeling continuous and fine-detailed signals in various tasks \citep{chan2021pi, ma2021pixel, yang2022implicit}.

\subsection{Model reduction methods}
Model reduction methods (also referred to as subspace simulation methods) are used for accelerating physics simulations by creating a lower dimensional representative subspace for the full space degrees of freedom in the discretization of choice. The subspace can be constructed by computing an appropriate subspace basis for nonlinear models \cite{klm01, bj05}. Extensions to accelerate force computations \cite{akj08} or utilize an adaptive combination of the full space and reduced subspace degrees of freedom \cite{tmdk15} have also been proposed. Furthermore, deep learning models have been integrated with these subspace simulation methods \edit{employing variational autoencoder \cite{fmdlj19} and deep autoencoder leveraging its high-order differentiability \cite{shen2021high}}. Recently, a framework to augment parametric skeletal models with subspace soft-tissue deformations has been proposed \cite{trpo21} to combine the benefits of data-driven skeletal models \cite{rtb17} and skinning-based subspace methods \cite{wjbk15}. \edit{Recently, reduced order models for material point method using implicit neural representations were proposed to construct low-dimensional manifolds of deformation fields \citep{chen2023model} as well as stress and affine fields \citep{zong2023neural}. The low-dimensional manifolds were subsequently employed in conjunction with projection-based dynamics.}

While our method and the class of model reduction methods share the common goal of simulation acceleration, we propose a complementary approach of using physics simulators augmented with deep learning for simulation super-resolution. Model reduction methods have been almost exclusively demonstrated only on linear or isotropic nonlinear constitutive models for passive bodies and require careful consideration to accommodate objects with varying shapes.


\edit{To the best of our knowledge, there has been no prior work on reduced-order modeling that accommodates anisotropic constitutive models for active biomechanical systems such as muscles. Incorporating anisotropic modeling and localized collision resolution into the lower-dimensional subspaces computed for model reduction methods, such as those proposed in \cite{fmdlj19, shen2021high}, is non-trivial. It requires a separate line of investigation and hinders their extensibility for accurate facial animation. In contrast, physics simulators are well-known for supporting anisotropic active models and resolving localized collisions \cite{cong2016art, sifakis2005automatic}. Our method utilizes a GPU-accelerated physics simulator capable of meeting both of these requirements. We demonstrate that our framework can achieve accurate and detailed facial animation without sacrificing speed.}

\section{Method}
\label{sec:method}

In this section, we present the specific design choices for our model architecture, aimed at learning to map from a low-resolution (LR) volumetric mesh to a high-resolution (HR) surface mesh depicting the same facial expression (Figure \ref{fig:pipeline}).
The input LR volumetric mesh contains 15,872 vertices and is derived from regular BCC (body-centered cubic) lattices for real-time simulation leveraging on its sparse and regular distribution of the vertices but with a compromise on accuracy and visual fidelity (Figure \ref{fig:meshes}c). On the other hand, the target HR mesh contains 35,637 vertices and is a triangular mesh conforming to a denser volumetric mesh capable of producing fine details of deformations but at a significantly slower simulation speed (Figure \ref{fig:meshes}b). More information about the data generation is outlined in Section \ref{sec:data_gen}.

We represent our input and output as a set of 3D displacement vectors from a rest pose stacked in an arbitrary yet consistent order.
We divide our pipeline into three modules for (1) feature encoding, (2) coordinate-based upsampling, and (3) surface reconstruction. The hyperparameters are specified in Appendix \ref{appendix:arch}.

\subsection{Feature encoding network} \label{ss:feat_enc}
The feature encoding network computes feature embedding for each input vector. We first concatenate each input displacement vector with a positional encoding $\in\R^{32}$ using sine and cosine functions as done in  Transformers \citep{vaswani2017attention}. 
Then, the concatenated input $\in \R^{D_0}$ (in our implementation, $D_0=35$) goes through the submodules of the feature encoding network. 

While deformations in the human face are primarily attributed to the activation and motion of the underlying muscles and bones respectively, they can also be a result of deformations in other parts of the face (e.g., a wide smile can cause the skin around the eyes to fold); therefore, the localized per-vertex information of deformation needs to be shared with other vertices. For this reason, we model the submodules of the feature encoding network with edge convolutional layers, dubbed \textit{EdgeConv}, introduced in DGCNN \citep{wang2019dynamic} which is capable of aggregating neighborhood information in feature space rather than coordinate space by dynamically constructing a $k$-NN graph in each layer.

We initialize the first $k$-NN graph of the network using geodesic distances based on the edge information of the LR mesh in the rest pose. The subsequent graphs are constructed on the fly in their learned feature spaces. The motivation is to encourage capturing \textit{local} spatial correlations in the first submodule and potentially \textit{global} feature correlations in the subsequent submodules (discussed more in Section \ref {ss:ablation_correlation}).

We apply max and average pooling on the intermediate outputs from EdgeConv to extract global features. They are repeated and concatenated with the outputs from EdgeConv and the preceding input encoding feature, which are then passed through a shared fully connected network.
We repeat the submodule $S=2$ times with the intermediate outputs from one module passed as input to the next. The output of the last submodule is concatenated with all of the previous $S$ intermediate features (including the position-encoded input) to construct the final encoded feature.
Specifically, denoting the output of the $s^{th}$ submodule for the $i^{th}$ LR mesh vertex as $\z^L_i \in \R^{D_s}$, the final encoded output has the dimension of $\z^L_i\in\R^{\sum_{s=0}^SD_s}$. In our implementation, we used $S=2$ with $D_1=64$ and $D_2=128$.

\subsection{Coordinate-based upsampling network}\label{ss:upsampling}
The upsampling network takes as the input a set of encoded per-vertex features from the LR mesh and outputs per-vertex features for the HR surface. To generalize over arbitrary and non-integer upsampling ratios, we propose to formulate the upsampling operation as a continuous local interpolation of the input features.

Formally, let the set of encoded features contributing to the upsampled $j^{th}$ feature be $\{\z^L_i\}_{i\in \mathcal{N}_j}$ where $\z^L_i$ denotes the encoded $i^{th}$ LR mesh feature, and $\mathcal{N}_j$ denotes a set of local interpolation neighbors for the $j^{th}$ feature. Then, the upsampling operation can be expressed as 
\begin{equation}
    \z_j^H = \sum_{i\in\mathcal{N}_j}{w_{ij} \z^L_i},
\end{equation}
where $w_{ij}$ indicates the contribution of the $i^{th}$ LR mesh feature to the $j^{th}$ HR mesh feature.
Different modeling options can be explored for defining the local neighbors set $\mathcal{N}_j$ (e.g., number and criteria of neighbors) and computing the interpolation weight $w_{ij}$ (e.g., inverse distance weighting (IDW), RBF, etc.), which we describe next.

\begin{figure}[h]
    \centering
    \copyrightbox[r]
    {\includegraphics[width=0.9\linewidth]{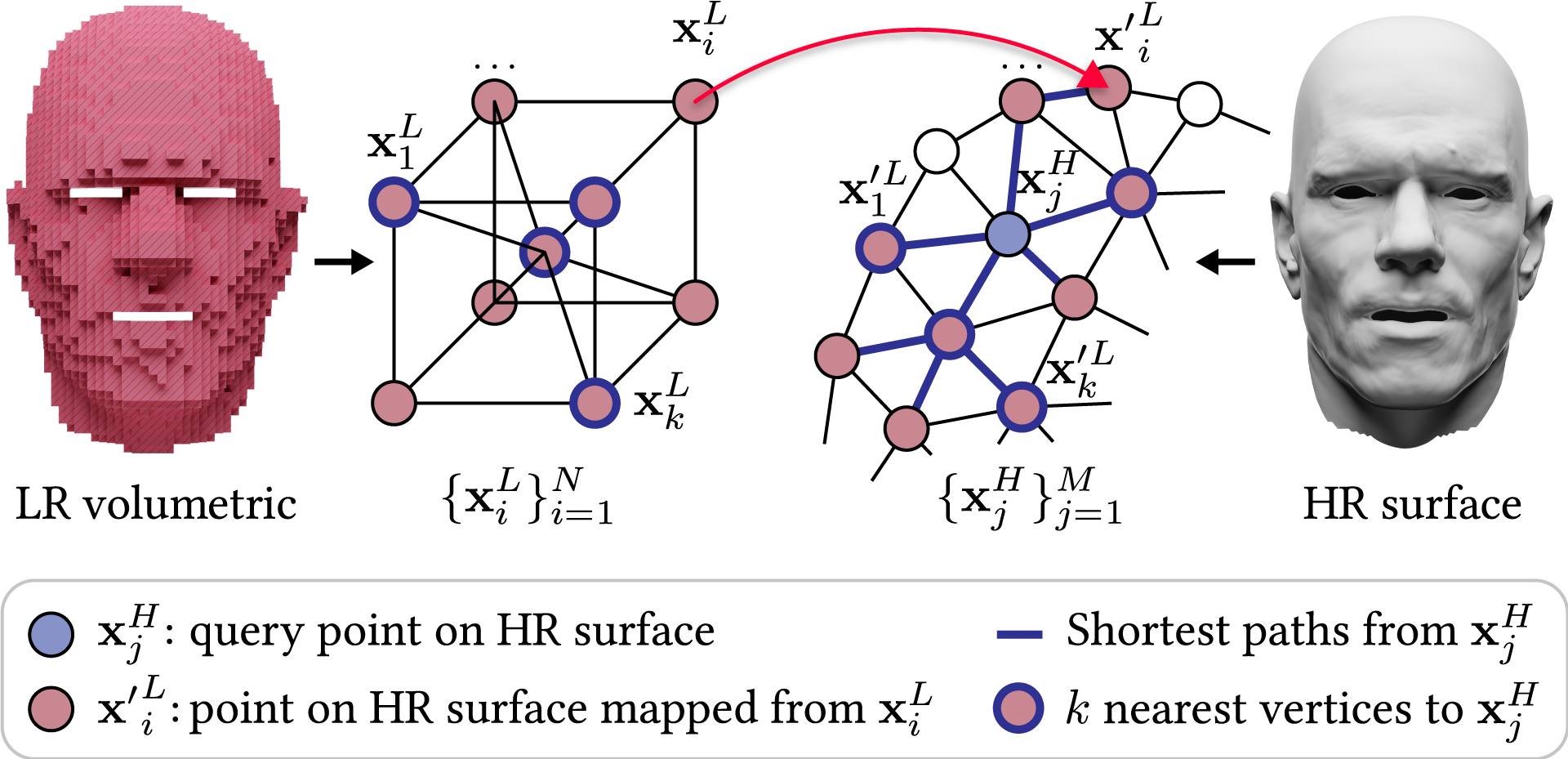}}
    {\nvidia{}}
    \caption{Illustration of finding the $k$ nearest vertices $\{\x^L_1, ..., \x^L_k\}$ (where $i,...,k \in \mathcal{N}_j$) on the LR mesh to the vertex $\x^H_j$ on the HR mesh using geodesic distances. \copyright NVIDIA}
    \label{fig:k_nearest}
\end{figure}

\paragraph{\textbf{Neighborhood locality}}
We define the local neighbors set $\mathcal{N}_j$ as the indices of the $k$ nearest LR mesh vertices from the $j^{th}$ HR mesh vertex in terms of geodesic distances (illustrated in the blue point cloud in center-bottom of Figure \ref{fig:pipeline}). Since the LR and HR vertices do not live on the same surface, we first map the LR vertices $\{\x^L_i\}$ to the HR vertices (we temporarily denote the resulting mapped vertices as $\{\x'^L_i\}$) using the linear assignment algorithm \citep{crouse2016implementing}. This finds the optimal one-to-one mapping between the LR and HR vertices by minimizing the mapping distance (Euclidean). Then, we use Dijkstra's algorithm to find the $k$ nearest mapped vertices $\{\x'^L_i\}$ (which directly corresponds to the original LR vertices $\{\mathbf{x}^L_i\}$) for every HR vertex using the edges of the HR surface mesh as paths (Figure \ref{fig:k_nearest}). 
The local neighbor information is pre-computed offline once.
In this work, we use $k=20$ and additionally explore the effects of different values of $k$ in Section \ref {ss:ablation}.

\paragraph{\textbf{Weighting function}}
The weighting function $w'_{ij}=f_\theta(\mathbf{u}_{ij})$ outputs the interpolation weight $w'_{ij}\in \mathbb{R}$ for the $i^{th}$ LR mesh vertex neighboring the $j^{th}$ HR mesh vertex, given some input vector $\mathbf{u}_{ij}$.

Conceptually, the HR surface mesh can be thought of as a discretization of a continuous and smooth limit-surface, i.e.~its vertices are approximations of the sampled points from the continuous surface. Thus, 
one could sample an infinite number of continuously varying features from any point on this surface. For this reason, we model $f_\theta$ as a trainable coordinate-based MLP where we employ SIREN \citep{sitzmann2020implicit} for its superiority in modeling continuous (and differentiable) functions.

As the input to $f_\theta(\mathbf{u}_{ij})$, we provide the spatial information using a concatenated vector of coordinates of the HR and LR mesh vertices ($\x^H_j, \x^L_i \in \R^3$, respectively) and their mutual Euclidean distance, written as
\begin{equation}
\mathbf{u}_{ij}=[\x^H_j, \x^L_i, ||\x^L_i-\x^H_j||_2].
\end{equation}
Then, we normalize the output weight $w'_{ij}$ across the local neighbors $\mathcal{N}_j$ using the softmax function $\sigma_{j}$ and obtain the final interpolation weight $w_{ij}$, expressed as
\begin{equation}
w_{ij} = \sigma_j(w'_{ij}|\{w'_{kj}\}_{k\in\mathcal{N}_j}) =\frac{e^{w'_{ij}}}{\sum_{k\in\mathcal{N}_j} e^{w'_{kj}}},
\end{equation}
for $j=1,...,M$ and $i\in\mathcal{N}_j$.

\subsection{Surface reconstruction network}\label{ss:recon}
The surface reconstruction network predicts the per-vertex displacements $\Delta \x_j^H$ from the upsampled features $\z_j^H$. Since $\z_j^H$ implicitly inherits coordinate information $\x_j^H$ from the upsampling network and to reconstruct fine deformation details on the HR surface, 
we also model the surface reconstruction network using SIREN \citep{sitzmann2020implicit} to exploit its ability to model high-frequency signals utilizing coordinate information.
As the last step, the predicted deformations are added to the HR mesh in its rest pose to reconstruct the final deformed HR surface.

We also note that we use a minimal modeling technique for the surface reconstruction network not only to reduce the computational overhead for processing a relatively large number of HR mesh vertices ($>$36k) but also because we assume all the information needed for the fine-detailed surface reconstruction is to be encoded in the LR mesh features. 

\subsection{Loss function}
We minimize the reconstruction loss $\mathcal{L}_{recon}$ between the predicted and ground-truth per-vertex deformations of the HR surface mesh denoted $\Delta \hat{\x}_j^H$ and $\Delta \x_j^H$, respectively: 
\begin{equation}
    \mathcal{L}_{recon} = \sum_{j=1}^M||\Delta \hat{\x}_j^H - \Delta {\x}_j^H ||_1.
\end{equation}

Moreover, we introduce the loss term $\mathcal{L}_{fn}$ for local smoothness which encourages the face normal of triangles on the predicted and target HR surface meshes (denoted $\hat{\n}_k$ and $\n_k$, respectively) to be equivalent in terms of cosine similarity: \begin{equation}
    \mathcal{L}_{fn} = \sum_{k=1}^F1-\frac{\hat{\n}_k \cdot \n_k}{||\hat{\n}_k||||\n_k||},
\end{equation}
where $F$ is the number of triangles on the HR surface mesh.

We also include the regularization term $\mathcal{L}_{reg}$ to encourage the encoded intermediate features $\{\{\bar{\z}_{s, i}\}_{i=1}^N\}_{s=1}^S$ (Figure \ref{fig:pipeline}) to center around 0, encouraging their prior to follow a multivariate normal distribution \citep{park2019deepsdf, chabra2020deep}: \begin{equation}
    \mathcal{L}_{reg} = \sum_{s=1}^S\sum_{i=1}^N||\bar{\z}_{s,i}||_F.
\end{equation}
We find that the face normal loss improves the visual fidelity of the reconstructed face and the regularization term helps prevent overfitting.

The final loss function $\mathcal{L}$ is written as
\begin{equation}\label{eq:loss}
    \mathcal{L}=\mathcal{L}_{recon} + \alpha \mathcal{L}_{fn} + \beta \mathcal{L}_{reg},
\end{equation}
where $\alpha$ and $\beta$ are the scalar weight terms whose values are reported in Table \ref{tab:spec} of the Appendix.

\section{Dataset Generation}
\label{sec:data_gen}
In this section, we outline the process for acquiring the mesh models and attachment of muscle fibers, as well as our simulation framework for synthesizing the dataset consisting of the low-resolution (LR) volumetric simulation mesh for flesh and the corresponding high-resolution (HR) surface mesh for the face as shown in Figure \ref{fig:meshes}.

\subsection{Acquisition of simulation models} \label{appendix:data_gen}
\begin{figure}[ht]
    \centering
    \copyrightbox[r]
    {\includegraphics[width=\linewidth]{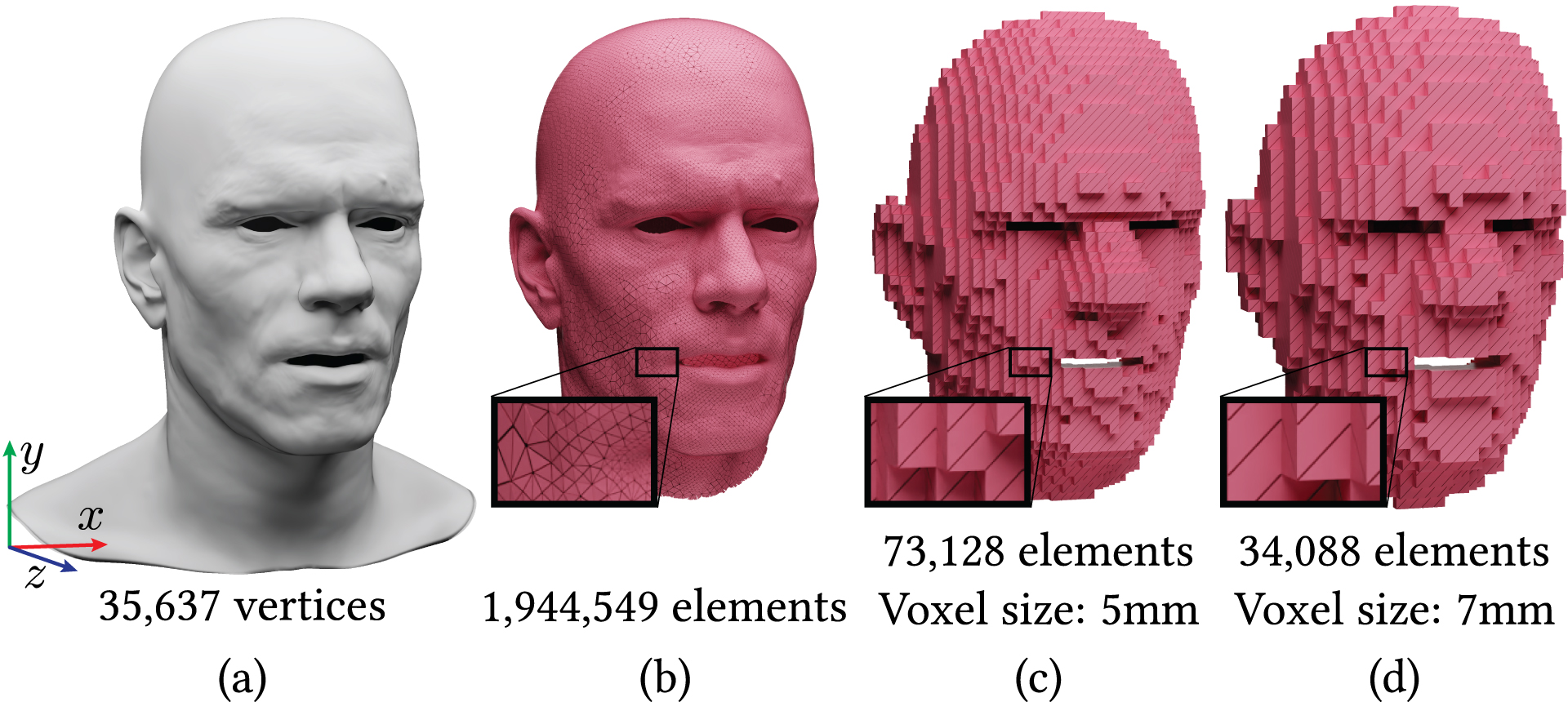}}
    {\nvidia{}}
    \caption{(a) High-resolution surface model in dimensions of $289.0 \times 342.7 \times291.1$ [mm] w.r.t. $x$, $y$, and $z$ axis, respectively, including the part of the shoulder, (b) high-resolution simulation model (\fpsHiSim{} simulation), (c) low-resolution simulation model (\fpsLoSim{} simulation) for the near-realtime end-to-end animation at \fpsEndToEnd{}, and (d) coarser low-resolution simulation model (\fpsSimCoarser{} simulation) for the true real-time end-to-end animation at \fpsEndToEndCoarser{}. \copyright NVIDIA}
    \label{fig:meshes}
\end{figure}

In this section, we explain the process for sculpturing our LR and HR simulation models ((b) and (c) in Figure \ref{fig:meshes}, respectively) which are then used for generating semantically corresponding facial animation dataset.

\paragraph{\textbf{Anatomical model}}
Following prior common approaches \cite{sifakis2005automatic, cong2015fully}, we construct an anatomically and biomechanically motivated simulation model of our subject's face. Given a HR neutral face mesh, we model the underlying anatomy including the cranium, mandible, teeth, and a comprehensive set of facial muscles with the aid of anatomical references. For each facial muscle, we calculate volumetric fiber directions by first tetrahedralizing the muscle and then applying the approach of \cite{choi2013skeletal}. Alternatively, a morphing approach such as \cite{ali-hamadi2013anatomy, cong2015fully} can also be employed to estimate the underlying anatomy. 

\paragraph{\textbf{High-resolution volumetric mesh}}
For our highest resolution model, we create a tetrahedral simulation mesh consisting of 1.9 million tetrahedra \cite{molino2003crystalline} (Figure \ref{fig:meshes}b) that conforms to the HR neutral face mesh (Figure \ref{fig:meshes}a) as well as the underlying skull. We opted for a conforming tetrahedralized simulation mesh in order to maximize deformation accuracy and minimize artificial stiffness often associated with non-conforming tetrahedra. The tradeoff is the potential for less well-conditioned tetrahedra and longer simulation times.

\paragraph{\textbf{Low-resolution volumetric mesh}}
For our LR model, we create a regular nonconforming tetrahedralized simulation mesh consisting of 73 thousand tetrahedra (Figure \ref{fig:meshes}c), to be used in an embedded simulation. We begin by voxelizing the HR conforming tetrahedron mesh at a coarse granularity and discarding tetrahedra outside the regions of the face most responsible for facial expression, including the neck and the back of the head. Then, we subdivide each voxel into eight regular tetrahedra. In constrast to our HR model, our nonconforming regular LR model consists of regular well-conditioned tetrahedra that enables us to target real-time simulation. In order to avoid merging the upper and lower lips with our coarse discretization, we separate the lips via linear blend skinning, pre-deforming the high-resolution conforming tetrahedralized simulation mesh by a small rotation of the jaw joint along its axis. This results in a rest configuration with the mouth slightly open; this necessary modeling discrepancy is among the factors that our super-resolution network must compensate for (and is largely successful in doing so).

\paragraph{\textbf{Muscle fibers and attachments}}
Following the prior approaches of \cite{sifakis2005automatic, cong2016art}, we rasterize the volumetric muscle fiber directions onto both the high- and low-resolution simulation meshes.
Then, we specify anatomically-motivated cranium and jaw attachments of the muscles on both simulation meshes via Dirichlet boundary conditions. Finally, the high-resolution neutral face mesh (containing 61,520 vertices) is embedded in both the high and low-resolution simulation mesh respectively via barycentric weights enabling us to deform the face mesh by interpolating vertex positions from the respective deformed simulation mesh.

\begin{figure}[h]
    \centering
    \copyrightbox[r]
    {\includegraphics[width=\linewidth]{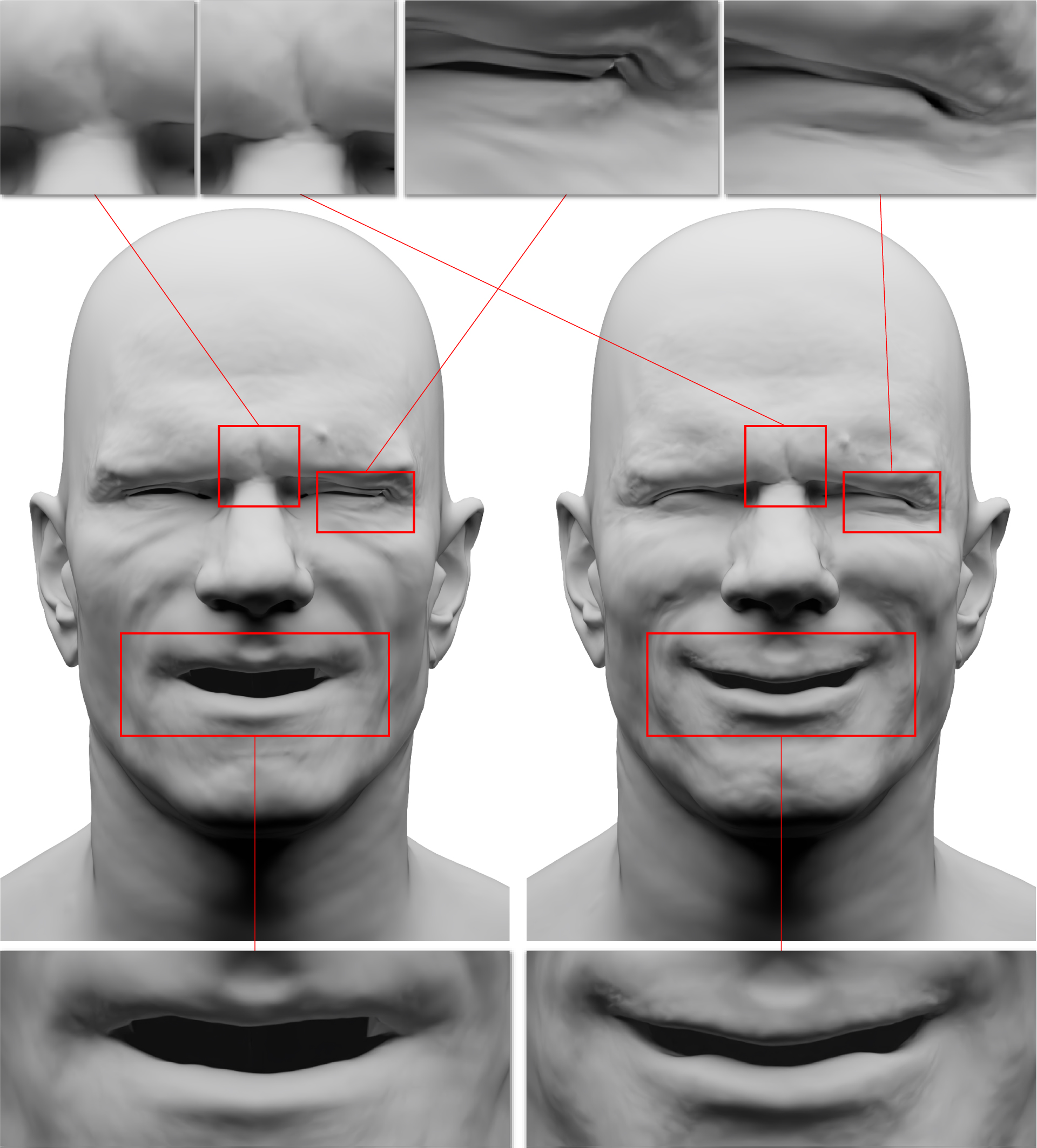}}
    {\nvidia{}}
    \caption{The face surface embedded in the non-conforming low-resolution volumetric mesh with 73 thousand tetrahedra (left) deviates significantly from the same surface simulated using a conforming high-resolution mesh with 1.9 million tetrahedra (right), even though both deformations are parameterized using the same blend shape weights and jaw transformation. We zoom into different regions of the face to highlight macro and microscopic discrepancies. \copyright NVIDIA}
    \label{fig:dx_zoom}
\end{figure}

\paragraph{\textbf{Discrepencies between high- and low-resolution surfaces}}
Figure \ref{fig:dx_zoom} illustrates the discrepancies between the surface embedded in the simulated LR mesh and the surface simulated using the conforming HR mesh. Even though the two performances show semantic similarities, there have both macroscopic (lips) and microscopic (forehead and eyes) differences owing to simulation resolution.

\subsection{Simulation framework}
We employ a CUDA-accelerated implementation of \cite{cong2016art} as our simulation framework for both resolutions. This framework endows the simulation mesh with the anisotropic constitutive model consisting of three components for modeling elasticity, incompressibility, and muscle contractions \cite{teran2003finite} as well as optional kinematic muscle tracks for additional expressivity and directability. Both the finite element forces and the track spring stiffnesses are parameterized to be invariant to mesh refinement in order to maintain consistent bulk behavior across resolutions. Given a set of control parameters and (optionally) kinematic muscle tracks, we calculate the deformation of the tetrahedralized simulation mesh using the quasistatic framework of \cite{teran2005robust}, factoring in object and self-collisions for the high-resolution simulation. In contrast, we forgo collision handling in our LR simulation for the sake of robustness and performance.

\paragraph{\textbf{High-resolution dataset}}\label{ss:sim_framework}
Prior to synthesizing our HR dataset, we ran simulations targeting a wide range of facial performance capture data as well as a set of 31 artist-sculpted blendshapes \cite{cong2016art} using our high-resolution anatomical model. This allowed us to validate that our simulation can accurately reproduce the performance range of the actor while also outputting a corresponding set of 31 kinematic muscle blendshapes. These kinematic muscle blendshapes are combined into a blendshape muscle rig which can be used to deform the kinematic muscle tracks and control the simulation. In addition, we also express the simulation control parameters in terms of the blendshape weights thus extending our simulation framework to be fully differentiable \cite{bao2019high}.

Using the Gauss-Newton optimization proposed in \cite{sifakis2005automatic} in conjunction with \cite{bao2019high}, we solve for four sequences of high-fidelity facial performance capture data corresponding to four different semantic themes (\textit{amazement}, \textit{anger}, \textit{fear}, and \textit{pain}) totaling 880 frames using our HR simulation mesh. This results in a simulated HR simulation and surface mesh, as well as time-varying blend shape weights and jaw transforms for each performance.

\paragraph{\textbf{Low-resolution dataset}}
Since our facial muscles are in correspondence between the HR and LR, we can use the same blend shape muscle rig to drive the LR simulation and synthesize a corresponding LR dataset. We use the blend shape weights and jaw transforms resulting from the HR optimization as input into our LR simulation and run the quasistatic solver to obtain the corresponding LR tetrahedral simulation mesh deformations across all four sequences. 
The discrepancies between the surfaces embedded in the simulated LR mesh and conforming HR mesh, respectively, are illustrated in Figure \ref{fig:dx_zoom} of Section \ref{appendix:data_gen}.

\section{Experiments and Evaluation}
\label{sec:experiments}
We report performance metrics in terms of reconstruction speed (Section \ref{ss:eval_speed}) and as well as quantitative and qualitative reconstruction errors (Section \ref{ss:eval_recon}).
We use the unseen performances in the test set to evaluate the generalization capacity of the trained model. We also evaluate our framework's ability to generalize to unseen dynamics and forces (Section \ref{ss:eval_beyond}).
Additionally, we present the experimental results pertaining to the utilization of blendshape inputs as a substitute for the low-resolution physics-based simulator in generating the input low-resolution tetrahedral mesh (Section \ref{ss:blendshape}).

We also conduct ablation experiments. In Section \ref{appendix:coarser}, we explore the trade-offs in the reconstruction performance of our model when trained using the coarser low-resolution volumetric mesh capable of attaining the \textit{true} real-time end-to-end animation at \fpsEndToEndCoarser{} as compared to our recommended \textit{near} real-time at \fpsEndToEnd{}. In Section \ref{ss:abl_architecture}, we explore how the submodules of our framework, namely Feature Encoding and Coordinate-based Upsampling modules, contribute to the reconstruction accuracy, and, in Section \ref{ss:locality_parameters}, evaluate the effects of using different interpolation neighbors $\mathcal{N}_j$ for the Coordinate-based Upsampling network and different neighbors $k$ for the $k$-NN graph from the Feature Encoding network. Then in Section \ref{ss:ablation_correlation}, we qualitatively evaluate the correlations among different parts of the face learned by the EdgeConv layers in the Feature Encoding submodules. 

\edit{In addition, we investigate our framework's capability to approximate self-collisions between the upper and lower lips in Section \ref{appendix:eval_self_col}, and we conduct ablation experiments to assess the impact of incorporating higher degrees of wrinkle details on the target surface mesh in Section \ref{appendix:wrinkles}.}

\subsection{Near-realtime high-resolution facial animations}\label{ss:eval_speed}
\paragraph{\textbf{Simulations speed}}
The average time to simulate the high-resolution conforming simulation with 1,944,549 tetrahedral elements is \timeHiSim{} per frame or a frame rate of \fpsHiSim{}.
Conversely, the average time to simulate the low-resolution embedding mesh with 73,128 tetrahedral elements is \timeLoSim{}, corresponding to \fpsLoSim{}, i.e.~\xfasterSim{} faster than the high-resolution simulation.
These simulation times are recorded on a workstation with a single GeForce RTX 4090 GPU.

\paragraph{\textbf{Super-resolution inference speed}}
To approximate the high-resolution surface from the low-resolution simulation, we need to infer the high-resolution displacements from our model. The computational overhead of our model inference on a single GeForce RTX 4090 GPU is \timeInfer{} per frame, corresponding to \fpsInfer{} for inference alone. 

\paragraph{\textbf{End-to-end speed and additional performance boosting}}
Consequently, our simulation super-resolution framework takes a total of \timeEndToEnd{} per frame, or \fpsEndToEnd{}, which implies that we achieve a speedup of \xfasterEndToEnd{} relative to the high-resolution simulation that takes \timeHiSim{} per frame (\fpsHiSim{}).
We emphasize that there are multiple ways to bridge the gap from near-realtime, e.g. \fpsEndToEnd{}, to true real-time, i.e. $24$ or more FPS.

First and foremost, using a coarser low-resolution simulation mesh can easily attain the true real-time end-to-end animation given tolerance to a minute trade-off in the quality of reconstructions which our current low-resolution mesh enjoy (we explore the trade-off in Section \ref{appendix:coarser}). Similarly, we can also achieve faster inference time by choosing to use fewer interpolation neighbors in the Coordinate-based Upsampling module but with a trade-off in the overall reconstruction accuracy (see Section \ref{ss:ablation}), as we identify the bottleneck of inference is the neighborhood information gathering step in the Coordinate-based Upsampling module. 

On the other hand, while adhering to the strict bar for the permissible reconstruction quality, we could pipeline the low-resolution simulation and inference steps using a 2 GPU workstation. In such a set up, we could achieve an end-to-end speed of \fpsLoSim{} after tolerating a single frame latency. Conversely, we could also move away from the inference library (we use ONNX Runtime for PyTorch) and implement custom inference kernels on GPUs that speed up computation. 

\begin{figure}[!htb]
    \centering
    \includegraphics[width=\linewidth]{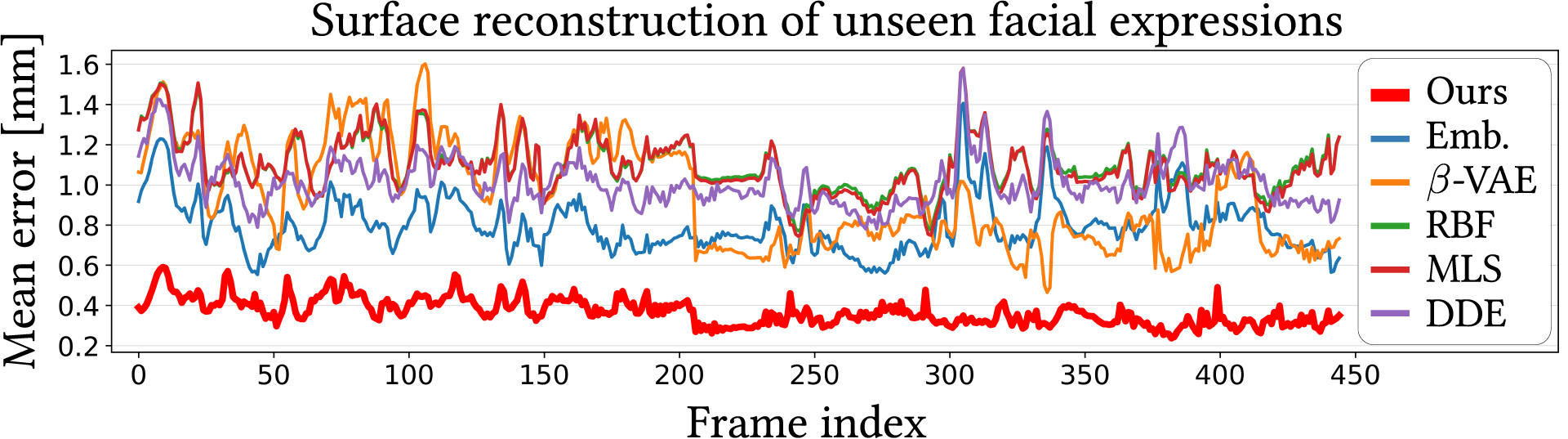}
    \caption{Frame-wise mean surface reconstruction error of unseen facial expressions for each tested model. Our method (in red line) achieved the lowest mean error across every test frame.}
    \label{fig:recon_err_plot}
\end{figure}
\setlength{\belowcaptionskip}{-10pt}
\begin{figure*}[htb!]
    \centering
    \copyrightbox[r]
    {\includegraphics[width=0.98\textwidth]{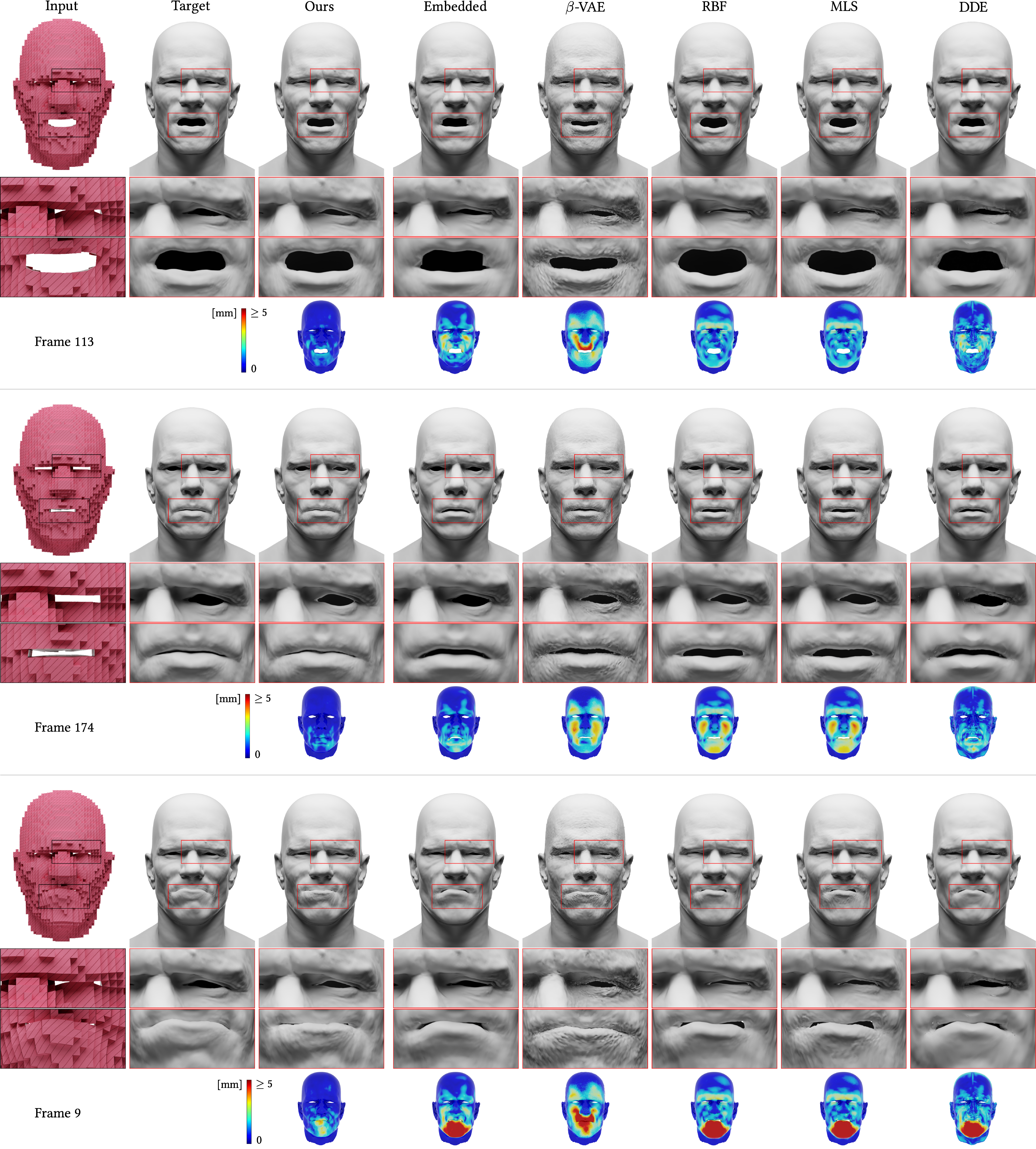}}
    {\nvidia{}}
    \caption{Our method can generalize to unseen facial expressions and reconstruct the target face with high accuracy compared to the standard embedded surface and other tested models ($\beta$-VAE, RBF, MLS, and DDE). The second and third rows show the left eye and mouth zoomed-in, respectively. The heatmaps visualizing the reconstruction errors are shown in the respective last rows. \edit{In the last row is frame 9 where our model has the largest reconstruction error across frames particularly near the lips (See Figure \ref{fig:recon_err_plot} for frame-wise mean errors). \copyright NVIDIA}}
    \label{fig:eval_recon}
\end{figure*}
\subsection{Generalization to unseen facial expressions}\label{ss:eval_recon}
Using the simulation data, generated as described in Section \ref{sec:data_gen}, we select the \textit{amazement} and \textit{pain} sequences for training (435 frames) and test on \textit{anger} and \textit{fear} sequences (445 frames), ensuring that the test set contains unseen performances. We use the trained model to infer the high-resolution face surface from unseen low-resolution volumetric mesh performances in the test set.

\paragraph{\textbf{Quantitative evaluation}}
As we have access to the high-resolution simulations of the test data, we can readily compute the reconstruction error in terms of per-point Euclidean distance between the reconstructed and the target (reference) mesh whose dimension is $179.8\times 257.3\times 164.5$ [mm] (Figure \ref{fig:meshes}).
We also set up other commonly used reconstruction methods to serve as comparisons for our method. We train a $\beta$-VAE \citep{higgins2022beta}, on the same data set to serve as a baseline generative neural framework comparison. We implement two of the commonly used surface reconstruction methods: the radial basis function (RBF) and moving least-square (MLS)-based methods as the representative global and local methods, respectively, where we employ the Gaussian function for RBF. Lastly, we compare with Deep Detail Enhancement (DDE) framework \cite{zhang2021deep} as the representative state-of-the-art super-resolution framework for 3D garment surfaces which uses normal maps to synthesize plausible wrinkle details on a coarse geometry. The formulations for RBF and MLS along with details on the $\beta$-VAE and DDE can be found in Section \ref{appendix:rbf}, \ref{appendix:mls}, \ref{appendix:beta-vae}, and \ref{appendix:dde}, respectively.

Our method outperformed the others and robustly achieved the lowest mean reconstruction errors per frame $<$0.59mm. We plot the frame-wise mean reconstruction errors of the comparisons to validate that our method has the least error for every test performance in Figure \ref{fig:recon_err_plot}. The evaluation result is summarized in Table \ref{table:eval_recon}.

\begin{table}[htb]
    \caption{Descriptive statistic measures 
    of mean surface reconstruction errors (in millimeters) on unseen facial expressions for each tested model.}
    \begin{tabularx}{\linewidth}{c || Y Y Y Y Y}
    \hline
    [mm] & Mean & Median & Std. & Max. & Min. \\
    \hline
    Ours & \textbf{0.37} & \textbf{0.36} & \textbf{0.07} & \textbf{0.59} & \textbf{0.24} \\
    \hline
    Embedded & 0.80 & 0.77 & 0.13 & 1.40 & 0.55 \\
    \hline
    $\beta$-VAE & 0.94 & 0.87 & 0.25 & 1.60 & 0.46 \\
    \hline
    RBF & 1.10 & 1.08 & 0.13 & 1.57 & 0.77 \\
    \hline
    MLS & 1.09 & 1.07 & 0.14 & 1.58 & 0.74 \\
    \hline
    DDE & 1.01 & 0.99 & 0.12 & 1.58 & 0.78 \\
    \hline
    \end{tabularx}
    \label{table:eval_recon}
\end{table}

\paragraph{\textbf{Qualitative evaluation}}
In Figure \ref{fig:eval_recon}, we evaluate the visual fidelity of the inferred face mesh by visualizing the reconstructed high-resolution surfaces and heatmaps of corresponding reconstruction errors for all the methods. Our method can infer the target facial expression from the input low-resolution volumetric mesh more faithfully than other methods, allowing us to conserve both the expression and the subtle deformation details that otherwise would have been compromised by using the low-resolution simulation.

\subsection{Generalization beyond parametric space}\label{ss:eval_beyond}
We test the ability of our framework to handle deformations that extend beyond the parametric space used in simulations. To evaluate, we simulate the low-resolution simulation mesh with unseen dynamics and 
external forces, respectively, and qualitatively evaluate the inference accuracy.

\begin{figure*}[!htb]
  \centering
    \copyrightbox[r]
    {\includegraphics[width=\linewidth]{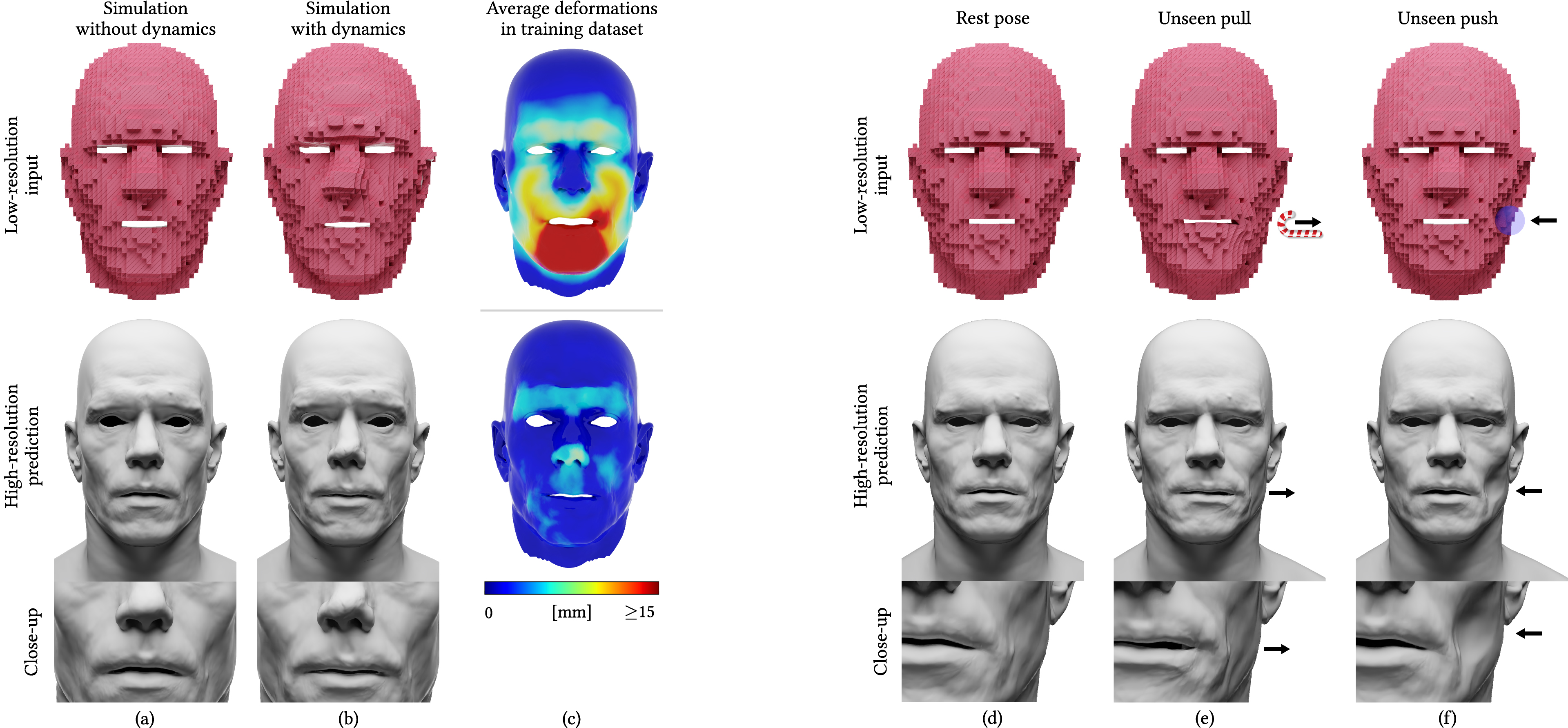}}
    {\nvidia{}}
    \caption{We test the ability of our framework to handle deformations that extend beyond the parametric space used in the simulation by visualizing the inferred surfaces from unseen dynamics (left) and unseen external forces (right) (Section \ref{ss:eval_beyond}). \copyright NVIDIA}
    \label{fig:unseen_physics}
\end{figure*}

\begin{figure*}[!htb]
    \centering
    \copyrightbox[r]
    {\includegraphics[width=\textwidth]{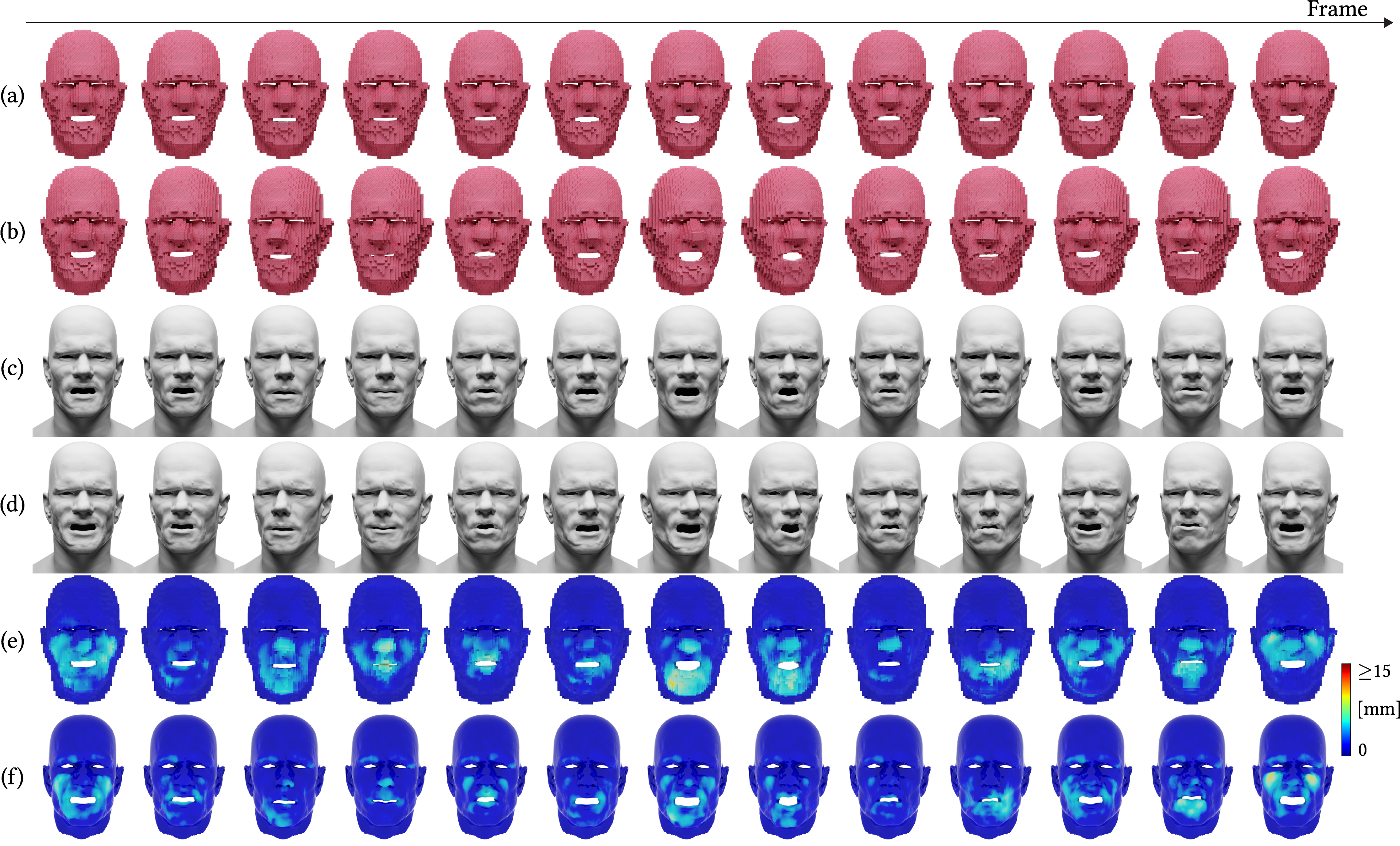}}
    {\nvidia{}}
    \caption{Visualization of sequential frames. From top to bottom: The low-resolution input meshes simulated using (a) a quasi-static and (b) dynamic scheme with left-and-right head spin motions. (c) The high-resolution target faces conforming to the high-resolution quasi-static simulation mesh. (d) Reconstructed surfaces inferred from (b). (e)/(f) The heatmaps showing deformation differences between the meshes $\{$(a), (b)$\}$ and $\{$(c), (d)$\}$, respectively. \copyright NVIDIA}
    \label{fig:dyn_seq}
\end{figure*}
\subsubsection{Unseen dynamics}\label{sss:eval_unseen_physics}
To evaluate our model's capability in generalizing to non-quasi-static simulations, we simulate the dynamics of the low-resolution simulation mesh using a semi-implicit backward Euler scheme. This allows us to model ballistic effects that are not present in our training dataset which was simulated under the quasi-static assumption. We further exaggerate the ballistic effects in the simulation by shaking the head back and forth in conjunction with the muscle contractions and jaw motion. 

We compare the reconstructed surface inferred from the input mesh with unseen dynamics (middle row of Figure \ref{fig:unseen_physics}b) and the reference surface conforming to the quasi-static simulation mesh (middle row of Figure \ref{fig:unseen_physics}a). Also, we visualize heatmaps showing average facial deformations across the training data (top row of Figure \ref{fig:unseen_physics}c) and the deformation differences between the predicted and reference surfaces, respectively (middle row of Figure \ref{fig:unseen_physics}c). We highlight that although the nose shows little or no deformations throughout the training data (thus, showing the nose as a dark blue region in the first heatmap), our model is capable of inferring them from the unseen input (showing as a lighter blue region in the second heatmap). 

Similarly, we visualize the dynamic simulations \edit{(with yaw rotation motions of the head)} and their reconstructions in a time sequence in Figure \ref{fig:dyn_seq} along with the heatmaps (Figure \ref{fig:dyn_seq}e-f) showing deformation differences between the quasi-static/dynamic simulation meshes (Figure \ref{fig:dyn_seq}a/b), and also the reference conforming quasi-static surface (Figure \ref{fig:dyn_seq}c) and the reconstructed surface inferred from the dynamic low-resolution simulation mesh (Figure \ref{fig:dyn_seq}d), respectively. Regions with distinctive facial deformations of the inferred faces (Figure \ref{fig:dyn_seq}e) are in line with the deformed regions of the input simulation meshes (Figure \ref{fig:dyn_seq}f), implying generalizations beyond the quasi-static simulation data.

\subsubsection{Unseen forces}\label{sss:eval_unseen_force}
We craft two quasi-static simulation examples with external forces applied \edit{on the rest pose mesh (Figure \ref{fig:unseen_physics}d)}. In the first example (Figure \ref{fig:unseen_physics}e), we apply a spring force pulling the side of the lips. This force can also be interpreted as a candy cane pulling on one side of the lips. In the second example (Figure \ref{fig:unseen_physics}f), we collide the low-resolution simulation mesh with a sphere, pushing the cheek inward. The low-resolution performances, reconciled by the simulator, are given as input to our framework. The predictions indicate that our framework is able to handle inputs that have deformations not seen in the training performances. 
\edit{Moreover, for side-by-side comparisons, we visualize the surface mesh embedded in the low-resolution simulation mesh in Appendix \ref{appendix:unseen_forces_emb}}.

\subsection{Experiments with blendshape inputs}\label{ss:blendshape}

Employing a low-resolution physics-based simulator for producing the input mesh is perfectly affordable and absorbs much of the nonlinearities in mapping from the simulation parameters (e.g., muscle activations) to the input mesh. Moreover, incorporating dynamics or external forces into the input mesh is a straightforward application for the physics-based simulator, providing an inherent advantage to its usage. Additionally, our super-resolution framework can produce intended facial expressions of the high-resolution surface mesh from its semantically corresponding low-resolution input while compensating for topological discrepancies and can extrapolate to unseen physical effects after being trained only on purely quasi-static simulations.

In this section, we further investigate whether our super-resolution framework can still predict the intended facial expressions from a non-physics-based low-resolution input animated using blendshapes.
Specifically, we conduct two experiments employing the blendshape system as a replacement for the low-resolution physics-based simulator. First, we construct volumetric blendshapes of our low-resolution input mesh and generate the training dataset using a \textit{blendshape animator}, instead of the physics simulator. We also go a step further and use the low-dimensional \textit{blendshape weights} to approximate the high-resolution facial performances by training a decoder-style neural network with around $628\times$ more trainable parameters than our method. The architecture of the neural network is specified in Appendix \ref{appendix:arch-blendshape-decoder}.
We highlight that in both approaches, incorporating dynamics or external forces into the input mesh presents significant challenges compared to the straightforward application of the low-resolution physics-based simulator, which inherently confers an advantage to its use.

In the following subsections, we describe our blendshape system setup used for constructing the volumetric blendshapes and weights for producing facial performances. Then, we provide the evaluation results of the two approaches.

\subsubsection{Construction of low-resolution tetrahedral mesh blendshapes}

For each blendshape in the blendshape muscle rig constructed in Section \ref{ss:sim_framework}, we set its weight to 1.0 and zero out the remaining weights in order to obtain the kinematic muscle deformation corresponding to solely that blendshape. Then, we run the quasi-static solver to obtain the muscle-driven deformation of the low-resolution tetrahedral mesh which is then stored as the corresponding low-resolution tetrahedral mesh blendshape.

\paragraph{\textbf{Volumetric blendshape animation as input}}
In the first scenario, we use the tetrahedral mesh animated using the blendshape weights constructed in Section \ref{ss:sim_framework} as input, as a replacement for the low-resolution physics-based simulator. We then re-initialize and train our existing neural network (Section \ref{sec:method}) to learn to predict the corresponding high-resolution surface mesh.

\begin{figure}[!htb]
    \centering
    \includegraphics[width=\linewidth]{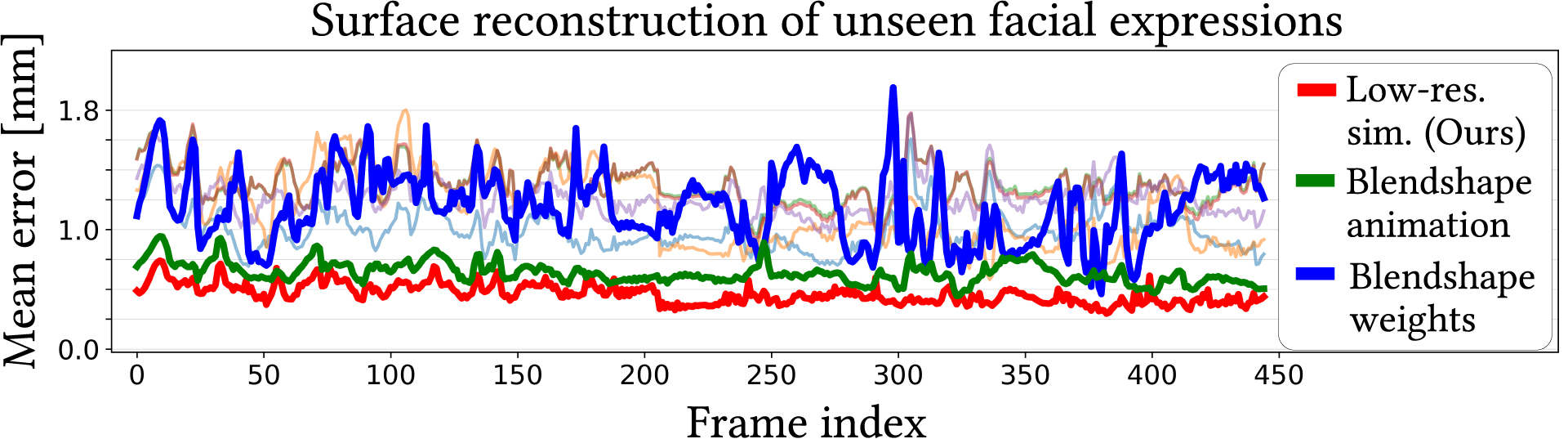}
    \caption{Frame-wise mean surface reconstruction error of unseen facial expressions of the three scenarios (using the low-resolution physics-based simulator, blendshape animator, and direclty using the blendshape weights) overlaid on the plot in Figure \ref{fig:recon_err_plot}.}
    \label{fig:blendshape_err_plot}
\end{figure}
\paragraph{\textbf{Blendshape weights as lower-dimensional input}}
In the second scenario, we directly use the blendshape weights of the facial performances as inputs, bypassing the use of the simulator. To achieve this, we construct a fully-connected neural network with ample capacity (443,840,125 trainable parameters) to learn the mapping from 38-dimensional blendshape weight vector (comprised of 31 blendshapes weights and a 7-dimensional vector for the rigid transformation of the jaw - quaternion and a translation vector) to the high-resolution surface mesh.

\subsubsection{Evaluation results}

\begin{table}[h]
    \vspace*{5mm}
    \caption{Descriptive statistic measures (normalized mean, median, standard deviation, and min/max values for each method) of mean surface reconstruction errors (in millimeters) on unseen facial expressions.}
    \begin{tabularx}{\linewidth}{Y||c c c c c}
    \hline
    [mm] & Mean & Median & Std. & Max. & Min. \\
    \hline
    Low-res. sim. (Ours) & \textbf{0.37} & \textbf{0.36} & \textbf{0.07} & \textbf{0.59} & \textbf{0.24} \\
    \hline
    Blendshape animation & 0.51 & 0.50 & 0.07 & 0.75 & 0.36 \\
    \hline
    Blendshape weights & 0.95 & 0.95 & 0.23 & 1.75 & 0.37 \\
    \hline
    \end{tabularx}
    \label{table:blendshape_recon}
\end{table}

We infer the high-resolution surface mesh in the test dataset and plot the framewise errors for both methods and ours utilizing the low-resolution physics-based simulator. We overlay the plots Figure \ref{fig:recon_err_plot} to highlight the overall difference. As shown in Figure \ref{fig:blendshape_err_plot} and detailed in Table \ref{table:blendshape_recon}, using the blendshape weights as inputs (in blue) yields the largest reconstruction error compared to the other two methods (in red and green). We explain the larger error by noting that the neural network, despite having 628$\times$ more learnable parameters than our method, must learn the blendshapes and produce accurate jaw transformations - tasks that the blendshape animator can easily produce.

On the other hand, using the input tetrahedral mesh produced by the blendshape animator (in green) leads to marginally higher error when compared to using the low-resolution physics-based simulator (in red). This finding aligns with our expectations, given that the physics-based simulator can generate an input mesh that more faithfully adheres to the target surface mesh, accommodating the highly nonlinear and intricate nature of the physics-based simulations.

\begin{figure}[!htb]
    \centering
    \copyrightbox[r]
    {\includegraphics[width=\linewidth]{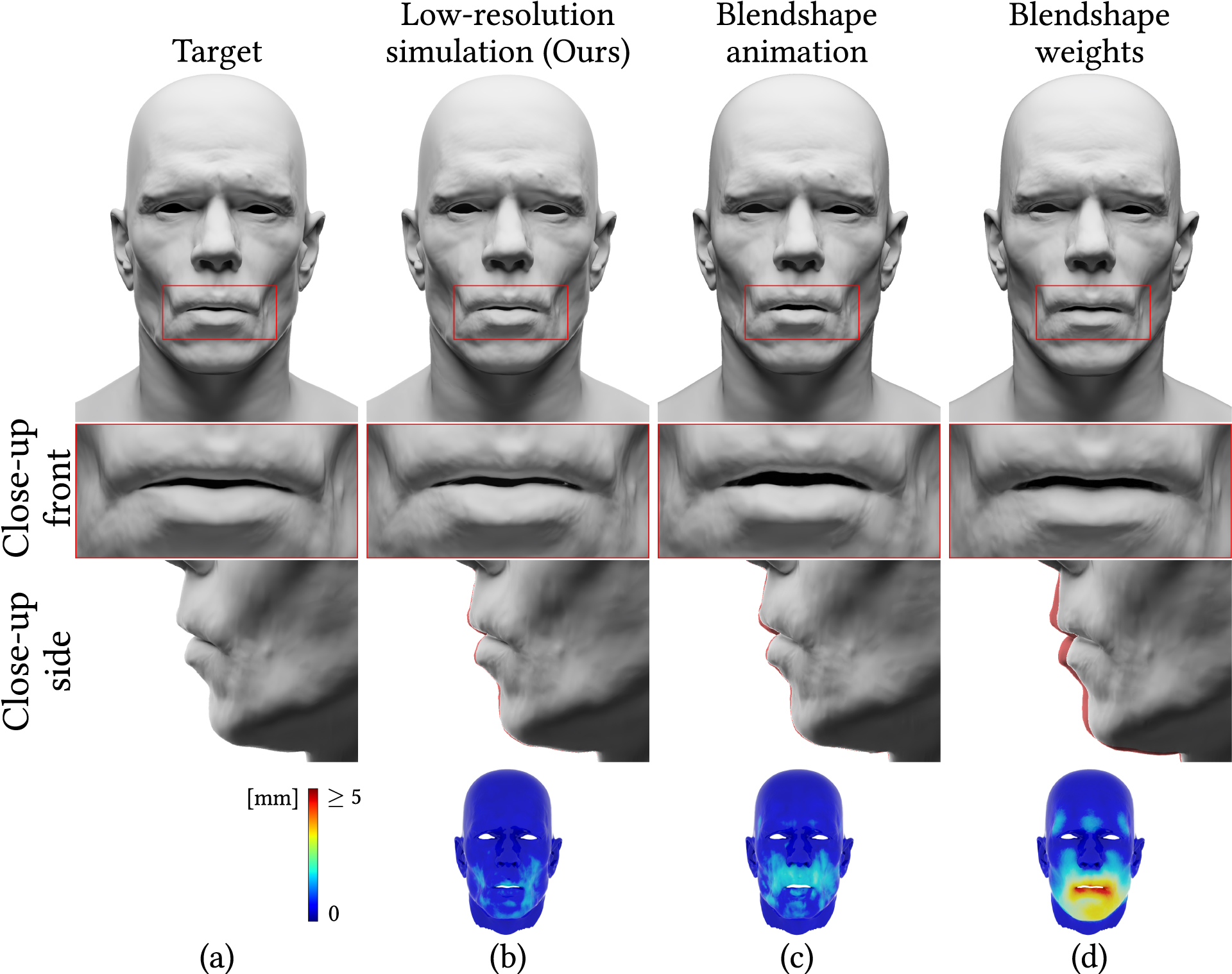}}
        {\nvidia{}}
    \caption{Visualization of (a) the target surface mesh and its reconstructions predicted by the three different methods using (b) the low-resolution physics-based simulator (ours), (c) blendshape animator, and (d) blendshape weights. The 3rd row shows close-up side views of the mouth where the target is shown as the red background and highlights the reconstruction difference. The reconstruction error heatmaps are shown in the last row. \copyright NVIDIA}
    \label{fig:blendshape_recons}
\end{figure}
Notably, relying on blendshape weights as inputs often leads to difficulties in generalizing to unseen jaw transformations. This is clearly observed in the close-up side view of the mouth in the 3rd row of Figure \ref{fig:blendshape_recons}d, where the red background highlights the reconstruction difference between the target mesh (Figure \ref{fig:blendshape_recons}a). Employing the blendshape animator helps to mitigate this issue by generating the low-resolution tetrahedral mesh with accurate jaw motions, as depicted in Figure \ref{fig:blendshape_recons}c. Nevertheless, using the low-resolution physics-based simulator demonstrates the superior performance in faithfully predicting the target facial deformations, particularly evident in the close-up front views of the mouth in the 2nd rows of Figure \ref{fig:blendshape_recons}a, b, and c.

\subsection{Additional experiments}\label{ss:ablation}
In this section, we compare the quality of reconstructed faces inferred by our model trained using the original low-resolution simulation mesh with 73k elements (Figure \ref{fig:meshes}c) and another one trained using a coarser low-resolution simulation mesh with 34k elements (Figure \ref{fig:meshes}d). The coarser mesh attains the \textit{true} real-time end-to-end animation at \fpsEndToEndCoarser{} (\fpsSimCoarser{}  simulation and \fpsInferCoarser{} inference) on the same hardware setup.

Furthermore, we evaluate the contributions of our Feature Encoding (Section \ref{ss:feat_enc}) and Coordinate-based Upsampling (Section \ref{ss:upsampling}) modules. We explore the effects of the key parameters in each of the two modules, namely, the neighbors $k$ in the feature encoding module and the interpolations neighbors in the upsampling module, respectively. Additionally, we qualitatively validate the correlations among different parts of the face learned by our feature encoding network.

\subsubsection{Comparison with coarser low-resolution simulation mesh} \label{appendix:coarser}
For training, we use the same hyperparameters as the training on the original low-resolution simulation mesh. Following the same procedure in Section \ref{ss:eval_recon}, we evaluate the surface reconstruction errors on the unseen facial expressions in the test dataset.

\begin{figure}[!htb]
    \centering
    \copyrightbox[r]
    {\includegraphics[height=\textwidth]{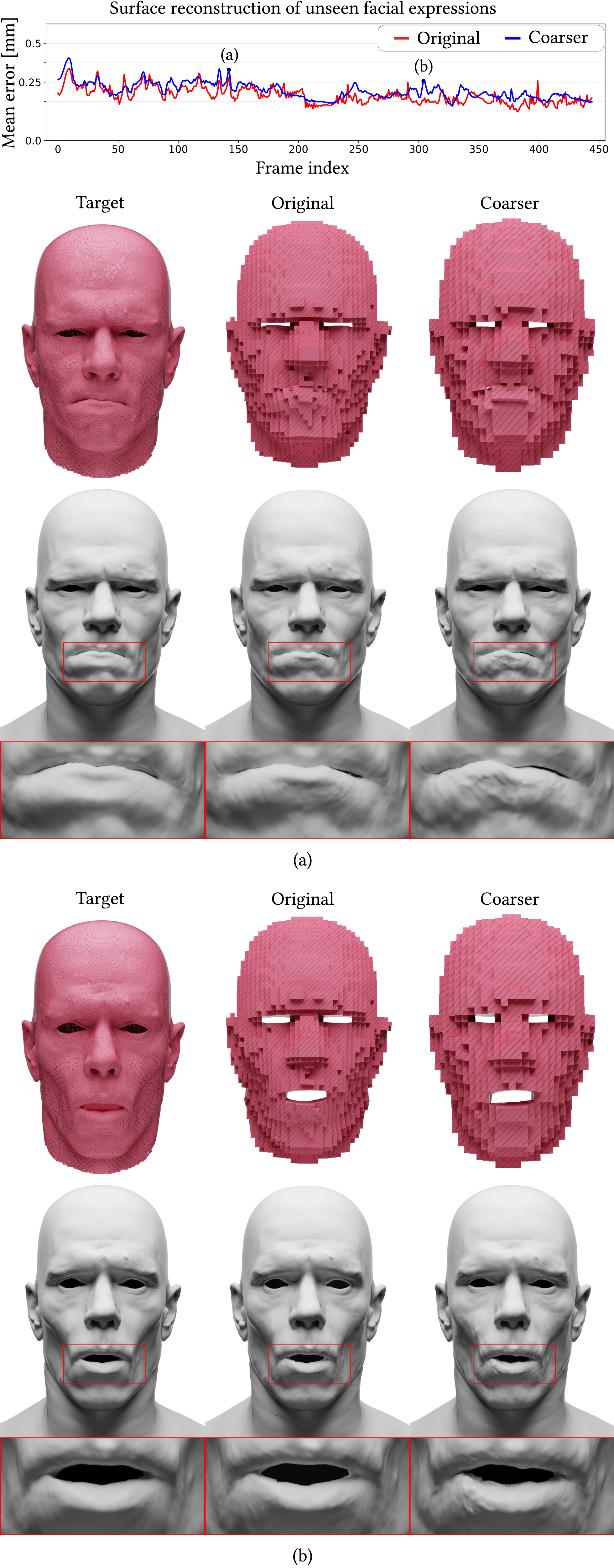}}
        {\nvidia{}}
    \caption{Comparisons of the surface reconstruction qualities by our model trained using the original low-resolution simulation mesh (73k elements) and a coarser mesh with half the resolution (34k elements), respectively. We visualize the reconstructed surfaces in (a) and (b). \copyright NVIDIA}
    \label{fig:recon_coarser}
\end{figure}

As shown in the error plot of Figure \ref{fig:recon_coarser}, using the coarser low-resolution mesh expectedly attains slightly larger reconstruction errors across most of the frames compared to the original mesh. We observe increased artifacts in the inferred surfaces especially around the mouth regions in Figure \ref{fig:recon_coarser}a-b. We highlight that, in practice, true real-time end-to-end animation is easily attainable had we tolerated a minute deterioration of the reconstruction quality which could become  unnoticeable to human eyes with different rendering techniques such as using texture map as opposed to a plain diffuse rendering. However, we choose to adhere to the current resolution for the robustness of generalization capabilities beyond the parametric space used in the simulation (e.g., unseen dynamics and external forces), given that true real-time animation is also attainable, in practice, had we tolerated one frame latency.

\begin{figure}[htb]
    \centering
    \copyrightbox[r]
    {\includegraphics[width=\linewidth]{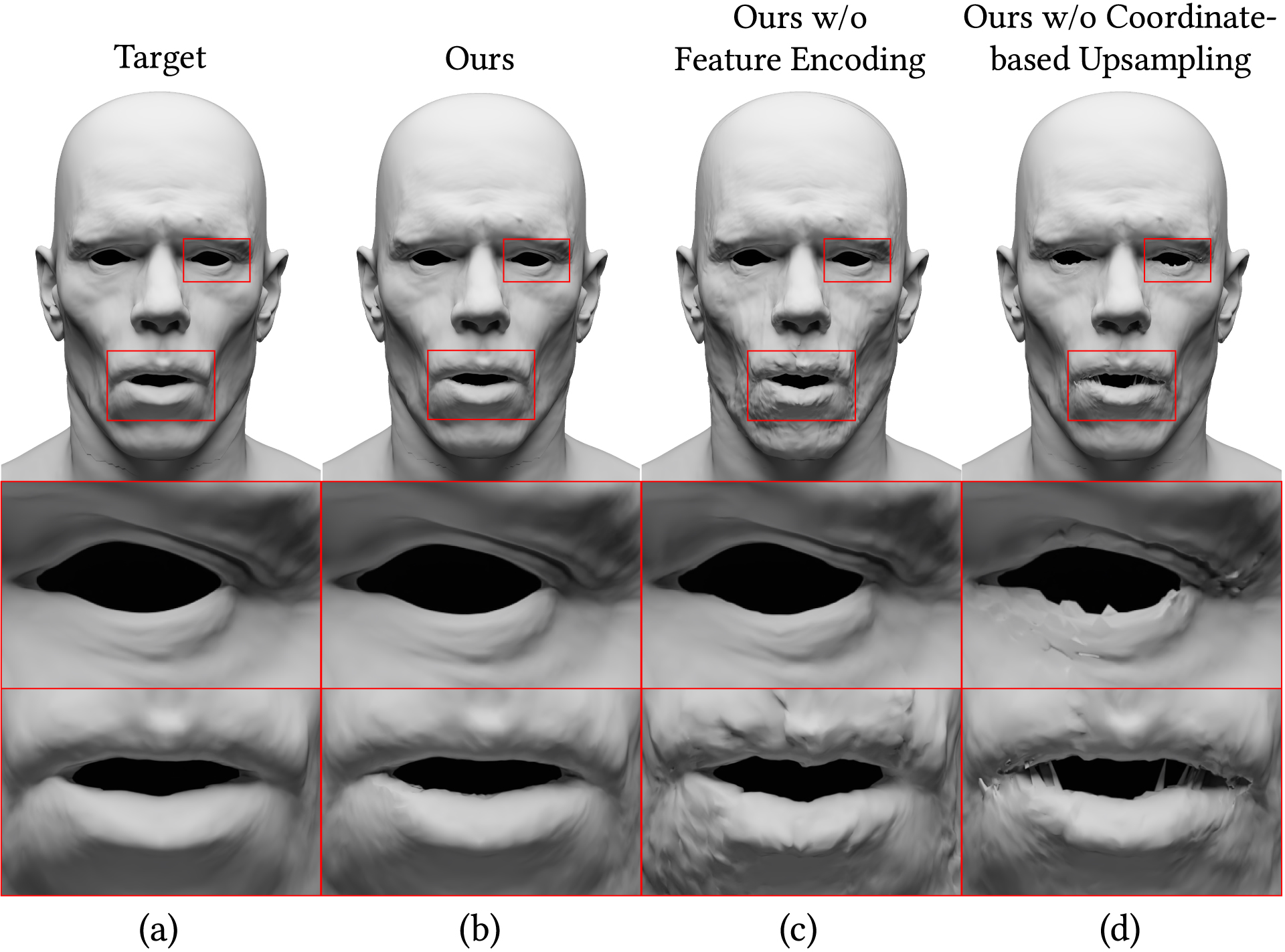}}
    {\nvidia{}}
    \caption{We visualize predictions on a test performance from 3 models - our proposed framework (b), model with feature encoding module excluded (c) and model with the coordinate-based upsampling module replaced (d). The same test performance, simulated in high resolution is visualized in (a). \copyright NVIDIA}
    \label{fig:ablation_architecture}
\end{figure}

\subsubsection{Contributions of Feature Encoding and Coordinate-based Upsampling modules}\label{ss:abl_architecture}
We evaluate the contributions of the Feature Encoding (FE) and Coordinate-based Upsampling (CU) modules by excluding them (one at a time). We compare the predictions on test performances. 

Specifically, we train 3 different models using the same dataset and hyperparameters for the same number of epochs (1000). The first model we train includes both the FE and CU modules (our proposed framework). The second model excludes the FE module and directly feeds the output of position-encoding to the CU module. In the third model, we reintroduce the FE module and exclude the CU module. To replace the CU module, we opt for a different and standard upsampling method (with a fixed upsampling ratio) that uses the transposed convolution operation, widely adopted in upsampling images for super-resolution \cite{yang2019deep}. To mimic the transposed convolution operator, we find 20 nearest LR mesh vertices from each HR mesh vertex in terms of Euclidean distance (same number as our neighbor interpolation in the CU module). We then compute weighted sums of the 20 LR mesh features for every HR mesh vertices. For a fair comparison, we learn these weights, similar to the weights learned in our CU module. 

From the 3 trained models, we compare the reconstruction error on the test dataset. As summarized in Table \ref{table:eval_recon_ablation}, our model which includes both the Feature Encoding and Coordinate-based Upsampling modules outperforms the other two variants which have been trained in the absence of the Feature Encoding and Coordinate-based Upsampling modules, respectively.

We qualitatively validate the visual fidelity of the performances reconstructed by the three models in Figure \ref{fig:ablation_architecture}. We observe that in the absence of the FE module, the model fails to reconstruct the parts of the face with larger deformations accurately (like the mouth area in Figure \ref{fig:ablation_architecture}c), and replacing the CU module leads to reconstruction artifacts and discontinuities in the high-resolution surface (Figure \ref{fig:ablation_architecture}d).

\subsubsection{Effects of different locality parameters}\label{ss:locality_parameters}
\paragraph{\textbf{Interpolation neighbors in Coordinate-based Upsampling}}
We explore the effects of using a different number of interpolation neighbors for defining the local neighbors set $\mathcal{N}_j$ in Section \ref{ss:upsampling}. For this experiment, we train our model using the same training dataset and hyperparameters for 500 epochs but vary the number of interpolation neighbors as 1, 3, 5, 10, and 20. We fix $k=5$ for the $k$-NN graph in the Feature Encoding module for these experiments. We plot the mean surface reconstruction error on the test dataset to study the effect of varying the number of interpolation neighbors on reconstruction accuracy.

\begin{figure}[!hbt]
    \centering
    \includegraphics[width=\linewidth]{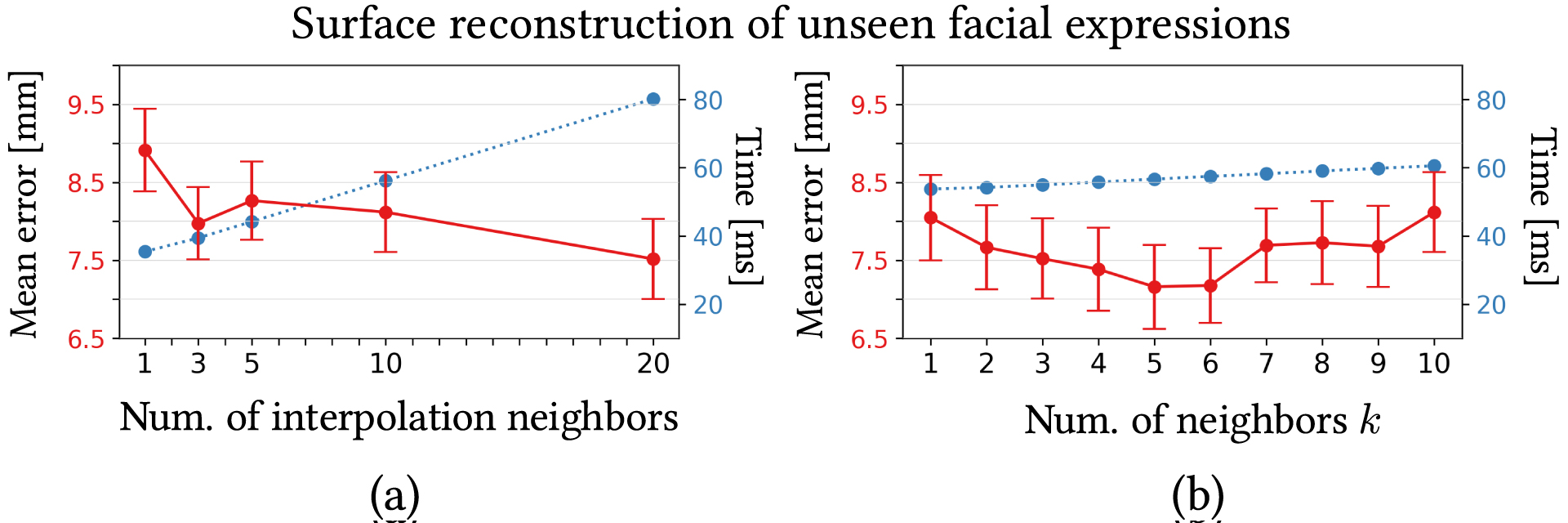}
    \caption{Surface reconstruction errors on unseen facial expressions (red plots) as a function of the number of interpolation neighbors (left) and the number of neighbors $k$ for the $k$-NN graphs in Feature Encoding submodules (right). The blue plots show the inference time per frame for each of the tested values.}
    \label{fig:eval_ablation}
\end{figure}

As shown in the plot in Figure \ref{fig:eval_ablation}a, we observe that using a higher number of interpolation neighbors achieves lower mean reconstruction error on unseen performances (shown in red). However, the trade-off is a linearly increasing time consumption for each inference (shown in blue). 

\begin{table}[!hbt]
    \caption{Descriptive statistic measures of surface reconstruction errors in the absence of our Feature Encoding (FE) and Coordinate-based Upsampling (CU) network.}
    \begin{tabularx}{\linewidth}{Y||Y Y Y Y Y}
    \hline
    [mm] & Ours & w/o FE & w/o CU\\
    \hline
    Mean&\textbf{0.38} &0.45 &0.59\\
    \hline
    Std.&\textbf{0.06} &0.07 &0.10\\
    \hline
    Median&\textbf{0.38} &0.45 &0.58\\
    \hline
    Max.&\textbf{0.64} &0.75 &1.11\\
    \hline
    Min.&\textbf{0.27} &0.33 &0.41\\
    \hline
    \end{tabularx}
    \label{table:eval_recon_ablation}
\end{table}
\paragraph{\textbf{Number of neighbors $k$ in Feature Encoding}}
We conduct another experiment to study the effect of varying the neighbors $k$ used in constructing the $k$-NN graph in the EdgeConv layer of the Feature Encoding module. We train our model for 500 epochs while varying $k$ from 1 to 10 in each experiment, and evaluate the mean surface reconstruction error on the test dataset. We fix the number of interpolation neighbors in the Coordinate-based Upsampling module to 10 for these experiments.
As shown in the plot in Figure \ref{fig:eval_ablation}b, we find that using $k=4, 5$ gives the minimum reconstruction error (shown in red) without a large trade-off in the inference time (shown in blue). 

\subsubsection{Correlations learned in Feature Encoding module}\label{ss:ablation_correlation}
We visualize the heatmaps of the feature similarities learned by the EdgeConv layer in the second Feature Encoding network submodule. This can reveal the correlations among different parts of the face learned from data. As outlined in Section \ref {ss:feat_enc}, we encourage the first submodule to learn local spatial correlations by constructing the $k$-NN graph in based on geodesic distances, and the second submodule to learn (potentially global) feature correlations in its learned feature space.

\begin{figure}[!ht]
    \centering
    \includegraphics[width=\linewidth]{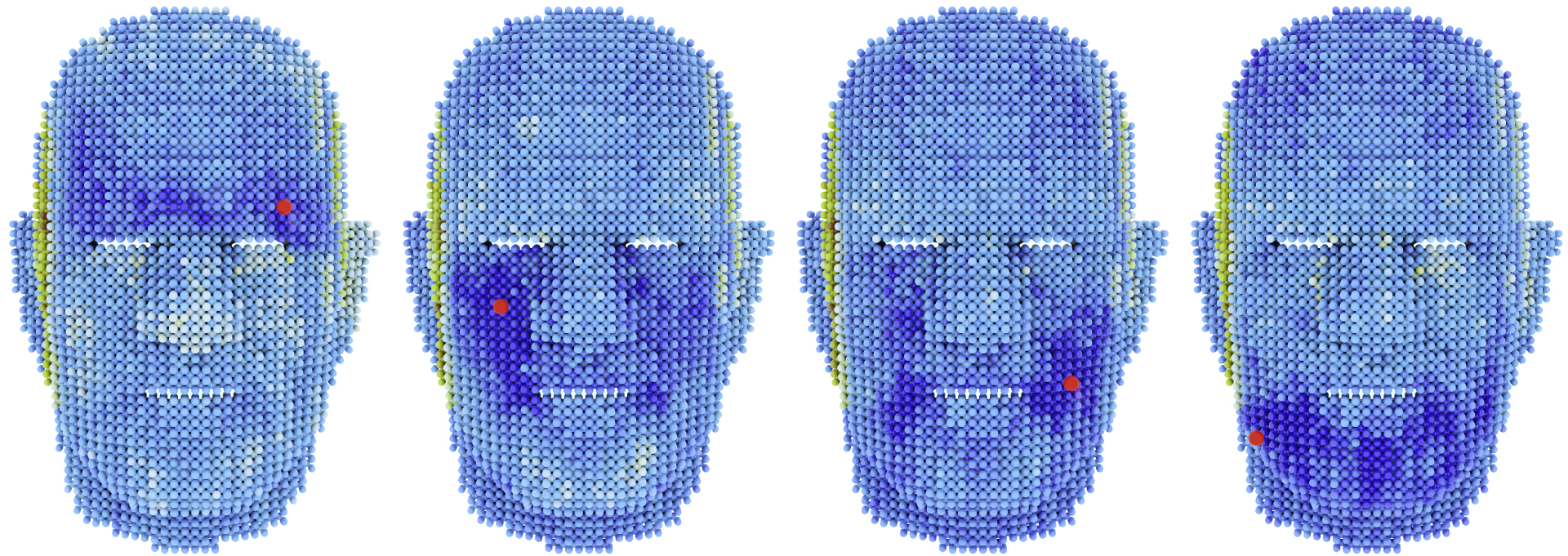}
    \caption{Correlations among different parts of the face learned in the second submodule of the Feature Encoding network. Similar colors and shades represent higher correlations with the queried (red) point.}
    \label{fig:edge_conv}
\end{figure}

Figure \ref{fig:edge_conv} shows the learned similarities for four selected frames where the red point in each image denotes a queried point, and the similar colors and shades represent higher similarities. We observe that the Feature Encoding module has captured the correlations among different parts of the face, such as the right part of the chin being correlated with the left part of the mouth (third image from the left).

\section{Conclusion}
\label{sec:conclusion}
We have proposed a data-driven deep neural network framework which, using as input a low-resolution simulation of facial expression, enhances its detail and visual fidelity to levels commensurate with that of a much more expensive, high-resolution simulation. The combined performance of the low-resolution simulator and the upsampling module itself is efficient enough to yield \fpsEndToEnd{} end-to-end, with the potential of the true real-time \fpsEndToEndCoarser{} end-to-end for a modest sacrifice of accuracy. We demonstrate that our super-resolution framework is able to convincingly bridge the visual quality gap between the real-time low-resolution and offline high-resolution simulations, even in instances where the two simulations have substantial differences due to discretization, modeling, and resolution disparities. 
Our super-resolution network successfully upsamples even deformations that go beyond the parametric poses exemplified in the training set (triggered by muscle action and bone motion), to include dynamics, external forces, and collision objects and constraints. Finally, we observe that our framework can approximate a degree of collision response purely via generalization from the training data. Our code is available on \edit{\ourgithub{}}

\subsection{Limitations and Future Work}
We have adopted a number of design choices that may consciously limit the scope of our work. We have chosen the output of our upsampling module to be the \textit{surface} of the face model, rather than a description that includes the interior of the high-resolution target simulation mesh. The same output is also purely geometry, as opposed to physical quantities such as volumetric strain tensor fields or action potentials (e.g. in the style of \cite{srinivasan2021learning, yang2022implicit}) which might have been useful for an extra simulation pass at the high resolution to incorporate additional effects. Both such choices are made to reduce the dependency of our system on any internal traits of the simulation engine that was used to produce the high-resolution training data, requiring only surfaces at high resolution for training (those could even have originated from performance acquisition, as opposed to simulation), and stay as close to the real-time regime as possible. 


Our super-resolution approach strives to recreate physical behaviors as exemplified at the high-resolution component of the training set; however, the degree at which such physical traits are conveyed is limited by how large and representative our training set is, and not enforced via explicit physics-based simulation at the high-resolution output. For example, traits such as volume preservation, strain limits, or contact/collision behavior are only approximated to the degree that the network can learn them from data, while a full-fledged simulator could provide stronger guarantees. Specifically, if the low-resolution simulation does not employ collision handling and the high-resolution simulator used for training does, it would be very challenging to resolve behaviors where the exact result of contact resolution is history dependent and admits multiple solutions. A typical example would be a facial motion that brings the lips into deep collision at low-resolution; at high-resolution, any result including the lips being pressed together, or sliding under one another in any order, would not have the benefit of history dependence or friction to naturally lead to one of the possible scenarios.

In future work, we wish to further investigate possibilities for boosting our method's efficacy of collision handling, by tuning the training loss to more directly emphasize collision avoidance (rather than just matching the target provided), and possibly augment the low-resolution simulation with cheap approximations to collisions (e.g., using proxy geometry and repulsive forces to create a ``soft'' collision response) to help disambiguate collision scenarios where multiple solutions are admissible. We would also investigate adding a temporal element to our prediction; this could be beneficial both as a way to enhance temporal consistency of our animation, and perhaps as a pathway to adding dynamic effects to the resulting animation (even if the low-resolution simulation was overdamped or quasistatic).
\edit{Lastly, our method is trained on the facial model of a single identity, overfitting on a specific face mesh. Extending our proposed simulation super-resolution framework to accommodate multiple identities is also an interesting direction for future work.}



\bibliographystyle{ACM-Reference-Format}
\bibliography{manuscript}


\begin{thebibliography}{87}


\ifx \showCODEN    \undefined \def \showCODEN     #1{\unskip}     \fi
\ifx \showDOI      \undefined \def \showDOI       #1{#1}\fi
\ifx \showISBNx    \undefined \def \showISBNx     #1{\unskip}     \fi
\ifx \showISBNxiii \undefined \def \showISBNxiii  #1{\unskip}     \fi
\ifx \showISSN     \undefined \def \showISSN      #1{\unskip}     \fi
\ifx \showLCCN     \undefined \def \showLCCN      #1{\unskip}     \fi
\ifx \shownote     \undefined \def \shownote      #1{#1}          \fi
\ifx \showarticletitle \undefined \def \showarticletitle #1{#1}   \fi
\ifx \showURL      \undefined \def \showURL       {\relax}        \fi
\providecommand\bibfield[2]{#2}
\providecommand\bibinfo[2]{#2}
\providecommand\natexlab[1]{#1}
\providecommand\showeprint[2][]{arXiv:#2}

\bibitem[Alexa et~al\mbox{.}(2001)]%
        {alexa2001point}
\bibfield{author}{\bibinfo{person}{Marc Alexa}, \bibinfo{person}{Johannes
  Behr}, \bibinfo{person}{Daniel Cohen-Or}, \bibinfo{person}{Shachar
  Fleishman}, \bibinfo{person}{David Levin}, {and} \bibinfo{person}{Claudio~T
  Silva}.} \bibinfo{year}{2001}\natexlab{}.
\newblock \showarticletitle{Point set surfaces}. In
  \bibinfo{booktitle}{\emph{Proceedings Visualization, 2001. VIS'01.}} IEEE,
  \bibinfo{pages}{21--29}.
\newblock


\bibitem[Alexa et~al\mbox{.}(2003)]%
        {alexa2003computing}
\bibfield{author}{\bibinfo{person}{Marc Alexa}, \bibinfo{person}{Johannes
  Behr}, \bibinfo{person}{Daniel Cohen-Or}, \bibinfo{person}{Shachar
  Fleishman}, \bibinfo{person}{David Levin}, {and} \bibinfo{person}{Claudio~T.
  Silva}.} \bibinfo{year}{2003}\natexlab{}.
\newblock \showarticletitle{Computing and rendering point set surfaces}.
\newblock \bibinfo{journal}{\emph{IEEE Transactions on visualization and
  computer graphics}} \bibinfo{volume}{9}, \bibinfo{number}{1}
  (\bibinfo{year}{2003}), \bibinfo{pages}{3--15}.
\newblock


\bibitem[Ali-Hamadi et~al\mbox{.}(2013)]%
        {ali-hamadi2013anatomy}
\bibfield{author}{\bibinfo{person}{Dicko Ali-Hamadi}, \bibinfo{person}{Tiantian
  Liu}, \bibinfo{person}{Benjamin Gilles}, \bibinfo{person}{Ladislav Kavan},
  \bibinfo{person}{Fran\c{c}ois Faure}, \bibinfo{person}{Olivier Palombi},
  {and} \bibinfo{person}{Marie-Paule Cani}.} \bibinfo{year}{2013}\natexlab{}.
\newblock \showarticletitle{Anatomy Transfer}.
\newblock \bibinfo{journal}{\emph{ACM Trans. Graph.}} \bibinfo{volume}{32},
  \bibinfo{number}{6}, Article \bibinfo{articleno}{188} (\bibinfo{date}{nov}
  \bibinfo{year}{2013}), \bibinfo{numpages}{8}~pages.
\newblock
\showISSN{0730-0301}
\urldef\tempurl%
\url{https://doi.org/10.1145/2508363.2508415}
\showDOI{\tempurl}


\bibitem[An et~al\mbox{.}(2008)]%
        {akj08}
\bibfield{author}{\bibinfo{person}{Steven~S. An}, \bibinfo{person}{Theodore
  Kim}, {and} \bibinfo{person}{Doug~L. James}.}
  \bibinfo{year}{2008}\natexlab{}.
\newblock \showarticletitle{Optimizing Cubature for Efficient Integration of
  Subspace Deformations}.
\newblock \bibinfo{journal}{\emph{ACM Trans. Graph.}} \bibinfo{volume}{27},
  \bibinfo{number}{5}, Article \bibinfo{articleno}{165} (\bibinfo{date}{dec}
  \bibinfo{year}{2008}), \bibinfo{numpages}{10}~pages.
\newblock
\showISSN{0730-0301}
\urldef\tempurl%
\url{https://doi.org/10.1145/1409060.1409118}
\showDOI{\tempurl}


\bibitem[Anjyo et~al\mbox{.}(2014)]%
        {anjyo2014scattered}
\bibfield{author}{\bibinfo{person}{Ken Anjyo}, \bibinfo{person}{John~P Lewis},
  {and} \bibinfo{person}{Fr{\'e}d{\'e}ric Pighin}.}
  \bibinfo{year}{2014}\natexlab{}.
\newblock \showarticletitle{Scattered data interpolation for computer
  graphics}.
\newblock In \bibinfo{booktitle}{\emph{ACM SIGGRAPH 2014 Courses}}.
  \bibinfo{pages}{1--69}.
\newblock


\bibitem[Bailey et~al\mbox{.}(2020)]%
        {bailey2020fast}
\bibfield{author}{\bibinfo{person}{Stephen~W Bailey}, \bibinfo{person}{Dalton
  Omens}, \bibinfo{person}{Paul Dilorenzo}, {and} \bibinfo{person}{James~F
  O'Brien}.} \bibinfo{year}{2020}\natexlab{}.
\newblock \showarticletitle{Fast and deep facial deformations}.
\newblock \bibinfo{journal}{\emph{ACM Transactions on Graphics (TOG)}}
  \bibinfo{volume}{39}, \bibinfo{number}{4} (\bibinfo{year}{2020}),
  \bibinfo{pages}{94--1}.
\newblock


\bibitem[Bao et~al\mbox{.}(2019)]%
        {bao2019high}
\bibfield{author}{\bibinfo{person}{Michael Bao}, \bibinfo{person}{Matthew
  Cong}, \bibinfo{person}{St{\'e}phane Grabli}, {and} \bibinfo{person}{Ronald
  Fedkiw}.} \bibinfo{year}{2019}\natexlab{}.
\newblock \showarticletitle{High-quality face capture using anatomical
  muscles}. In \bibinfo{booktitle}{\emph{Proceedings of the IEEE/CVF Conference
  on Computer Vision and Pattern Recognition}}. \bibinfo{pages}{10802--10811}.
\newblock


\bibitem[Barbi\v{c} and James(2005)]%
        {bj05}
\bibfield{author}{\bibinfo{person}{Jernej Barbi\v{c}} {and}
  \bibinfo{person}{Doug~L. James}.} \bibinfo{year}{2005}\natexlab{}.
\newblock \showarticletitle{Real-Time Subspace Integration for St.
  Venant-Kirchhoff Deformable Models}.
\newblock \bibinfo{journal}{\emph{ACM Trans. Graph.}} \bibinfo{volume}{24},
  \bibinfo{number}{3} (\bibinfo{date}{jul} \bibinfo{year}{2005}),
  \bibinfo{pages}{982–990}.
\newblock
\showISSN{0730-0301}
\urldef\tempurl%
\url{https://doi.org/10.1145/1073204.1073300}
\showDOI{\tempurl}


\bibitem[Bender et~al\mbox{.}(2013)]%
        {bender2013position}
\bibfield{author}{\bibinfo{person}{Jan Bender}, \bibinfo{person}{Matthias
  M{\"u}ller}, \bibinfo{person}{Miguel~A Otaduy}, {and}
  \bibinfo{person}{Matthias Teschner}.} \bibinfo{year}{2013}\natexlab{}.
\newblock \showarticletitle{Position-based Methods for the Simulation of Solid
  Objects in Computer Graphics.}. In \bibinfo{booktitle}{\emph{Eurographics
  (State of the Art Reports)}}. \bibinfo{pages}{1--22}.
\newblock


\bibitem[Berger et~al\mbox{.}(2017)]%
        {berger2017survey}
\bibfield{author}{\bibinfo{person}{Matthew Berger}, \bibinfo{person}{Andrea
  Tagliasacchi}, \bibinfo{person}{Lee~M Seversky}, \bibinfo{person}{Pierre
  Alliez}, \bibinfo{person}{Gael Guennebaud}, \bibinfo{person}{Joshua~A
  Levine}, \bibinfo{person}{Andrei Sharf}, {and} \bibinfo{person}{Claudio~T
  Silva}.} \bibinfo{year}{2017}\natexlab{}.
\newblock \showarticletitle{A survey of surface reconstruction from point
  clouds}. In \bibinfo{booktitle}{\emph{Computer Graphics Forum}},
  Vol.~\bibinfo{volume}{36}. Wiley Online Library, \bibinfo{pages}{301--329}.
\newblock


\bibitem[Bergou et~al\mbox{.}(2007)]%
        {bergou2007tracks}
\bibfield{author}{\bibinfo{person}{Mikl{\'o}s Bergou}, \bibinfo{person}{Saurabh
  Mathur}, \bibinfo{person}{Max Wardetzky}, {and} \bibinfo{person}{Eitan
  Grinspun}.} \bibinfo{year}{2007}\natexlab{}.
\newblock \showarticletitle{Tracks: toward directable thin shells}.
\newblock \bibinfo{journal}{\emph{ACM Transactions on Graphics (TOG)}}
  \bibinfo{volume}{26}, \bibinfo{number}{3} (\bibinfo{year}{2007}),
  \bibinfo{pages}{50--es}.
\newblock


\bibitem[Berretti et~al\mbox{.}(2012)]%
        {berretti2012superfaces}
\bibfield{author}{\bibinfo{person}{Stefano Berretti},
  \bibinfo{person}{Alberto~Del Bimbo}, {and} \bibinfo{person}{Pietro Pala}.}
  \bibinfo{year}{2012}\natexlab{}.
\newblock \showarticletitle{Superfaces: A super-resolution model for 3D faces}.
  In \bibinfo{booktitle}{\emph{European Conference on Computer Vision}}.
  Springer, \bibinfo{pages}{73--82}.
\newblock


\bibitem[Berretti et~al\mbox{.}(2014)]%
        {berretti2014face}
\bibfield{author}{\bibinfo{person}{Stefano Berretti}, \bibinfo{person}{Pietro
  Pala}, {and} \bibinfo{person}{Alberto Del~Bimbo}.}
  \bibinfo{year}{2014}\natexlab{}.
\newblock \showarticletitle{Face recognition by super-resolved 3D models from
  consumer depth cameras}.
\newblock \bibinfo{journal}{\emph{IEEE transactions on information forensics
  and security}} \bibinfo{volume}{9}, \bibinfo{number}{9}
  (\bibinfo{year}{2014}), \bibinfo{pages}{1436--1449}.
\newblock


\bibitem[Bondi et~al\mbox{.}(2016)]%
        {bondi2016reconstructing}
\bibfield{author}{\bibinfo{person}{Enrico Bondi}, \bibinfo{person}{Pietro
  Pala}, \bibinfo{person}{Stefano Berretti}, {and} \bibinfo{person}{Alberto
  Del~Bimbo}.} \bibinfo{year}{2016}\natexlab{}.
\newblock \showarticletitle{Reconstructing high-resolution face models from
  kinect depth sequences}.
\newblock \bibinfo{journal}{\emph{IEEE Transactions on Information Forensics
  and Security}} \bibinfo{volume}{11}, \bibinfo{number}{12}
  (\bibinfo{year}{2016}), \bibinfo{pages}{2843--2853}.
\newblock


\bibitem[Bouaziz et~al\mbox{.}(2014)]%
        {bouaziz2014projective}
\bibfield{author}{\bibinfo{person}{Sofien Bouaziz}, \bibinfo{person}{Sebastian
  Martin}, \bibinfo{person}{Tiantian Liu}, \bibinfo{person}{Ladislav Kavan},
  {and} \bibinfo{person}{Mark Pauly}.} \bibinfo{year}{2014}\natexlab{}.
\newblock \showarticletitle{Projective dynamics: Fusing constraint projections
  for fast simulation}.
\newblock \bibinfo{journal}{\emph{ACM transactions on graphics (TOG)}}
  \bibinfo{volume}{33}, \bibinfo{number}{4} (\bibinfo{year}{2014}),
  \bibinfo{pages}{1--11}.
\newblock


\bibitem[Carr et~al\mbox{.}(2001)]%
        {carr2001reconstruction}
\bibfield{author}{\bibinfo{person}{Jonathan~C Carr}, \bibinfo{person}{Richard~K
  Beatson}, \bibinfo{person}{Jon~B Cherrie}, \bibinfo{person}{Tim~J Mitchell},
  \bibinfo{person}{W~Richard Fright}, \bibinfo{person}{Bruce~C McCallum}, {and}
  \bibinfo{person}{Tim~R Evans}.} \bibinfo{year}{2001}\natexlab{}.
\newblock \showarticletitle{Reconstruction and representation of 3D objects
  with radial basis functions}. In \bibinfo{booktitle}{\emph{Proceedings of the
  28th annual conference on Computer graphics and interactive techniques}}.
  \bibinfo{pages}{67--76}.
\newblock


\bibitem[Chabra et~al\mbox{.}(2020)]%
        {chabra2020deep}
\bibfield{author}{\bibinfo{person}{Rohan Chabra}, \bibinfo{person}{Jan~E
  Lenssen}, \bibinfo{person}{Eddy Ilg}, \bibinfo{person}{Tanner Schmidt},
  \bibinfo{person}{Julian Straub}, \bibinfo{person}{Steven Lovegrove}, {and}
  \bibinfo{person}{Richard Newcombe}.} \bibinfo{year}{2020}\natexlab{}.
\newblock \showarticletitle{Deep local shapes: Learning local sdf priors for
  detailed 3d reconstruction}. In \bibinfo{booktitle}{\emph{European Conference
  on Computer Vision}}. Springer, \bibinfo{pages}{608--625}.
\newblock


\bibitem[Chan et~al\mbox{.}(2021)]%
        {chan2021pi}
\bibfield{author}{\bibinfo{person}{Eric~R Chan}, \bibinfo{person}{Marco
  Monteiro}, \bibinfo{person}{Petr Kellnhofer}, \bibinfo{person}{Jiajun Wu},
  {and} \bibinfo{person}{Gordon Wetzstein}.} \bibinfo{year}{2021}\natexlab{}.
\newblock \showarticletitle{pi-gan: Periodic implicit generative adversarial
  networks for 3d-aware image synthesis}. In
  \bibinfo{booktitle}{\emph{Proceedings of the IEEE/CVF conference on computer
  vision and pattern recognition}}. \bibinfo{pages}{5799--5809}.
\newblock


\bibitem[Chen et~al\mbox{.}(2023)]%
        {chen2023model}
\bibfield{author}{\bibinfo{person}{Peter~Yichen Chen},
  \bibinfo{person}{Maurizio~M Chiaramonte}, \bibinfo{person}{Eitan Grinspun},
  {and} \bibinfo{person}{Kevin Carlberg}.} \bibinfo{year}{2023}\natexlab{}.
\newblock \showarticletitle{Model reduction for the material point method via
  an implicit neural representation of the deformation map}.
\newblock \bibinfo{journal}{\emph{J. Comput. Phys.}}  \bibinfo{volume}{478}
  (\bibinfo{year}{2023}), \bibinfo{pages}{111908}.
\newblock


\bibitem[Chen et~al\mbox{.}(2021)]%
        {chen2021learning}
\bibfield{author}{\bibinfo{person}{Yinbo Chen}, \bibinfo{person}{Sifei Liu},
  {and} \bibinfo{person}{Xiaolong Wang}.} \bibinfo{year}{2021}\natexlab{}.
\newblock \showarticletitle{Learning continuous image representation with local
  implicit image function}. In \bibinfo{booktitle}{\emph{Proceedings of the
  IEEE/CVF conference on computer vision and pattern recognition}}.
  \bibinfo{pages}{8628--8638}.
\newblock


\bibitem[Choi and Blemker(2013)]%
        {choi2013skeletal}
\bibfield{author}{\bibinfo{person}{Hon~Fai Choi} {and}
  \bibinfo{person}{Silvia~S. Blemker}.} \bibinfo{year}{2013}\natexlab{}.
\newblock \showarticletitle{Skeletal Muscle Fascicle Arrangements Can Be
  Reconstructed Using a Laplacian Vector Field Simulation}.
\newblock \bibinfo{journal}{\emph{PLOS ONE}} \bibinfo{volume}{8},
  \bibinfo{number}{10} (\bibinfo{date}{10} \bibinfo{year}{2013}),
  \bibinfo{pages}{1--7}.
\newblock
\urldef\tempurl%
\url{https://doi.org/10.1371/journal.pone.0077576}
\showDOI{\tempurl}


\bibitem[Cong et~al\mbox{.}(2015)]%
        {cong2015fully}
\bibfield{author}{\bibinfo{person}{Matthew Cong}, \bibinfo{person}{Michael
  Bao}, \bibinfo{person}{Jane~L E}, \bibinfo{person}{Kiran~S Bhat}, {and}
  \bibinfo{person}{Ronald Fedkiw}.} \bibinfo{year}{2015}\natexlab{}.
\newblock \showarticletitle{Fully automatic generation of anatomical face
  simulation models}. In \bibinfo{booktitle}{\emph{Proceedings of the 14th ACM
  SIGGRAPH/Eurographics Symposium on Computer Animation}}.
  \bibinfo{pages}{175--183}.
\newblock


\bibitem[Cong et~al\mbox{.}(2016)]%
        {cong2016art}
\bibfield{author}{\bibinfo{person}{Matthew Cong}, \bibinfo{person}{Kiran~S.
  Bhat}, {and} \bibinfo{person}{Ronald Fedkiw}.}
  \bibinfo{year}{2016}\natexlab{}.
\newblock \showarticletitle{Art-Directed Muscle Simulation for High-End Facial
  Animation}. In \bibinfo{booktitle}{\emph{Proceedings of the ACM
  SIGGRAPH/Eurographics Symposium on Computer Animation}} (Zurich, Switzerland)
  \emph{(\bibinfo{series}{SCA '16})}. \bibinfo{publisher}{Eurographics
  Association}, \bibinfo{address}{Goslar, DEU}, \bibinfo{pages}{119–127}.
\newblock
\showISBNx{9783905674613}


\bibitem[Crouse(2016)]%
        {crouse2016implementing}
\bibfield{author}{\bibinfo{person}{David~F Crouse}.}
  \bibinfo{year}{2016}\natexlab{}.
\newblock \showarticletitle{On implementing 2D rectangular assignment
  algorithms}.
\newblock \bibinfo{journal}{\emph{IEEE Trans. Aerospace Electron. Systems}}
  \bibinfo{volume}{52}, \bibinfo{number}{4} (\bibinfo{year}{2016}),
  \bibinfo{pages}{1679--1696}.
\newblock


\bibitem[Curless and Levoy(1996)]%
        {curless1996volumetric}
\bibfield{author}{\bibinfo{person}{Brian Curless} {and} \bibinfo{person}{Marc
  Levoy}.} \bibinfo{year}{1996}\natexlab{}.
\newblock \showarticletitle{A volumetric method for building complex models
  from range images}. In \bibinfo{booktitle}{\emph{Proceedings of the 23rd
  annual conference on Computer graphics and interactive techniques}}.
  \bibinfo{pages}{303--312}.
\newblock


\bibitem[Esmaeilzadeh et~al\mbox{.}(2020)]%
        {esmaeilzadeh2020meshfreeflownet}
\bibfield{author}{\bibinfo{person}{Soheil Esmaeilzadeh},
  \bibinfo{person}{Kamyar Azizzadenesheli}, \bibinfo{person}{Karthik
  Kashinath}, \bibinfo{person}{Mustafa Mustafa}, \bibinfo{person}{Hamdi~A
  Tchelepi}, \bibinfo{person}{Philip Marcus}, \bibinfo{person}{Mr Prabhat},
  \bibinfo{person}{Anima Anandkumar}, {et~al\mbox{.}}}
  \bibinfo{year}{2020}\natexlab{}.
\newblock \showarticletitle{Meshfreeflownet: A physics-constrained deep
  continuous space-time super-resolution framework}. In
  \bibinfo{booktitle}{\emph{SC20: International Conference for High Performance
  Computing, Networking, Storage and Analysis}}. IEEE, \bibinfo{pages}{1--15}.
\newblock


\bibitem[Fleishman et~al\mbox{.}(2005)]%
        {fleishman2005robust}
\bibfield{author}{\bibinfo{person}{Shachar Fleishman}, \bibinfo{person}{Daniel
  Cohen-Or}, {and} \bibinfo{person}{Cl{\'a}udio~T Silva}.}
  \bibinfo{year}{2005}\natexlab{}.
\newblock \showarticletitle{Robust moving least-squares fitting with sharp
  features}.
\newblock \bibinfo{journal}{\emph{ACM transactions on graphics (TOG)}}
  \bibinfo{volume}{24}, \bibinfo{number}{3} (\bibinfo{year}{2005}),
  \bibinfo{pages}{544--552}.
\newblock


\bibitem[Fulton et~al\mbox{.}(2019)]%
        {fmdlj19}
\bibfield{author}{\bibinfo{person}{Lawson Fulton}, \bibinfo{person}{Vismay
  Modi}, \bibinfo{person}{David Duvenaud}, \bibinfo{person}{David I.~W. Levin},
  {and} \bibinfo{person}{Alec Jacobson}.} \bibinfo{year}{2019}\natexlab{}.
\newblock \showarticletitle{Latent-space Dynamics for Reduced Deformable
  Simulation}.
\newblock \bibinfo{journal}{\emph{Computer Graphics Forum}}
  \bibinfo{volume}{38}, \bibinfo{number}{2} (\bibinfo{year}{2019}),
  \bibinfo{pages}{379--391}.
\newblock
\urldef\tempurl%
\url{https://doi.org/10.1111/cgf.13645}
\showDOI{\tempurl}
\showeprint{https://onlinelibrary.wiley.com/doi/pdf/10.1111/cgf.13645}


\bibitem[Hauth and Etzmuss(2001)]%
        {hauth2001high}
\bibfield{author}{\bibinfo{person}{Michael Hauth} {and} \bibinfo{person}{Olaf
  Etzmuss}.} \bibinfo{year}{2001}\natexlab{}.
\newblock \showarticletitle{A high performance solver for the animation of
  deformable objects using advanced numerical methods}. In
  \bibinfo{booktitle}{\emph{Computer Graphics Forum}},
  Vol.~\bibinfo{volume}{20}. Wiley Online Library, \bibinfo{pages}{319--328}.
\newblock


\bibitem[Higgins et~al\mbox{.}(2022)]%
        {higgins2022beta}
\bibfield{author}{\bibinfo{person}{Irina Higgins}, \bibinfo{person}{Loic
  Matthey}, \bibinfo{person}{Arka Pal}, \bibinfo{person}{Christopher Burgess},
  \bibinfo{person}{Xavier Glorot}, \bibinfo{person}{Matthew Botvinick},
  \bibinfo{person}{Shakir Mohamed}, {and} \bibinfo{person}{Alexander
  Lerchner}.} \bibinfo{year}{2022}\natexlab{}.
\newblock \showarticletitle{beta-VAE: Learning Basic Visual Concepts with a
  Constrained Variational Framework}. In
  \bibinfo{booktitle}{\emph{International Conference on Learning
  Representations}}.
\newblock


\bibitem[Jiang et~al\mbox{.}(2020)]%
        {jiang2020local}
\bibfield{author}{\bibinfo{person}{Chiyu Jiang}, \bibinfo{person}{Avneesh Sud},
  \bibinfo{person}{Ameesh Makadia}, \bibinfo{person}{Jingwei Huang},
  \bibinfo{person}{Matthias Nie{\ss}ner}, \bibinfo{person}{Thomas Funkhouser},
  {et~al\mbox{.}}} \bibinfo{year}{2020}\natexlab{}.
\newblock \showarticletitle{Local implicit grid representations for 3d scenes}.
  In \bibinfo{booktitle}{\emph{Proceedings of the IEEE/CVF Conference on
  Computer Vision and Pattern Recognition}}. \bibinfo{pages}{6001--6010}.
\newblock


\bibitem[Kazhdan et~al\mbox{.}(2006)]%
        {kazhdan2006poisson}
\bibfield{author}{\bibinfo{person}{Michael Kazhdan}, \bibinfo{person}{Matthew
  Bolitho}, {and} \bibinfo{person}{Hugues Hoppe}.}
  \bibinfo{year}{2006}\natexlab{}.
\newblock \showarticletitle{Poisson surface reconstruction}. In
  \bibinfo{booktitle}{\emph{Proceedings of the fourth Eurographics symposium on
  Geometry processing}}, Vol.~\bibinfo{volume}{7}.
\newblock


\bibitem[Kazhdan and Hoppe(2013)]%
        {kazhdan2013screened}
\bibfield{author}{\bibinfo{person}{Michael Kazhdan} {and}
  \bibinfo{person}{Hugues Hoppe}.} \bibinfo{year}{2013}\natexlab{}.
\newblock \showarticletitle{Screened poisson surface reconstruction}.
\newblock \bibinfo{journal}{\emph{ACM Transactions on Graphics (ToG)}}
  \bibinfo{volume}{32}, \bibinfo{number}{3} (\bibinfo{year}{2013}),
  \bibinfo{pages}{1--13}.
\newblock


\bibitem[Kharevych et~al\mbox{.}(2006)]%
        {kharevych2006geometric}
\bibfield{author}{\bibinfo{person}{Liliya Kharevych}, \bibinfo{person}{W Wei},
  \bibinfo{person}{Yiying Tong}, \bibinfo{person}{Eva Kanso},
  \bibinfo{person}{Jerrold~E Marsden}, \bibinfo{person}{Peter Schr{\"o}der},
  {and} \bibinfo{person}{Matthieu Desbrun}.} \bibinfo{year}{2006}\natexlab{}.
\newblock \bibinfo{booktitle}{\emph{Geometric, variational integrators for
  computer animation}}.
\newblock \bibinfo{publisher}{Eurographics Association}.
\newblock


\bibitem[Kim et~al\mbox{.}(2008)]%
        {kim2008wavelet}
\bibfield{author}{\bibinfo{person}{Theodore Kim}, \bibinfo{person}{Nils
  Th{\"u}rey}, \bibinfo{person}{Doug James}, {and} \bibinfo{person}{Markus
  Gross}.} \bibinfo{year}{2008}\natexlab{}.
\newblock \showarticletitle{Wavelet turbulence for fluid simulation}.
\newblock \bibinfo{journal}{\emph{ACM Transactions on Graphics (TOG)}}
  \bibinfo{volume}{27}, \bibinfo{number}{3} (\bibinfo{year}{2008}),
  \bibinfo{pages}{1--6}.
\newblock


\bibitem[Kingma and Ba(2014)]%
        {kingma2014adam}
\bibfield{author}{\bibinfo{person}{Diederik~P Kingma} {and}
  \bibinfo{person}{Jimmy Ba}.} \bibinfo{year}{2014}\natexlab{}.
\newblock \showarticletitle{Adam: A method for stochastic optimization}.
\newblock \bibinfo{journal}{\emph{arXiv preprint arXiv:1412.6980}}
  (\bibinfo{year}{2014}).
\newblock


\bibitem[Krysl et~al\mbox{.}(2001)]%
        {klm01}
\bibfield{author}{\bibinfo{person}{P. Krysl}, \bibinfo{person}{S. Lall}, {and}
  \bibinfo{person}{J.~E. Marsden}.} \bibinfo{year}{2001}\natexlab{}.
\newblock \showarticletitle{Dimensional model reduction in non-linear finite
  element dynamics of solids and structures}.
\newblock \bibinfo{journal}{\emph{Internat. J. Numer. Methods Engrg.}}
  \bibinfo{volume}{51}, \bibinfo{number}{4} (\bibinfo{year}{2001}),
  \bibinfo{pages}{479--504}.
\newblock
\urldef\tempurl%
\url{https://doi.org/10.1002/nme.167}
\showDOI{\tempurl}
\showeprint{https://onlinelibrary.wiley.com/doi/pdf/10.1002/nme.167}


\bibitem[Li et~al\mbox{.}(2021)]%
        {li20213d}
\bibfield{author}{\bibinfo{person}{Jiaxin Li}, \bibinfo{person}{Feiyu Zhu},
  \bibinfo{person}{Xiao Yang}, {and} \bibinfo{person}{Qijun Zhao}.}
  \bibinfo{year}{2021}\natexlab{}.
\newblock \showarticletitle{3D face point cloud super-resolution network}. In
  \bibinfo{booktitle}{\emph{2021 IEEE International Joint Conference on
  Biometrics (IJCB)}}. IEEE, \bibinfo{pages}{1--8}.
\newblock


\bibitem[Li et~al\mbox{.}(2019)]%
        {li2019pu}
\bibfield{author}{\bibinfo{person}{Ruihui Li}, \bibinfo{person}{Xianzhi Li},
  \bibinfo{person}{Chi-Wing Fu}, \bibinfo{person}{Daniel Cohen-Or}, {and}
  \bibinfo{person}{Pheng-Ann Heng}.} \bibinfo{year}{2019}\natexlab{}.
\newblock \showarticletitle{Pu-gan: a point cloud upsampling adversarial
  network}. In \bibinfo{booktitle}{\emph{Proceedings of the IEEE/CVF
  international conference on computer vision}}. \bibinfo{pages}{7203--7212}.
\newblock


\bibitem[Liang et~al\mbox{.}(2014)]%
        {liang20143d}
\bibfield{author}{\bibinfo{person}{Shu Liang}, \bibinfo{person}{Ira
  Kemelmacher-Shlizerman}, {and} \bibinfo{person}{Linda~G Shapiro}.}
  \bibinfo{year}{2014}\natexlab{}.
\newblock \showarticletitle{3d face hallucination from a single depth frame}.
  In \bibinfo{booktitle}{\emph{2014 2nd International Conference on 3D
  Vision}}, Vol.~\bibinfo{volume}{1}. IEEE, \bibinfo{pages}{31--38}.
\newblock


\bibitem[Liu et~al\mbox{.}(2013)]%
        {liu2013fast}
\bibfield{author}{\bibinfo{person}{Tiantian Liu}, \bibinfo{person}{Adam~W
  Bargteil}, \bibinfo{person}{James~F O'Brien}, {and} \bibinfo{person}{Ladislav
  Kavan}.} \bibinfo{year}{2013}\natexlab{}.
\newblock \showarticletitle{Fast simulation of mass-spring systems}.
\newblock \bibinfo{journal}{\emph{ACM Transactions on Graphics (TOG)}}
  \bibinfo{volume}{32}, \bibinfo{number}{6} (\bibinfo{year}{2013}),
  \bibinfo{pages}{1--7}.
\newblock


\bibitem[Liu et~al\mbox{.}(1995)]%
        {liu1995reproducing}
\bibfield{author}{\bibinfo{person}{Wing~Kam Liu}, \bibinfo{person}{Sukky Jun},
  {and} \bibinfo{person}{Yi~Fei Zhang}.} \bibinfo{year}{1995}\natexlab{}.
\newblock \showarticletitle{Reproducing kernel particle methods}.
\newblock \bibinfo{journal}{\emph{International journal for numerical methods
  in fluids}} \bibinfo{volume}{20}, \bibinfo{number}{8-9}
  (\bibinfo{year}{1995}), \bibinfo{pages}{1081--1106}.
\newblock


\bibitem[Ma et~al\mbox{.}(2021)]%
        {ma2021pixel}
\bibfield{author}{\bibinfo{person}{Shugao Ma}, \bibinfo{person}{Tomas Simon},
  \bibinfo{person}{Jason Saragih}, \bibinfo{person}{Dawei Wang},
  \bibinfo{person}{Yuecheng Li}, \bibinfo{person}{Fernando De~La~Torre}, {and}
  \bibinfo{person}{Yaser Sheikh}.} \bibinfo{year}{2021}\natexlab{}.
\newblock \showarticletitle{Pixel codec avatars}. In
  \bibinfo{booktitle}{\emph{Proceedings of the IEEE/CVF Conference on Computer
  Vision and Pattern Recognition}}. \bibinfo{pages}{64--73}.
\newblock


\bibitem[Macklin et~al\mbox{.}(2016)]%
        {macklin2016xpbd}
\bibfield{author}{\bibinfo{person}{Miles Macklin}, \bibinfo{person}{Matthias
  M{\"u}ller}, {and} \bibinfo{person}{Nuttapong Chentanez}.}
  \bibinfo{year}{2016}\natexlab{}.
\newblock \showarticletitle{XPBD: position-based simulation of compliant
  constrained dynamics}. In \bibinfo{booktitle}{\emph{Proceedings of the 9th
  International Conference on Motion in Games}}. \bibinfo{pages}{49--54}.
\newblock


\bibitem[Mescheder et~al\mbox{.}(2019)]%
        {mescheder2019occupancy}
\bibfield{author}{\bibinfo{person}{Lars Mescheder}, \bibinfo{person}{Michael
  Oechsle}, \bibinfo{person}{Michael Niemeyer}, \bibinfo{person}{Sebastian
  Nowozin}, {and} \bibinfo{person}{Andreas Geiger}.}
  \bibinfo{year}{2019}\natexlab{}.
\newblock \showarticletitle{Occupancy networks: Learning 3d reconstruction in
  function space}. In \bibinfo{booktitle}{\emph{Proceedings of the IEEE/CVF
  conference on computer vision and pattern recognition}}.
  \bibinfo{pages}{4460--4470}.
\newblock


\bibitem[Mildenhall et~al\mbox{.}(2021)]%
        {mildenhall2021nerf}
\bibfield{author}{\bibinfo{person}{Ben Mildenhall}, \bibinfo{person}{Pratul~P
  Srinivasan}, \bibinfo{person}{Matthew Tancik}, \bibinfo{person}{Jonathan~T
  Barron}, \bibinfo{person}{Ravi Ramamoorthi}, {and} \bibinfo{person}{Ren Ng}.}
  \bibinfo{year}{2021}\natexlab{}.
\newblock \showarticletitle{Nerf: Representing scenes as neural radiance fields
  for view synthesis}.
\newblock \bibinfo{journal}{\emph{Commun. ACM}} \bibinfo{volume}{65},
  \bibinfo{number}{1} (\bibinfo{year}{2021}), \bibinfo{pages}{99--106}.
\newblock


\bibitem[Molino et~al\mbox{.}(2003)]%
        {molino2003crystalline}
\bibfield{author}{\bibinfo{person}{Neil Molino}, \bibinfo{person}{Robert
  Bridson}, \bibinfo{person}{Joseph Teran}, {and} \bibinfo{person}{Ronald
  Fedkiw}.} \bibinfo{year}{2003}\natexlab{}.
\newblock \showarticletitle{A crystalline, red green strategy for meshing
  highly deformable objects with tetrahedra.}. In
  \bibinfo{booktitle}{\emph{IMR}}. Citeseer, \bibinfo{pages}{103--114}.
\newblock


\bibitem[M{\"u}ller et~al\mbox{.}(2007)]%
        {muller2007position}
\bibfield{author}{\bibinfo{person}{Matthias M{\"u}ller}, \bibinfo{person}{Bruno
  Heidelberger}, \bibinfo{person}{Marcus Hennix}, {and} \bibinfo{person}{John
  Ratcliff}.} \bibinfo{year}{2007}\natexlab{}.
\newblock \showarticletitle{Position based dynamics}.
\newblock \bibinfo{journal}{\emph{Journal of Visual Communication and Image
  Representation}} \bibinfo{volume}{18}, \bibinfo{number}{2}
  (\bibinfo{year}{2007}), \bibinfo{pages}{109--118}.
\newblock


\bibitem[Nagai et~al\mbox{.}(2009)]%
        {nagai2009smoothing}
\bibfield{author}{\bibinfo{person}{Yukie Nagai}, \bibinfo{person}{Yutaka
  Ohtake}, {and} \bibinfo{person}{Hiromasa Suzuki}.}
  \bibinfo{year}{2009}\natexlab{}.
\newblock \showarticletitle{Smoothing of partition of unity implicit surfaces
  for noise robust surface reconstruction}. In
  \bibinfo{booktitle}{\emph{Computer Graphics Forum}},
  Vol.~\bibinfo{volume}{28}. Wiley Online Library, \bibinfo{pages}{1339--1348}.
\newblock


\bibitem[Nasrollahi and Moeslund(2014)]%
        {nasrollahi2014super}
\bibfield{author}{\bibinfo{person}{Kamal Nasrollahi} {and}
  \bibinfo{person}{Thomas~B Moeslund}.} \bibinfo{year}{2014}\natexlab{}.
\newblock \showarticletitle{Super-resolution: a comprehensive survey}.
\newblock \bibinfo{journal}{\emph{Machine vision and applications}}
  \bibinfo{volume}{25}, \bibinfo{number}{6} (\bibinfo{year}{2014}),
  \bibinfo{pages}{1423--1468}.
\newblock


\bibitem[Ohtake et~al\mbox{.}(2005a)]%
        {ohtake2005sparse}
\bibfield{author}{\bibinfo{person}{Yutaka Ohtake}, \bibinfo{person}{Alexander
  Belyaev}, {and} \bibinfo{person}{Marc Alexa}.}
  \bibinfo{year}{2005}\natexlab{a}.
\newblock \showarticletitle{Sparse low-degree implicit surfaces with
  applications to high quality rendering, feature extraction, and smoothing}.
  In \bibinfo{booktitle}{\emph{Proc. Symp. Geometry Processing}}.
  \bibinfo{pages}{149--158}.
\newblock


\bibitem[Ohtake et~al\mbox{.}(2005b)]%
        {ohtake20053d}
\bibfield{author}{\bibinfo{person}{Yutaka Ohtake}, \bibinfo{person}{Alexander
  Belyaev}, {and} \bibinfo{person}{Hans-Peter Seidel}.}
  \bibinfo{year}{2005}\natexlab{b}.
\newblock \showarticletitle{3D scattered data interpolation and approximation
  with multilevel compactly supported RBFs}.
\newblock \bibinfo{journal}{\emph{Graphical Models}} \bibinfo{volume}{67},
  \bibinfo{number}{3} (\bibinfo{year}{2005}), \bibinfo{pages}{150--165}.
\newblock


\bibitem[Pan et~al\mbox{.}(2006)]%
        {pan2006super}
\bibfield{author}{\bibinfo{person}{Gang Pan}, \bibinfo{person}{Shi Han},
  \bibinfo{person}{Zhaohui Wu}, {and} \bibinfo{person}{Yueming Wang}.}
  \bibinfo{year}{2006}\natexlab{}.
\newblock \showarticletitle{Super-resolution of 3d face}. In
  \bibinfo{booktitle}{\emph{European Conference on Computer Vision}}. Springer,
  \bibinfo{pages}{389--401}.
\newblock


\bibitem[Park et~al\mbox{.}(2019)]%
        {park2019deepsdf}
\bibfield{author}{\bibinfo{person}{Jeong~Joon Park}, \bibinfo{person}{Peter
  Florence}, \bibinfo{person}{Julian Straub}, \bibinfo{person}{Richard
  Newcombe}, {and} \bibinfo{person}{Steven Lovegrove}.}
  \bibinfo{year}{2019}\natexlab{}.
\newblock \showarticletitle{Deepsdf: Learning continuous signed distance
  functions for shape representation}. In \bibinfo{booktitle}{\emph{Proceedings
  of the IEEE/CVF conference on computer vision and pattern recognition}}.
  \bibinfo{pages}{165--174}.
\newblock


\bibitem[Peng et~al\mbox{.}(2005)]%
        {peng2005learning}
\bibfield{author}{\bibinfo{person}{Shiqi Peng}, \bibinfo{person}{Gang Pan},
  {and} \bibinfo{person}{Zhaohui Wu}.} \bibinfo{year}{2005}\natexlab{}.
\newblock \showarticletitle{Learning-based super-resolution of 3D face model}.
  In \bibinfo{booktitle}{\emph{IEEE International Conference on Image
  Processing 2005}}, Vol.~\bibinfo{volume}{2}. IEEE, \bibinfo{pages}{II--382}.
\newblock


\bibitem[Qian et~al\mbox{.}(2021)]%
        {qian2021pu}
\bibfield{author}{\bibinfo{person}{Guocheng Qian}, \bibinfo{person}{Abdulellah
  Abualshour}, \bibinfo{person}{Guohao Li}, \bibinfo{person}{Ali Thabet}, {and}
  \bibinfo{person}{Bernard Ghanem}.} \bibinfo{year}{2021}\natexlab{}.
\newblock \showarticletitle{Pu-gcn: Point cloud upsampling using graph
  convolutional networks}. In \bibinfo{booktitle}{\emph{Proceedings of the
  IEEE/CVF Conference on Computer Vision and Pattern Recognition}}.
  \bibinfo{pages}{11683--11692}.
\newblock


\bibitem[Qian et~al\mbox{.}(2020)]%
        {qian2020pugeo}
\bibfield{author}{\bibinfo{person}{Yue Qian}, \bibinfo{person}{Junhui Hou},
  \bibinfo{person}{Sam Kwong}, {and} \bibinfo{person}{Ying He}.}
  \bibinfo{year}{2020}\natexlab{}.
\newblock \showarticletitle{PUGeo-Net: A geometry-centric network for 3D point
  cloud upsampling}. In \bibinfo{booktitle}{\emph{European conference on
  computer vision}}. Springer, \bibinfo{pages}{752--769}.
\newblock


\bibitem[Romero et~al\mbox{.}(2017)]%
        {rtb17}
\bibfield{author}{\bibinfo{person}{Javier Romero}, \bibinfo{person}{Dimitrios
  Tzionas}, {and} \bibinfo{person}{Michael~J. Black}.}
  \bibinfo{year}{2017}\natexlab{}.
\newblock \showarticletitle{Embodied Hands: Modeling and Capturing Hands and
  Bodies Together}.
\newblock \bibinfo{journal}{\emph{ACM Trans. Graph.}} \bibinfo{volume}{36},
  \bibinfo{number}{6}, Article \bibinfo{articleno}{245} (\bibinfo{date}{nov}
  \bibinfo{year}{2017}), \bibinfo{numpages}{17}~pages.
\newblock
\showISSN{0730-0301}
\urldef\tempurl%
\url{https://doi.org/10.1145/3130800.3130883}
\showDOI{\tempurl}


\bibitem[Saito et~al\mbox{.}(2019)]%
        {saito2019pifu}
\bibfield{author}{\bibinfo{person}{Shunsuke Saito}, \bibinfo{person}{Zeng
  Huang}, \bibinfo{person}{Ryota Natsume}, \bibinfo{person}{Shigeo Morishima},
  \bibinfo{person}{Angjoo Kanazawa}, {and} \bibinfo{person}{Hao Li}.}
  \bibinfo{year}{2019}\natexlab{}.
\newblock \showarticletitle{Pifu: Pixel-aligned implicit function for
  high-resolution clothed human digitization}. In
  \bibinfo{booktitle}{\emph{Proceedings of the IEEE/CVF International
  Conference on Computer Vision}}. \bibinfo{pages}{2304--2314}.
\newblock


\bibitem[Shen et~al\mbox{.}(2021)]%
        {shen2021high}
\bibfield{author}{\bibinfo{person}{S Shen}, \bibinfo{person}{Y Yang},
  \bibinfo{person}{T Shao}, \bibinfo{person}{H Wang}, \bibinfo{person}{C
  Jiang}, \bibinfo{person}{L Lan}, {and} \bibinfo{person}{K Zhou}.}
  \bibinfo{year}{2021}\natexlab{}.
\newblock \showarticletitle{High-order Differentiable Autoencoder for Nonlinear
  Model Reduction}.
\newblock \bibinfo{journal}{\emph{ACM Transactions on Graphics}}
  \bibinfo{volume}{40}, \bibinfo{number}{4} (\bibinfo{year}{2021}).
\newblock


\bibitem[Sifakis et~al\mbox{.}(2005)]%
        {sifakis2005automatic}
\bibfield{author}{\bibinfo{person}{Eftychios Sifakis}, \bibinfo{person}{Igor
  Neverov}, {and} \bibinfo{person}{Ronald Fedkiw}.}
  \bibinfo{year}{2005}\natexlab{}.
\newblock \showarticletitle{Automatic determination of facial muscle
  activations from sparse motion capture marker data}.
\newblock In \bibinfo{booktitle}{\emph{ACM SIGGRAPH 2005 Papers}}.
  \bibinfo{pages}{417--425}.
\newblock


\bibitem[Sitzmann et~al\mbox{.}(2020)]%
        {sitzmann2020implicit}
\bibfield{author}{\bibinfo{person}{Vincent Sitzmann}, \bibinfo{person}{Julien
  Martel}, \bibinfo{person}{Alexander Bergman}, \bibinfo{person}{David
  Lindell}, {and} \bibinfo{person}{Gordon Wetzstein}.}
  \bibinfo{year}{2020}\natexlab{}.
\newblock \showarticletitle{Implicit neural representations with periodic
  activation functions}.
\newblock \bibinfo{journal}{\emph{Advances in Neural Information Processing
  Systems}}  \bibinfo{volume}{33} (\bibinfo{year}{2020}),
  \bibinfo{pages}{7462--7473}.
\newblock


\bibitem[Srinivasan et~al\mbox{.}(2021)]%
        {srinivasan2021learning}
\bibfield{author}{\bibinfo{person}{Sangeetha~Grama Srinivasan},
  \bibinfo{person}{Qisi Wang}, \bibinfo{person}{Junior Rojas},
  \bibinfo{person}{Gergely Kl{\'a}r}, \bibinfo{person}{Ladislav Kavan}, {and}
  \bibinfo{person}{Eftychios Sifakis}.} \bibinfo{year}{2021}\natexlab{}.
\newblock \showarticletitle{Learning active quasistatic physics-based models
  from data}.
\newblock \bibinfo{journal}{\emph{ACM Transactions on Graphics (TOG)}}
  \bibinfo{volume}{40}, \bibinfo{number}{4} (\bibinfo{year}{2021}),
  \bibinfo{pages}{1--14}.
\newblock


\bibitem[Stam(2009)]%
        {stam2009nucleus}
\bibfield{author}{\bibinfo{person}{Jos Stam}.} \bibinfo{year}{2009}\natexlab{}.
\newblock \showarticletitle{Nucleus: Towards a unified dynamics solver for
  computer graphics}. In \bibinfo{booktitle}{\emph{2009 11th IEEE International
  Conference on Computer-Aided Design and Computer Graphics}}. IEEE,
  \bibinfo{pages}{1--11}.
\newblock


\bibitem[Stern and Grinspun(2009)]%
        {stern2009implicit}
\bibfield{author}{\bibinfo{person}{Ari Stern} {and} \bibinfo{person}{Eitan
  Grinspun}.} \bibinfo{year}{2009}\natexlab{}.
\newblock \showarticletitle{Implicit-explicit variational integration of highly
  oscillatory problems}.
\newblock \bibinfo{journal}{\emph{Multiscale Modeling \& Simulation}}
  \bibinfo{volume}{7}, \bibinfo{number}{4} (\bibinfo{year}{2009}),
  \bibinfo{pages}{1779--1794}.
\newblock


\bibitem[Su et~al\mbox{.}(2013)]%
        {su2013energy}
\bibfield{author}{\bibinfo{person}{Jonathan Su}, \bibinfo{person}{Rahul Sheth},
  {and} \bibinfo{person}{Ronald Fedkiw}.} \bibinfo{year}{2013}\natexlab{}.
\newblock \showarticletitle{Energy conservation for the simulation of
  deformable bodies.}
\newblock \bibinfo{journal}{\emph{IEEE Trans. Vis. Comput. Graph.}}
  \bibinfo{volume}{19}, \bibinfo{number}{2} (\bibinfo{year}{2013}),
  \bibinfo{pages}{189--200}.
\newblock


\bibitem[Tancik et~al\mbox{.}(2020)]%
        {tancik2020fourier}
\bibfield{author}{\bibinfo{person}{Matthew Tancik}, \bibinfo{person}{Pratul
  Srinivasan}, \bibinfo{person}{Ben Mildenhall}, \bibinfo{person}{Sara
  Fridovich-Keil}, \bibinfo{person}{Nithin Raghavan}, \bibinfo{person}{Utkarsh
  Singhal}, \bibinfo{person}{Ravi Ramamoorthi}, \bibinfo{person}{Jonathan
  Barron}, {and} \bibinfo{person}{Ren Ng}.} \bibinfo{year}{2020}\natexlab{}.
\newblock \showarticletitle{Fourier features let networks learn high frequency
  functions in low dimensional domains}.
\newblock \bibinfo{journal}{\emph{Advances in Neural Information Processing
  Systems}}  \bibinfo{volume}{33} (\bibinfo{year}{2020}),
  \bibinfo{pages}{7537--7547}.
\newblock


\bibitem[Tapia et~al\mbox{.}(2021)]%
        {trpo21}
\bibfield{author}{\bibinfo{person}{Javier Tapia}, \bibinfo{person}{Cristian
  Romero}, \bibinfo{person}{Jesús Pérez}, {and} \bibinfo{person}{Miguel~A.
  Otaduy}.} \bibinfo{year}{2021}\natexlab{}.
\newblock \showarticletitle{Parametric Skeletons with Reduced Soft-Tissue
  Deformations}.
\newblock \bibinfo{journal}{\emph{Computer Graphics Forum}}
  \bibinfo{volume}{40}, \bibinfo{number}{6} (\bibinfo{year}{2021}),
  \bibinfo{pages}{34--46}.
\newblock
\urldef\tempurl%
\url{https://doi.org/10.1111/cgf.14199}
\showDOI{\tempurl}
\showeprint{https://onlinelibrary.wiley.com/doi/pdf/10.1111/cgf.14199}


\bibitem[Teng et~al\mbox{.}(2015)]%
        {tmdk15}
\bibfield{author}{\bibinfo{person}{Yun Teng}, \bibinfo{person}{Mark Meyer},
  \bibinfo{person}{Tony DeRose}, {and} \bibinfo{person}{Theodore Kim}.}
  \bibinfo{year}{2015}\natexlab{}.
\newblock \showarticletitle{Subspace Condensation: Full Space Adaptivity for
  Subspace Deformations}.
\newblock \bibinfo{journal}{\emph{ACM Trans. Graph.}} \bibinfo{volume}{34},
  \bibinfo{number}{4}, Article \bibinfo{articleno}{76} (\bibinfo{date}{jul}
  \bibinfo{year}{2015}), \bibinfo{numpages}{9}~pages.
\newblock
\showISSN{0730-0301}
\urldef\tempurl%
\url{https://doi.org/10.1145/2766904}
\showDOI{\tempurl}


\bibitem[Teran et~al\mbox{.}(2003)]%
        {teran2003finite}
\bibfield{author}{\bibinfo{person}{Joseph Teran}, \bibinfo{person}{Sylvia
  Blemker}, \bibinfo{person}{V~Ng~Thow Hing}, {and} \bibinfo{person}{Ronald
  Fedkiw}.} \bibinfo{year}{2003}\natexlab{}.
\newblock \showarticletitle{Finite volume methods for the simulation of
  skeletal muscle}. In \bibinfo{booktitle}{\emph{Proceedings of the 2003 ACM
  SIGGRAPH/Eurographics symposium on Computer animation}}.
  \bibinfo{pages}{68--74}.
\newblock


\bibitem[Teran et~al\mbox{.}(2005)]%
        {teran2005robust}
\bibfield{author}{\bibinfo{person}{Joseph Teran}, \bibinfo{person}{Eftychios
  Sifakis}, \bibinfo{person}{Geoffrey Irving}, {and} \bibinfo{person}{Ronald
  Fedkiw}.} \bibinfo{year}{2005}\natexlab{}.
\newblock \showarticletitle{Robust quasistatic finite elements and flesh
  simulation}. In \bibinfo{booktitle}{\emph{Proceedings of the 2005 ACM
  SIGGRAPH/Eurographics symposium on Computer animation}}.
  \bibinfo{pages}{181--190}.
\newblock


\bibitem[Turk and O'brien(2002)]%
        {turk2002modelling}
\bibfield{author}{\bibinfo{person}{Greg Turk} {and} \bibinfo{person}{James~F
  O'brien}.} \bibinfo{year}{2002}\natexlab{}.
\newblock \showarticletitle{Modelling with implicit surfaces that interpolate}.
\newblock \bibinfo{journal}{\emph{ACM Transactions on Graphics (TOG)}}
  \bibinfo{volume}{21}, \bibinfo{number}{4} (\bibinfo{year}{2002}),
  \bibinfo{pages}{855--873}.
\newblock


\bibitem[Vaswani et~al\mbox{.}(2017)]%
        {vaswani2017attention}
\bibfield{author}{\bibinfo{person}{Ashish Vaswani}, \bibinfo{person}{Noam
  Shazeer}, \bibinfo{person}{Niki Parmar}, \bibinfo{person}{Jakob Uszkoreit},
  \bibinfo{person}{Llion Jones}, \bibinfo{person}{Aidan~N Gomez},
  \bibinfo{person}{{\L}ukasz Kaiser}, {and} \bibinfo{person}{Illia
  Polosukhin}.} \bibinfo{year}{2017}\natexlab{}.
\newblock \showarticletitle{Attention is all you need}.
\newblock \bibinfo{journal}{\emph{Advances in neural information processing
  systems}}  \bibinfo{volume}{30} (\bibinfo{year}{2017}).
\newblock


\bibitem[Wang et~al\mbox{.}(2015)]%
        {wjbk15}
\bibfield{author}{\bibinfo{person}{Yu Wang}, \bibinfo{person}{Alec Jacobson},
  \bibinfo{person}{Jernej Barbi\v{c}}, {and} \bibinfo{person}{Ladislav Kavan}.}
  \bibinfo{year}{2015}\natexlab{}.
\newblock \showarticletitle{Linear Subspace Design for Real-Time Shape
  Deformation}.
\newblock \bibinfo{journal}{\emph{ACM Trans. Graph.}} \bibinfo{volume}{34},
  \bibinfo{number}{4}, Article \bibinfo{articleno}{57} (\bibinfo{date}{jul}
  \bibinfo{year}{2015}), \bibinfo{numpages}{11}~pages.
\newblock
\showISSN{0730-0301}
\urldef\tempurl%
\url{https://doi.org/10.1145/2766952}
\showDOI{\tempurl}


\bibitem[Wang et~al\mbox{.}(2019)]%
        {wang2019dynamic}
\bibfield{author}{\bibinfo{person}{Yue Wang}, \bibinfo{person}{Yongbin Sun},
  \bibinfo{person}{Ziwei Liu}, \bibinfo{person}{Sanjay~E Sarma},
  \bibinfo{person}{Michael~M Bronstein}, {and} \bibinfo{person}{Justin~M
  Solomon}.} \bibinfo{year}{2019}\natexlab{}.
\newblock \showarticletitle{Dynamic graph cnn for learning on point clouds}.
\newblock \bibinfo{journal}{\emph{Acm Transactions On Graphics (tog)}}
  \bibinfo{volume}{38}, \bibinfo{number}{5} (\bibinfo{year}{2019}),
  \bibinfo{pages}{1--12}.
\newblock


\bibitem[Xie et~al\mbox{.}(2018)]%
        {xie2018tempogan}
\bibfield{author}{\bibinfo{person}{You Xie}, \bibinfo{person}{Erik Franz},
  \bibinfo{person}{Mengyu Chu}, {and} \bibinfo{person}{Nils Thuerey}.}
  \bibinfo{year}{2018}\natexlab{}.
\newblock \showarticletitle{tempoGAN: A temporally coherent, volumetric GAN for
  super-resolution fluid flow}.
\newblock \bibinfo{journal}{\emph{ACM Transactions on Graphics (TOG)}}
  \bibinfo{volume}{37}, \bibinfo{number}{4} (\bibinfo{year}{2018}),
  \bibinfo{pages}{1--15}.
\newblock


\bibitem[Yang et~al\mbox{.}(2022)]%
        {yang2022implicit}
\bibfield{author}{\bibinfo{person}{Lingchen Yang}, \bibinfo{person}{Byungsoo
  Kim}, \bibinfo{person}{Gaspard Zoss}, \bibinfo{person}{Baran G{\"o}zc{\"u}},
  \bibinfo{person}{Markus Gross}, {and} \bibinfo{person}{Barbara Solenthaler}.}
  \bibinfo{year}{2022}\natexlab{}.
\newblock \showarticletitle{Implicit neural representation for physics-driven
  actuated soft bodies}.
\newblock \bibinfo{journal}{\emph{ACM Transactions on Graphics (TOG)}}
  \bibinfo{volume}{41}, \bibinfo{number}{4} (\bibinfo{year}{2022}),
  \bibinfo{pages}{1--10}.
\newblock


\bibitem[Yang et~al\mbox{.}(2019)]%
        {yang2019deep}
\bibfield{author}{\bibinfo{person}{Wenming Yang}, \bibinfo{person}{Xuechen
  Zhang}, \bibinfo{person}{Yapeng Tian}, \bibinfo{person}{Wei Wang},
  \bibinfo{person}{Jing-Hao Xue}, {and} \bibinfo{person}{Qingmin Liao}.}
  \bibinfo{year}{2019}\natexlab{}.
\newblock \showarticletitle{Deep learning for single image super-resolution: A
  brief review}.
\newblock \bibinfo{journal}{\emph{IEEE Transactions on Multimedia}}
  \bibinfo{volume}{21}, \bibinfo{number}{12} (\bibinfo{year}{2019}),
  \bibinfo{pages}{3106--3121}.
\newblock


\bibitem[Ye et~al\mbox{.}(2021)]%
        {ye2021meta}
\bibfield{author}{\bibinfo{person}{Shuquan Ye}, \bibinfo{person}{Dongdong
  Chen}, \bibinfo{person}{Songfang Han}, \bibinfo{person}{Ziyu Wan}, {and}
  \bibinfo{person}{Jing Liao}.} \bibinfo{year}{2021}\natexlab{}.
\newblock \showarticletitle{Meta-PU: An arbitrary-scale upsampling network for
  point cloud}.
\newblock \bibinfo{journal}{\emph{IEEE transactions on visualization and
  computer graphics}} (\bibinfo{year}{2021}).
\newblock


\bibitem[Yifan et~al\mbox{.}(2019)]%
        {yifan2019patch}
\bibfield{author}{\bibinfo{person}{Wang Yifan}, \bibinfo{person}{Shihao Wu},
  \bibinfo{person}{Hui Huang}, \bibinfo{person}{Daniel Cohen-Or}, {and}
  \bibinfo{person}{Olga Sorkine-Hornung}.} \bibinfo{year}{2019}\natexlab{}.
\newblock \showarticletitle{Patch-based progressive 3d point set upsampling}.
  In \bibinfo{booktitle}{\emph{Proceedings of the IEEE/CVF Conference on
  Computer Vision and Pattern Recognition}}. \bibinfo{pages}{5958--5967}.
\newblock


\bibitem[Yu et~al\mbox{.}(2018a)]%
        {yu2018ec}
\bibfield{author}{\bibinfo{person}{Lequan Yu}, \bibinfo{person}{Xianzhi Li},
  \bibinfo{person}{Chi-Wing Fu}, \bibinfo{person}{Daniel Cohen-Or}, {and}
  \bibinfo{person}{Pheng-Ann Heng}.} \bibinfo{year}{2018}\natexlab{a}.
\newblock \showarticletitle{Ec-net: an edge-aware point set consolidation
  network}. In \bibinfo{booktitle}{\emph{Proceedings of the European conference
  on computer vision (ECCV)}}. \bibinfo{pages}{386--402}.
\newblock


\bibitem[Yu et~al\mbox{.}(2018b)]%
        {yu2018pu}
\bibfield{author}{\bibinfo{person}{Lequan Yu}, \bibinfo{person}{Xianzhi Li},
  \bibinfo{person}{Chi-Wing Fu}, \bibinfo{person}{Daniel Cohen-Or}, {and}
  \bibinfo{person}{Pheng-Ann Heng}.} \bibinfo{year}{2018}\natexlab{b}.
\newblock \showarticletitle{Pu-net: Point cloud upsampling network}. In
  \bibinfo{booktitle}{\emph{Proceedings of the IEEE conference on computer
  vision and pattern recognition}}. \bibinfo{pages}{2790--2799}.
\newblock


\bibitem[Zhang et~al\mbox{.}(2020)]%
        {zhang20203d}
\bibfield{author}{\bibinfo{person}{Fan Zhang}, \bibinfo{person}{Junli Zhao},
  \bibinfo{person}{Liang Wang}, {and} \bibinfo{person}{Fuqing Duan}.}
  \bibinfo{year}{2020}\natexlab{}.
\newblock \showarticletitle{3d face model super-resolution based on radial
  curve estimation}.
\newblock \bibinfo{journal}{\emph{Applied Sciences}} \bibinfo{volume}{10},
  \bibinfo{number}{3} (\bibinfo{year}{2020}), \bibinfo{pages}{1047}.
\newblock


\bibitem[Zhang et~al\mbox{.}(2021)]%
        {zhang2021deep}
\bibfield{author}{\bibinfo{person}{Meng Zhang}, \bibinfo{person}{Tuanfeng
  Wang}, \bibinfo{person}{Duygu Ceylan}, {and} \bibinfo{person}{Niloy~J
  Mitra}.} \bibinfo{year}{2021}\natexlab{}.
\newblock \showarticletitle{Deep detail enhancement for any garment}. In
  \bibinfo{booktitle}{\emph{Computer Graphics Forum}},
  Vol.~\bibinfo{volume}{40}. Wiley Online Library, \bibinfo{pages}{399--411}.
\newblock


\bibitem[Zhang et~al\mbox{.}(2022)]%
        {zhang2022point}
\bibfield{author}{\bibinfo{person}{Yan Zhang}, \bibinfo{person}{Wenhan Zhao},
  \bibinfo{person}{Bo Sun}, \bibinfo{person}{Ying Zhang}, {and}
  \bibinfo{person}{Wen Wen}.} \bibinfo{year}{2022}\natexlab{}.
\newblock \showarticletitle{Point Cloud Upsampling Algorithm: A Systematic
  Review}.
\newblock \bibinfo{journal}{\emph{Algorithms}} \bibinfo{volume}{15},
  \bibinfo{number}{4} (\bibinfo{year}{2022}), \bibinfo{pages}{124}.
\newblock


\bibitem[Zong et~al\mbox{.}(2023)]%
        {zong2023neural}
\bibfield{author}{\bibinfo{person}{Zeshun Zong}, \bibinfo{person}{Xuan Li},
  \bibinfo{person}{Minchen Li}, \bibinfo{person}{Maurizio~M Chiaramonte},
  \bibinfo{person}{Wojciech Matusik}, \bibinfo{person}{Eitan Grinspun},
  \bibinfo{person}{Kevin Carlberg}, \bibinfo{person}{Chenfanfu Jiang}, {and}
  \bibinfo{person}{Peter~Yichen Chen}.} \bibinfo{year}{2023}\natexlab{}.
\newblock \showarticletitle{Neural Stress Fields for Reduced-order
  Elastoplasticity and Fracture}. In \bibinfo{booktitle}{\emph{SIGGRAPH Asia
  2023 Conference Papers}}. \bibinfo{pages}{1--11}.
\newblock


\bibitem[Zurdo et~al\mbox{.}(2012)]%
        {zurdo2012animating}
\bibfield{author}{\bibinfo{person}{Javier~S Zurdo}, \bibinfo{person}{Juan~P
  Brito}, {and} \bibinfo{person}{Miguel~A Otaduy}.}
  \bibinfo{year}{2012}\natexlab{}.
\newblock \showarticletitle{Animating wrinkles by example on non-skinned
  cloth}.
\newblock \bibinfo{journal}{\emph{IEEE Transactions on Visualization and
  Computer Graphics}} \bibinfo{volume}{19}, \bibinfo{number}{1}
  (\bibinfo{year}{2012}), \bibinfo{pages}{149--158}.
\newblock


\end{thebibliography}

\pagebreak
\clearpage
\appendix
\section{Appendix}

\subsection{Additional information of our framework} \label{appendix:arch}
\subsubsection{Neural-network architecture}
We report the specifications of parameters in the implemented model in Table \ref{tab:spec}, whose definitions and uses are as introduced in Section \ref{sec:method}. Our model is comprised of 706,871 trainable parameters.
\begin{table}[h]
    \caption{Specifications of parameters in the implemented model.}
    \label{tab:spec}
    \centering
    \begin{tabularx}{\linewidth}{Y||Y}
        \hline
        Notation & Value\\
        \hline
        $N$ (num. of LR volumetric mesh vertices) & 15,872\\
        \hline
        $M$ (num. HR surface mesh vertices) & 35,637\\
        \hline
        S (num. of submodule layers in Feature Encoding network) & 2\\
        \hline
        $D_0$ & 35\\
        \hline
        $D_1$ & 64\\
        \hline
        $D_2$ & 128\\
        \hline
        $\alpha$ in Eq. (\ref{eq:loss}) & 0.001\\
        \hline
        $\beta$ in Eq. (\ref{eq:loss}) & gradually increased from 0.001 to 20\\
        \hline
        $k$ neighbors in the $k$-NN graphs from Feature Encoding networks & 5\\
        \hline
        Interpolation neighbors for Coordinate-based Upsampling & 20\\
        \hline
    \end{tabularx}
\end{table}

\subsubsection{Training Statistics}
Each training epoch takes 45s on a workstation with 2 NVLink-connected NVIDIA RTX A6000 GPUs, for a batch size of 6. We trained the model for 2800 epochs (which took about 35 hours on the 2 GPU workstation). We used Adam \cite{kingma2014adam} to optimize the loss with a learning rate of 1e-4.

\subsection{Additional information of compared models}
\subsubsection{Radial Basis Function (RBF)} \label{appendix:rbf}
Following the standard RBF techniques \cite{anjyo2014scattered},
we formulate our surface reconstruction based on RBF interpolation to predict the deformation vectors $\{\Delta \x^H_j\}_{j=1}^M$ for vertices on the HR surface mesh $\{\x^H_j\}_{j=1}^M$.

Each deformation vector of the LR mesh can be approximated as
\begin{align}
    \Delta \x^L_i = \sum_{k=1}^N \mathbf{w}_k \phi(||\x^L_i-\x^L_k||_2),
\end{align}
where $\{\mathbf{w}_k\in\R^3\}$ is the set of weights we wish to find, and $\phi(||\x^L_i-\x^L_k||_2)\in \R$ is the radial function centered at $\x^L_k$ modeled as the Gaussian function
\begin{equation}
\phi(R) = e^{-R^2/\sigma^2_{RBF}}
\end{equation}
We compute the distance measure $R(\cdot)$ geodesically following the method in Section \ref{ss:upsampling}, and use $\sigma_{RBF}^2=25$. The weights $\{\mathbf{w}_k\}$ then can be obtained by solving the following linear system in each frame:
\begin{equation}
    \underbrace{\begin{bmatrix}
    \phi_{1,1} & \hdots & \phi_{1,N}\\
    \vdots & \ddots &\vdots \\
    \phi_{N,1} & \hdots & \phi_{N,N}
    \end{bmatrix}}_{=: \text{ $\Phi$}}
    \begin{bmatrix}
        \mathbf{w}^T_1\\
        \vdots\\
        \mathbf{w}^T_N
    \end{bmatrix}
    = \begin{bmatrix}
        \Delta \x^L_1\\
        \vdots\\
        \Delta \x^L_N
    \end{bmatrix},
\end{equation} 
where $\Phi$ is invertible for the given Gaussian radial function.

Finally, the deformation vectors $\{\Delta \x^H_j\}^M_{j=1}$ of the HR surface mesh is calculated as
\begin{equation}
    \begin{bmatrix}
        \Delta \x^H_1\\
        \vdots\\
        \Delta \x^H_M
    \end{bmatrix}
    = \begin{bmatrix}
        \phi(||\x^H_1-\x^L_1||_2) & \hdots & \phi(||\x^H_1-\x^L_N||_2) \\
        \vdots &\ddots&\vdots \\
        \phi(||\x^H_M-\x^L_1||_2) &\hdots & \phi(||\x^H_M-\x^L_N||_2) \\
    \end{bmatrix}
    \begin{bmatrix}
        \mathbf{w}^T_1\\
        \vdots\\
        \mathbf{w}^T_N
    \end{bmatrix}.
\end{equation}

\subsubsection{Moving Least-Square (MLS)} \label{appendix:mls}
Similarly, following the standard MLS technique for approximating scalar functions \citep{anjyo2014scattered, liu1995reproducing} we formulate our MLS-based surface reconstruction as approximating each component of displacement vectors $[\Delta x^H_j, \Delta y^H_j, \Delta z^H_j] \in \R^3$ for every vertex on the HR surface mesh $\{\x^H_j\}_{j=1}^M$. 

The approximation is a linear combination of polynomials of degree $r$ (we use $r=2$) which, using the $y$ component (i.e., $\Delta y^H_j$) for an example, can be written as
\begin{equation}\label{eq:mls_delta}
    \Delta y^H_j = \mathbf{b}^T(y^L_k)\mathbf{c}(\x^H_j),
\end{equation}
where $\mathbf{b}(y) =: [1, y, y^2, ..., y^r]\in \R^{r+1}$ is the basis function, and $\mathbf{c}(\x^H_j)=[c_0, c_1, ..., c_r]\in \R^{r+1}$ is a vector of unknown coefficients dependent on $\x^H_j$, which we wish to find.

The coefficients can be obtained by solving the following weighted least-square problem:
\begin{equation}\label{eq:mls}
    \mathbf{c}(\x^H_j) = \arg \min_{\mathbf{c}\in \R^{r+1}} \sum_{k\in \mathcal{N}_j} w_k(||\x^H_j - \x^L_k ||_2) \left( \mathbf{b}^T(y^L_k)\mathbf{c}-\Delta y^L_k \right)^2,
\end{equation}
where $\mathcal{N}_j$ is a set of indices of LR mesh vertices neighboring $\x^H_j$ (we use the same 20 neighbors as defined in Section \ref{ss:upsampling}), and $w_k(R)$ is a weighting function modeled as
\begin{equation}
w_k(R)=e^{-R^2/\sigma_{MLS}^2},
\end{equation}
where we use the geodesic distance between $\x^H_j$ and $\x^L_k$ for the distance measure $R(\cdot)$ (as computed in Section \ref{ss:upsampling}), and use $\sigma_{MLS}^2=200$.

Then, $\mathbf{c}(\x^H_j)$ can be computed by differentiating Eq. (\ref{eq:mls}) w.r.t. $\mathbf{c}$ and setting it to zero:
\begin{align}\begin{split}
    &\frac{\partial}{\partial \mathbf{c}}\left( \sum_{k\in\mathcal{N}_j} w_k(||\x^H_j-\x^L_k||_2)\left(\mathbf{b}^T(y^L_k)\mathbf{c}-\Delta y^L_k\right)^2 \right) \Biggr\rvert_{\mathbf{c}(\x^H_j)}=0\\
   \Leftrightarrow & \underbrace{\left[ \sum_{k\in \mathcal{N}_j} w_k(||\x^H_j-\x^L_k||_2) \mathbf{b} (y^L_k)\mathbf{b}^T(y^L_k) \right]}_{=: \text{ $M$}} \mathbf{c}(\x^H_j)\\
    &= \underbrace{\sum_{k\in\mathcal{N}_j} w_k( || \x^H_j - \x^L_{k} ||_2 ) \Delta y^L_k \mathbf{b}(y^L_k)}_{=: \text{ $\mathbf{d}$}},
\end{split}\end{align}
and solving $\mathbf{c}=M^{-1}\mathbf{d}$, where the matrix $M$ is invertible for a non-negative value of $w_k(D)$.
For numerical stability, we re-center the polynomial basis around $\x^H_j$ \citep{liu1995reproducing}, replacing $\mathbf{b}(y^L_{k})$ with $\mathbf{b}(y^L_{k}-y^H_j)$ which reduces Eq. (\ref{eq:mls_delta}) to
\begin{equation}
    \Delta y^H_j = c_0.
\end{equation}

This process is repeated for each of $x$, $y$, $z$ components (i.e., $\Delta x^H_j, \Delta y^H_j$, and $\Delta z^H_j$) for every vertex on the HR mesh $\{\x^H_j\}^M_{j=1}$.

\subsubsection{$\beta$-Variational Auto Encoder}
\label{appendix:beta-vae}
We train a $\beta$-Variational Auto Encoder ($\beta$-VAE) \citep{higgins2022beta} to predict high-resolution displacements using low-resolution displacements as input to serve as a baseline generative neural network. The $\beta$-VAE has 2 fully connected layers in the encoder and 3 fully connected layers in the decoder. The encoder has 2 hidden layers with 1024 neurons in the first layer and 512 neurons in the second layer. The output of the encoder is composed of 256 neurons (128 neurons for the mean and 128 neurons for the variance). The decoder has 3 hidden layers with 256, 1024, and 4096 neurons. All the hidden layers use Leaky RELU activations. During every training epoch, the mean and variance output from the encoder are used to compute latent parameters by sampling from a normal distribution. To train the weights of this network, we compute the loss on the output displacements (L2-norm) and the KL-Divergence of the latent parameters. The former penalizes reconstruction error while the latter encourages disentanglement between latent parameters. The KL-Divergence term is also scaled by a hyperparameter $\beta$ which controls the degree of disentanglement between the latent parameters. We fixed $\beta$ to be 0.01 for this dataset and used Adam \citep{kingma2014adam} to train the network weights, with a learning rate of 1e-4. Since the input and output dimensions of our $\beta$-VAE are different, we do not design identical encoder and decoder architectures. We use the same partition for the train and test sets as our method. 

\subsubsection{Deep Detail Enhancement framework}\label{appendix:dde}
We compare with Deep Detail Enhancement (DDE) framework \cite{zhang2021deep} as the representative state-of-the-art method for synthesizing plausible wrinkle details on a coarse garment geometry based on normal maps. For implementation, we first bake two UV normal maps of size 512$\times$512 for each of the surface mesh embedded in the low-resolution (LR) simulation mesh (e.g., left image of Figure \ref{fig:dx_zoom}) and the surface conforming to the high-resolution (HR) simulation mesh (e.g., right image of Figure \ref{fig:dx_zoom}) on a frame-by-frame basis. Then, we train the DDE network (with U-Net architecture) to predict the HR normal map from its LR counterpart, baked from the training dataset.
We train on the full-size normal maps rather than randomly subsampled patches as in the original work and omit training of the garment material classifier since we have only one type of mesh, the face. Also, we added one layers of downsampling and upsampling, respectively, given our input dimension is larger compared to the original work (128$\times$128) and also follow the same energy-minimization method to recover 3D surfaces from the normal maps, initialized with the coarse embedded mesh.

\subsubsection{Decoder-style neural network for blendshape weights input}\label{appendix:arch-blendshape-decoder}
The decoder-style neural network in Section \ref{ss:blendshape} learns to predict per-vertex deformations of the high-resolution surface mesh (35,637 vertices) from the 38-dimensional input blendshape weights. Its architecture is comprised of fully-connected layers (\texttt{Linear(input dimension, output dimension)}) and Leaky-ReLU activations (\texttt{LeakyReLU(negative slope)}) with 443,840,125 trainable parameters. After the last layer, the vector of shape (106911,1) is reshaped to (35637,3) to obtain the per-vertex deformations.

\begin{tabularx}{\linewidth}{Y}
    \hline
    \texttt{Linear(38, 256)-LeakyReLU(0.01)} \\
    \hline 
    \texttt{Linear(256, 1024)-LeakyReLU(0.01)} \\
    \hline 
    \texttt{Linear(1024, 4096)-LeakyReLU(0.01)} \\
    \hline 
    \texttt{Linear(4096, 106911)}\\
    \hline
\end{tabularx}

\subsection{Approximate resolution of self-collision}\label{appendix:eval_self_col}
\begin{figure}[!htb]
    \centering
    \copyrightbox[r]
    {\includegraphics[width=\linewidth]{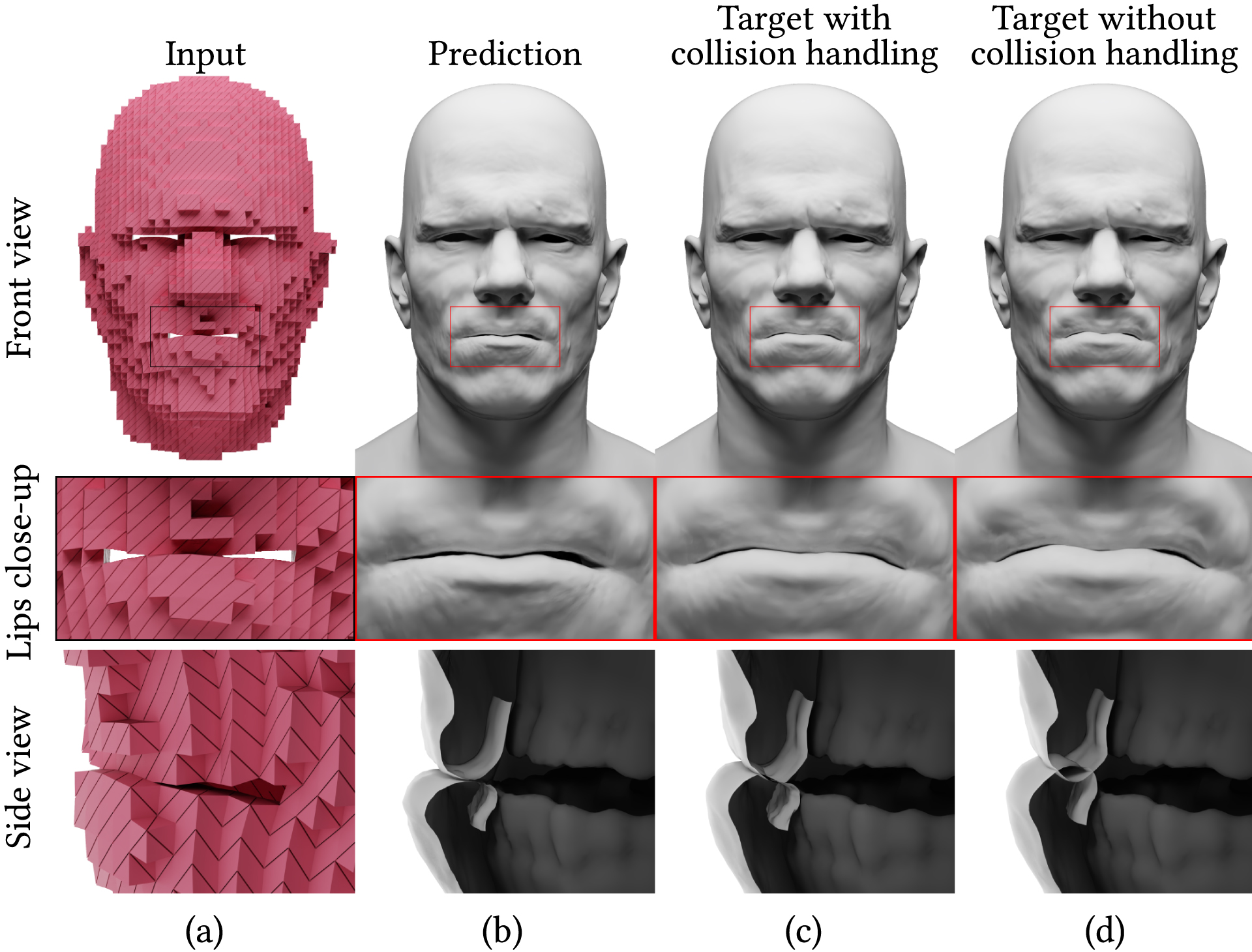}}
    {\nvidia{}}
    \caption{An example of \textit{complete} collision resolution: The prediction of our framework (b) on a test performance (a) has collisions resolved. The performance (when simulated in high resolution) with and without collision handling is shown in (c) and (d), respectively. Notice that when the penetration is low, collisions are resolved in the prediction. \copyright NVIDIA}
    \label{fig:eval_lips_success}
\end{figure}

\begin{figure}[!htb]
    \centering
    \copyrightbox[r]
    {\includegraphics[width=\linewidth]{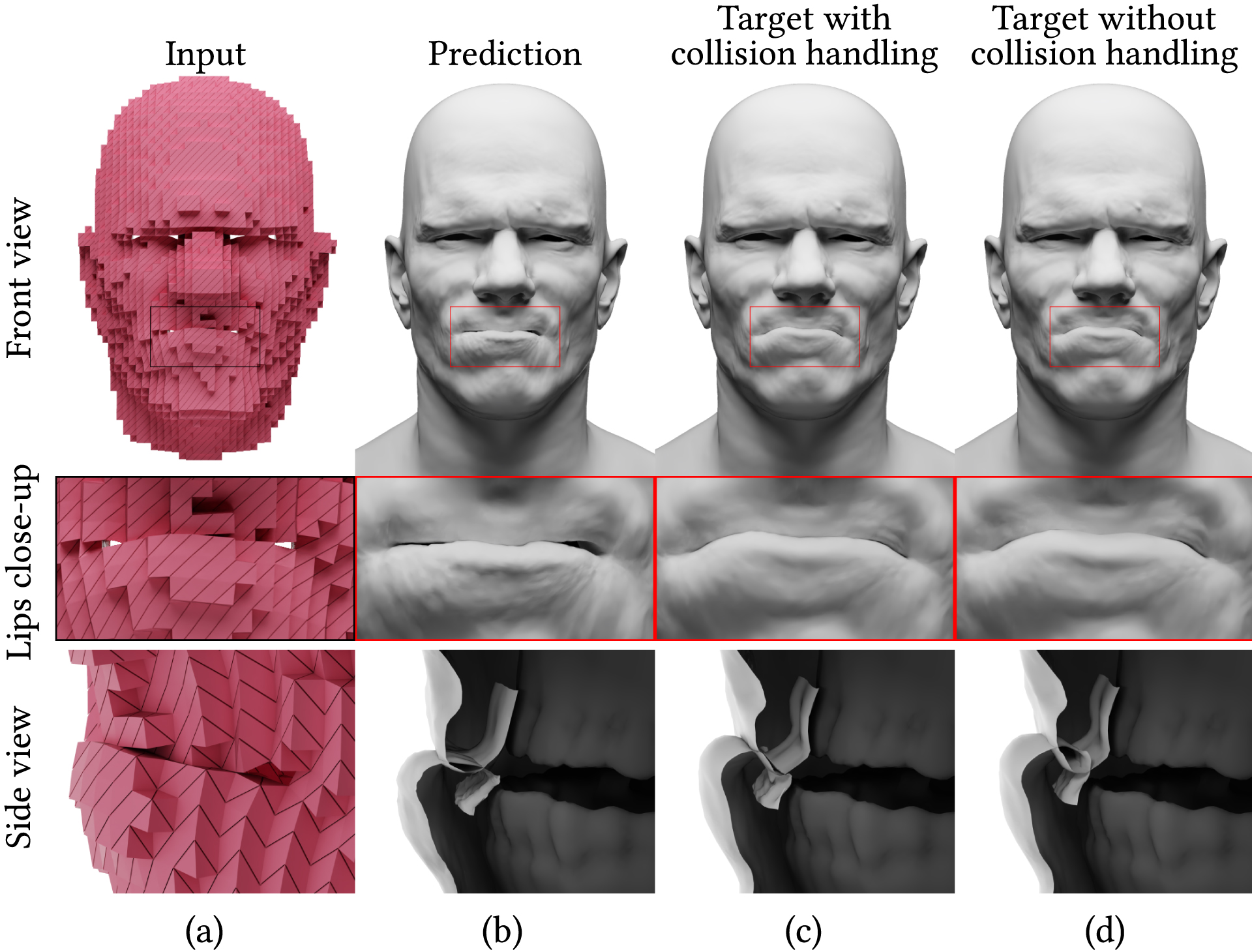}}
    {\nvidia{}}
    \caption{An example of \textit{partial} collision resolution: The prediction of our framework (b) on a test performance (a) has collisions partially resolved. The performance (when simulated in high resolution) with and without collision handling is shown in (c) and (d), respectively. Notice that when the penetration is higher, collisions are partially resolved in the prediction. \copyright NVIDIA}
    \label{fig:eval_lips_fail}
\end{figure}

We validate the qualitative performance of self-collisions by visualizing and comparing the predictions on the test set with two variants of the high-resolution surface, collision handling applied in the simulation (Figures \ref{fig:eval_lips_success}c and \ref{fig:eval_lips_fail}c) and omitted in the simulation (Figures \ref{fig:eval_lips_success}d and \ref{fig:eval_lips_fail}d). As mentioned in Section \ref{sec:data_gen}, we do not resolve self-collisions in the low-resolution simulations, but only in the high-resolution simulations. We observe that the trained model is able to predict high-resolution performances with partial collision resolution, depending on the degree of collision (or penetration). Figure \ref{fig:eval_lips_success} illustrates one such test set performance where the prediction from our model (Figure \ref{fig:eval_lips_success}b) does not have lip self-collisions when the penetration is low (Figure \ref{fig:eval_lips_success}d). Conversely, when the penetration is high, as shown in Figure \ref{fig:eval_lips_fail}d, the prediction has collisions partially resolved (Figure \ref{fig:eval_lips_fail}b). We also highlight that we do not include any additional penalty for collisions during training (which is an avenue for future work), and the model has approximated partial collision resolution from the high-resolution performances in the training dataset.

\subsection{Unseen external forces - embedded surface}\label{appendix:unseen_forces_emb}
\begin{figure}[]
\begin{subfigure}{\linewidth}
    \copyrightbox[r]
    {\includegraphics[width=\linewidth]{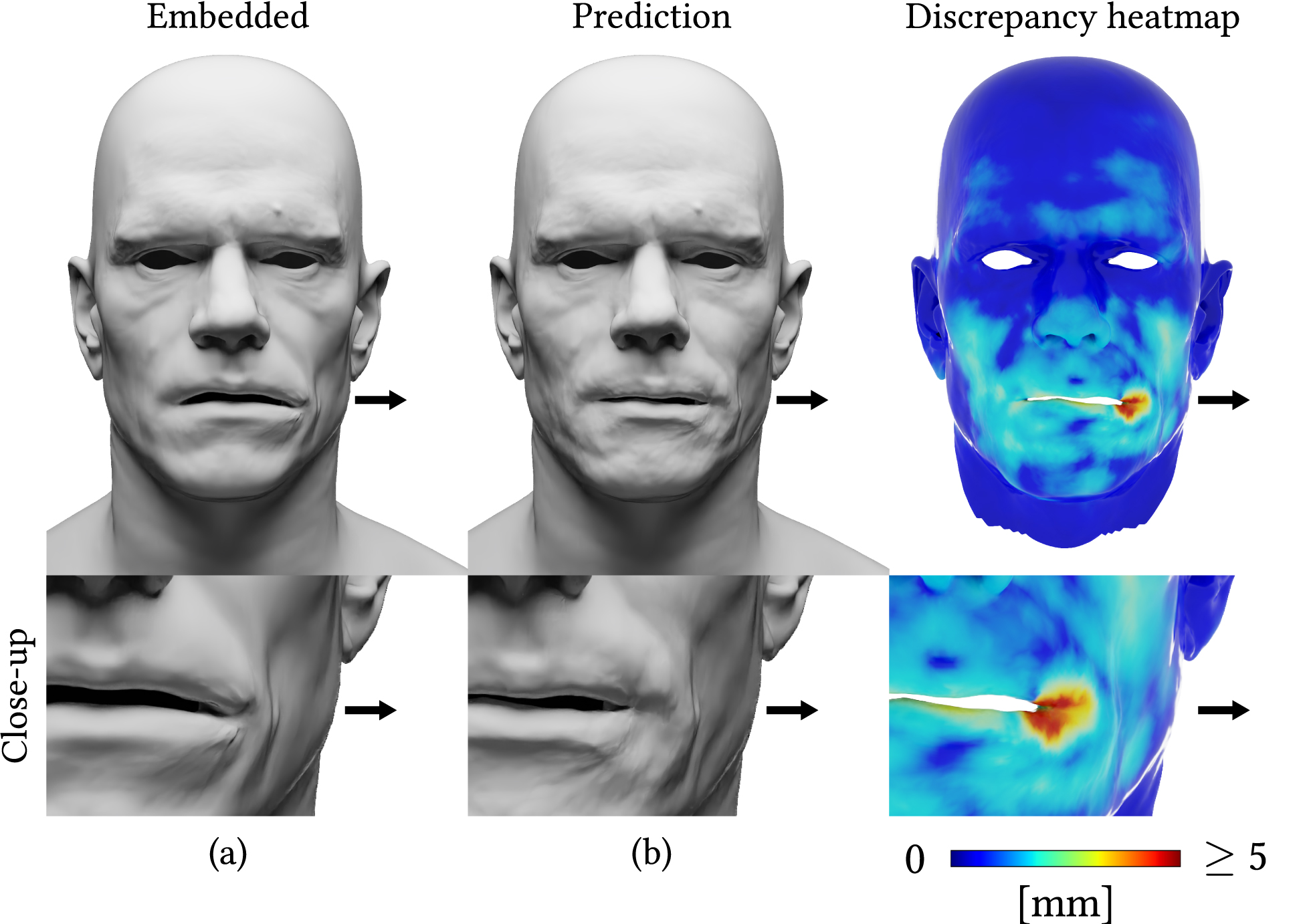}}
    {\nvidia{}}
\end{subfigure}
\bigskip
\begin{subfigure}{\linewidth}
    \copyrightbox[r]
    {\includegraphics[width=\linewidth]{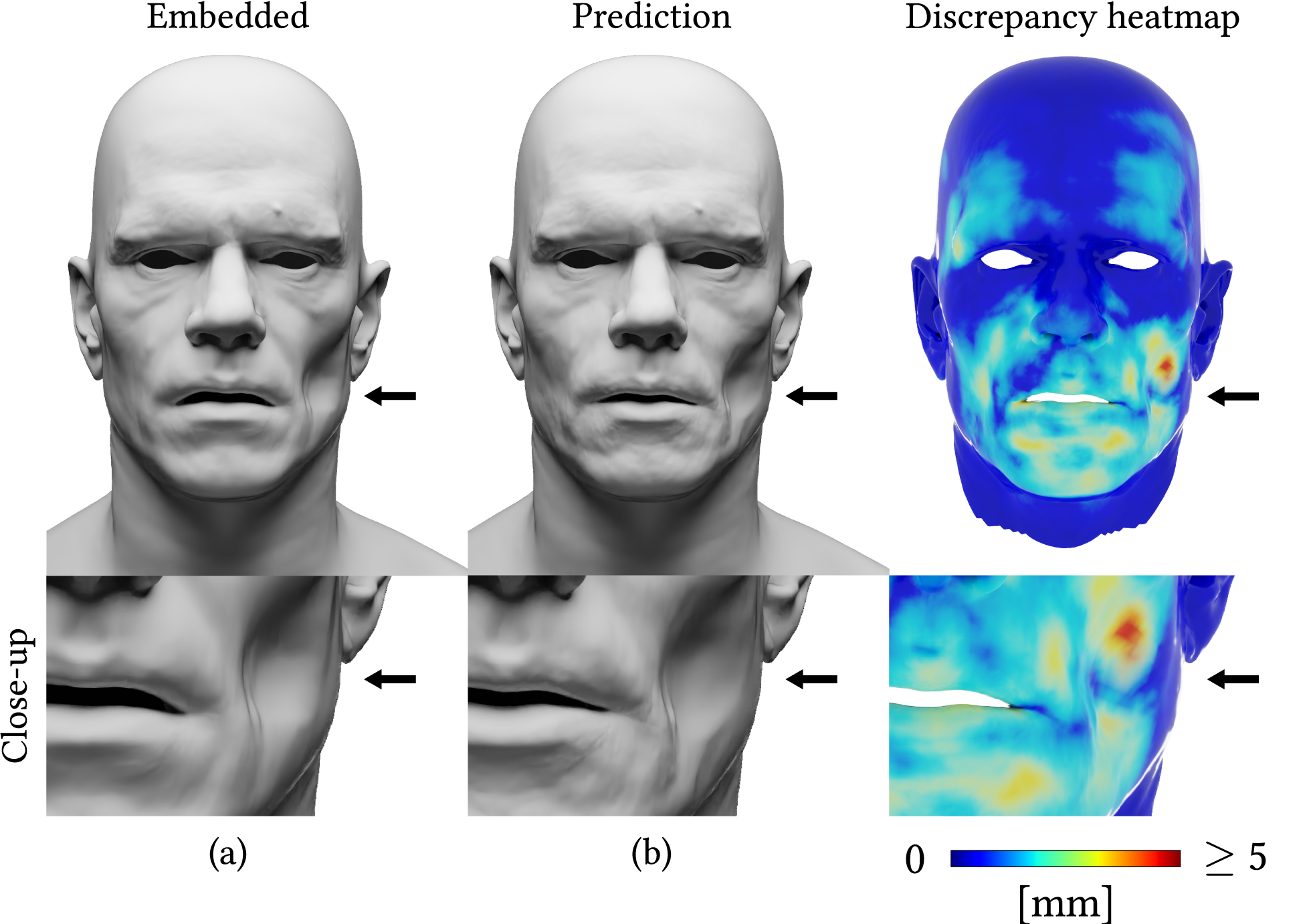}}
    {\nvidia{}}
\end{subfigure}
\vspace{-4em}
\caption{Visualization of the surface mesh embedded in the low-resolution simulation mesh undergoing unseen external forces (a) and our prediction of the target mesh (b), respectively (see Section \ref{appendix:unseen_forces_emb}). \copyright NVIDIA}
\label{fig:unseen_physics_heatmap}
\end{figure}

\edit{In Figure \ref{fig:unseen_physics_heatmap} (in addition to Figure \ref{fig:unseen_physics}), we visualize the surface mesh (Figure \ref{fig:unseen_physics_heatmap}a) embedded in the low-resolution simulation mesh undergoing unseen external forces for side-by-side comparisons with the predicted mesh (Figure \ref{fig:unseen_physics_heatmap}b).
We also visualize heatmaps showing deformation discrepancies between the embedded and predicted meshes by computing their per-point Euclidean distances.
Note that the embedded mesh is \textit{not} the target mesh for prediction but only provided as a visual reference.}

\begin{figure}[htb]
    \vspace{2em}
    \centering
    \includegraphics[width=\linewidth]{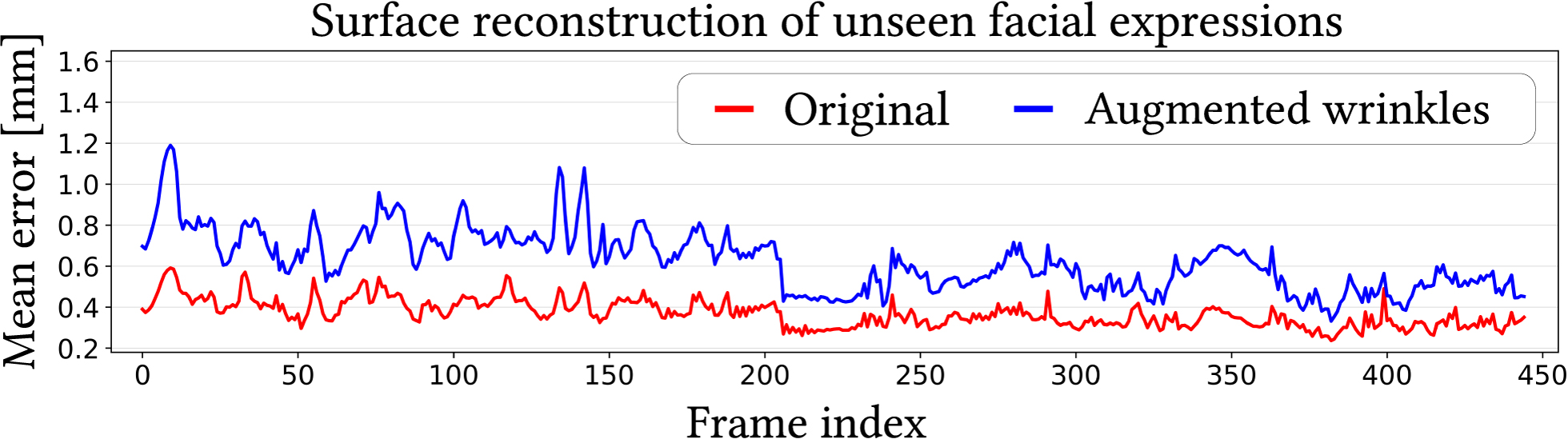}
    \caption{Frame-wise mean surface reconstruction error of unseen facial expressions without (i.e., original) vs. with augmented wrinkles.}
    \label{fig:wrinkles_loss}
\end{figure}

\begin{table}[htb]
    \vspace{1em}
\caption{Descriptive statistic measures of frame-wise mean surface reconstruction errors on unseen facial expressions without (i.e., original) vs. with augmented wrinkles.}
    \begin{tabularx}{\linewidth}{c||Y | Y}
    \hline
    [mm] & Original & Augmented Wrinkles\\
    \hline
    Mean&0.37 &0.62 \\
    \hline
    Median&0.36 &0.61 \\
    \hline
    Std.&0.07 &0.15 \\
    \hline
    Max.&0.59 &1.19 \\
    \hline
    Min.&0.24 &0.33 \\
    \hline
    \end{tabularx}
    \label{table:wrinkles}
\end{table}

\begin{figure*}
    \centering
    \copyrightbox[r]
    {\includegraphics[width=\linewidth]{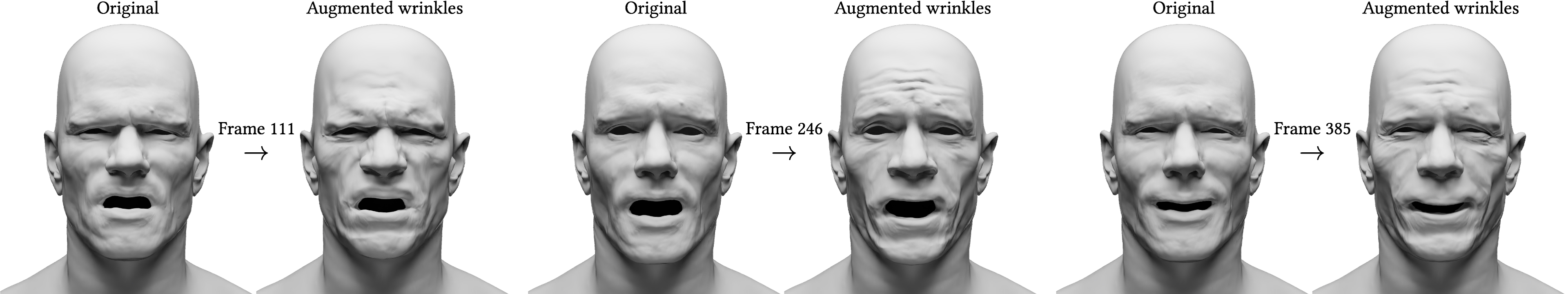}}
    {\nvidia{}}
    \caption{The augmented wrinkles (right) are incorporated by applying wrinkle blendshapes onto the original surface mesh (left). \copyright NVIDIA}
    \label{fig:wrinkles_before_after}
\end{figure*}


\begin{figure*}[!h]
    \vspace{2em}
    \centering
    \copyrightbox[r]
    {\includegraphics[width=\linewidth]{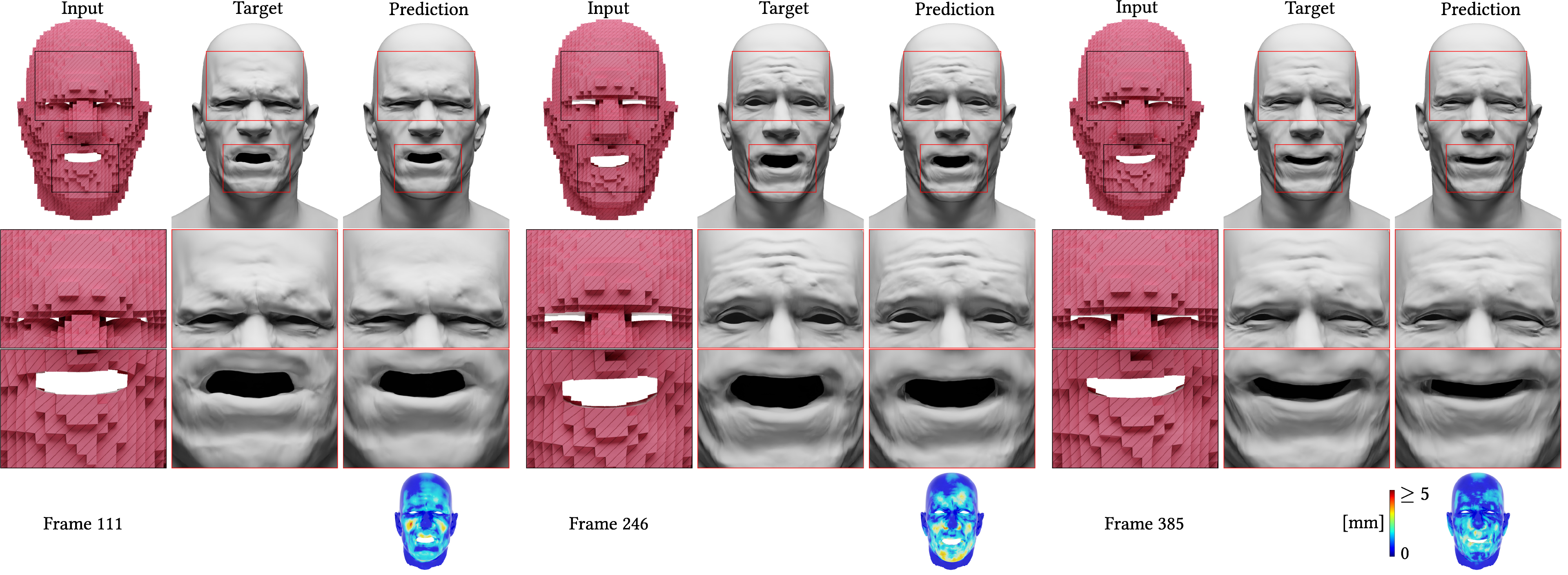}}
    {\nvidia{}}
    \caption{Reconstructions of the unseen facial expressions trained on the dataset with augmented wrinkles (see Section \ref{appendix:wrinkles}). \copyright NVIDIA}
    \label{fig:wrinkles_recon}
\end{figure*}

\subsection{Experiment with augmented wrinkles}\label{appendix:wrinkles}
\edit{We evaluate the quality of reconstructed faces inferred by our model trained using two types of target surface meshes: the original surface mesh and one with additional wrinkle details. The augmented wrinkles are incorporated by applying wrinkle blendshapes onto the original surface mesh (see Figure \ref{fig:wrinkles_before_after}).}
\edit{Using the same low-resolution volumetric input mesh and hyperparameters from Section \ref{ss:eval_recon}, we train our model until convergence to predict the wrinkle-augmented high-resolution surface mesh.}

\edit{We visually compare the predicted meshes generated by our model with the target mesh in Figure \ref{fig:wrinkles_recon}. Our model effectively captures visually reasonable details of augmented wrinkles, particularly in areas around the forehead, eyes, and mouth, where wrinkles are most pronounced.
Additionally, we plot the frame-wise mean reconstruction errors in Figure \ref{fig:wrinkles_loss} and provide a summary in Table \ref{table:wrinkles}. 
While the mean errors increased overall, we consider this reasonable considering the additional high-resolution details our model must infer given the equivalent capacity of our neural network model.}

\end{document}